\documentclass[twocolumn]{aastex631}
\usepackage{hyperref,amsmath}

\newcommand\g{\; {\rm g}}
\newcommand\s{\; {\rm s}}
\newcommand\pcc{\;{\rm cm}^{-3}}
\newcommand\Msun{\; {\rm M}_{\odot}}
\newcommand\kms{\; {\rm km}\,{\rm s}^{-1}}
\newcommand\ergpers{\; {\rm erg}\,{\rm s}^{-1}}
\newcommand\erg{\; {\rm erg}}

\newcommand\cm{\;{\rm cm}}

\newcommand\Myr{\;{\rm Myr}}

\newcommand\pc{\;{\rm pc}}
\newcommand\kpc{\;{\rm kpc}}
\newcommand\sfrunit{\Msun \pc^{-2} \Myr^{-1}}

\newcommand\Punit{\pcc\,{\rm K}}
\newcommand\surfunit{\Msun\,{\rm pc^{-2}}}

\newcommand\K{\;{\rm K}}

\newcommand\simgt{\lower.5ex\hbox{$\; \buildrel > \over \sim \;$}}
\newcommand\simlt{\lower.5ex\hbox{$\; \buildrel < \over \sim \;$}}
\newcommand\pderiv[2]{\frac{\partial {#1}}{\partial {#2}}}

\newcommand\torb{t_{\rm orb}}
\newcommand{\tdep}{t_{\rm dep}}
\newcommand{\tdyn}{t_{\rm ver}}

\newcommand{\Upsth}{\Upsilon_{\rm th}}
\newcommand{\Upsturb}{\Upsilon_{\rm turb}}
\newcommand{\Upstot}{\Upsilon_{\rm tot}}

\newcommand{\PDE}{P_{\rm DE}}
\newcommand{\Pth}{P_{\rm th}}
\newcommand{\Pthtwo}{P_{\rm th,2p}}

\newcommand{\Pturb}{P_{\rm turb}}
\newcommand{\Pturbtwo}{P_{\rm turb,2p}}

\newcommand{\Pmag}{P_{\rm mag}}
\newcommand{\Pimag}{\Pi_{\rm mag}}
\newcommand{\Ptot}{P_{\rm tot}}
\newcommand{\Ptottwo}{P_{\rm tot,2p}}

\newcommand{\seff}{\sigma_{\rm eff}}

\newcommand{\SSFR}{\Sigma_\mathrm{SFR}}
\newcommand{\Sgas}{\Sigma_\mathrm{gas}}

\defcitealias{OML_10}{OML10}
\defcitealias{OS_11}{OS11}
\defcitealias{KKO_2011}{KKO11}
\defcitealias{KOK_2013}{KOK13}
\defcitealias{Kim_Ostriker2017}{KO17}

\accepted{\today}

\shorttitle{Pressure-Regulated, Feedback-Modulated Star Formation}
\shortauthors{Ostriker \& Kim}

\graphicspath{{../}{figures/}}

\begin{document}

\title{Pressure-Regulated, Feedback-Modulated Star Formation In Disk Galaxies}

\author[0000-0002-0509-9113]{Eve C. Ostriker}
\affiliation{Department of Astrophysical Sciences, Princeton University, Princeton, NJ 08544, USA}

\author[0000-0003-2896-3725]{Chang-Goo Kim}
\affiliation{Department of Astrophysical Sciences, Princeton University, Princeton, NJ 08544, USA}

\email{eco@astro.princeton.edu,cgkim@astro.princeton.edu}

\correspondingauthor{Eve C. Ostriker}

\begin{abstract}
The star formation rate (SFR) in galactic disks depends on both the
quantity of available interstellar medium (ISM) gas and its physical
state.  Conversely, the ISM's physical state depends on the SFR,
because the ``feedback'' energy and momentum injected by
recently-formed massive stars is crucial to offsetting losses from
turbulent dissipation and radiative cooling.  The ISM's physical state
also responds to the gravitational field that confines it, with
increased weight driving higher pressure.  In a quasi-steady state, it
is expected that the mean total pressure of different thermal phases
will match each other, that the component pressures and total pressure
will satisfy thermal and dynamical equilibrium requirements, and that
the SFR will adjust as needed to provide the requisite stellar
radiation and supernova feedback.  The pressure-regulated,
feedback-modulated (PRFM) theory of the star-forming ISM formalizes
these ideas, leading to a prediction that the SFR per unit area,
$\SSFR$, will scale nearly linearly with ISM weight $\cal W$. In terms
of large-scale gas surface density $\Sgas$, stellar plus dark matter
density $\rho_{\rm sd}$, and effective ISM velocity dispersion
$\sigma_{\rm eff}$, an observable weight estimator is ${\cal
  W}\approx\PDE=\pi G\Sgas^2/2+\Sgas(2G\rho_{\rm sd})^{1/2}\sigma_{\rm
  eff}$, and this is predicted to match the total midplane pressure
$\Ptot$.  Using a suite of multiphase magnetohydrodynamic simulations
run with the TIGRESS computational framework, we test the principles
of the PRFM model and calibrate the total feedback yield $\Upstot
=\Ptot/\SSFR \sim 1000\kms$, as well as its components.  We compare
results from TIGRESS to theory, previous numerical simulations, and
observations, finding excellent agreement.

\end{abstract}

\keywords{ Star formation (1569), Interstellar medium (847),
Stellar feedback (1602), Magnetohydrodynamical simulations (1966)}

\section{Introduction}\label{sec:intro}

\subsection{Feedback and Star Formation/ISM Regulation}

The importance of star formation ``feedback'' to energetics of the
interstellar medium (ISM) has been appreciated throughout the modern
history  of astronomy
\citep[see e.g.~Spitzer's chapter in][for a mid-20th-century view]{Middlehurst_1968}, with the idea of
star formation self-regulation a corollary: 
the energy returned by stars to their surroundings may prevent or limit
gravitational collapse and further
star formation.  From observations, it is evident that young,
massive stars impart large amounts of energy to their near and far
environments through radiation, winds, and supernovae (SNe); and that the
generic outcome of energy injection is heating, acceleration,
and dispersal of gas
over an increased volume.

In  theoretical work, investigation of star formation feedback and its 
implications for self-regulation has developed in scope and complexity 
through the years.  Idealized spherical solutions for
dynamical evolution of the ISM driven by sources of stellar energy
include the expansion of an \ion{H}{2} region \citep{Spitzer_1978},
the expansion of a single SN remnant adiabatically
\citep{Taylor_1950,Sedov_1959} and with cooling
\citep[e.g.][]{Ostriker_McKee1988}, 
generalizations of this in various limits 
for continuous energy input modeling stellar wind sources \citep[e.g.][]{Avedisova_1972,Steigman_1975,Weaver_1977,Lancaster_2021}
or a series
of SNe \citep[e.g.][]{McCray_1987,Kim_Ostriker_Raileanu2017}.
Early theory of  semi-confined (``blister'')
\ion{H}{2} regions also provided estimates of evaporation rates and outflow
velocities of ionized gas \citep{Whitworth_1979}.
Radiation pressure on dust leads to gradients of radiation and gas pressure
within \ion{H}{2} regions and increases the net force on the surrounding shell
\citep{Draine_2011},
and spherical solutions for \ion{H}{2} region expansion accounting for this
have been developed \citep[e.g.][]{Martinez_2014,KKO_2016,Rahner_2017,Akimkin_2017}.
These idealized spherical and hemispherical \ion{H}{2} region solutions 
have also been used to estimate giant molecular cloud (GMC)
lifetimes and star formation
efficiencies as limited by feedback
\citep[e.g.][]{Elmegreen_1983,Franco_1994,Matzner_2002,Krumholz_Matzner2009,Fall_2010,MQT_2010,KKO_2016}.
In recent years, numerical simulations have been applied to model more realistically effects of stellar energy inputs in turbulent,
self-gravitating models of individual GMCs that are highly inhomogeneous.
Under most circumstances photoionization and radiation
pressure effects dominate on GMC scales, and radiation hydrodynamic simulations
\citep{Kim_JG2018,Kim_JG_Ostriker_Filippova2021,He_Ricotti2019,Fukushima_2020,Fukushima_2021}
have demonstrated that clouds are dispersed on realistic
timescales and yield realistic lifetime star formation efficiencies,
when compared to empirical
estimates based on spatial correlations of molecular gas and star formation
tracers \citep[e.g.][]{Chevance2020a,KimChevance21}.   
{A recent comprehensive review of feedback effects in GMCs
  is given in \citet{PPVII_GMC2022}.}

In galaxies with significant gas content (spirals and dwarfs),
the interstellar medium
(ISM) takes on a disk configuration on large scales.  These disk
galaxies are clearly
long-lived systems with ongoing star formation. Thus, unlike the
situation for individual star-forming molecular clouds, feedback from
star formation does not disperse the whole of the ISM disk on a dynamical
timescale (although significant ISM
 material can be carried away in galactic winds -- see e.g.~reviews of 
 \citealt{Veilleux_2005,Veilleux_2020}).  Rather, the feedback returned from
young, massive stars contributes to the overall energetic state of the ISM,
with important consequences for dynamics on a range of scales.

For ISM disks in rotationally-supported galaxies,
a major focus over the years has been on consideration
of large-scale (exceeding the ISM scale height $h_{\rm gas}$)
gravitational instabilities and how they are limited.  
Starting with \citet{Goldreich_Lynden-Bell1965},
many theoretical investigations have framed star formation
regulation in terms of processes that maintain a minimum effective
velocity dispersion and therefore keep the Toomre parameter $Q\equiv \kappa c_{\rm eff}/(\pi G \Sgas)$ close
to unity; here $\kappa$ is the epicyclic frequency, $c_{\rm eff}$ is the effective sound speed, and $\Sgas$ is the large-scale gas surface density
(e.g. \citet{Binney_Tremaine2008} Chapter 6; see also \citet{Toomre_1964} for the corresponding analysis for a stellar disk).
{If $Q$ is close to unity,}
large-scale self-gravitating instability is expected to be marginal; see 
  \citet{Elmegreen_2002,McKee_Ostriker2007} for reviews of earlier work, and below for some more recent contributions.\footnote{Connected to this,
a suggestion of long standing is that
self-gravitating instabilities lead to conversion of
potential energy to turbulence while driving secular gaseous inflow
\citep[e.g.][]{von_Weizsacker_1951,Goldreich_Lynden-Bell1965,Fleck_1981,Elmegreen_etal2003,Agertz_2009,Krumholz_etal2018}.
Self-gravitating instabilities (as well as magnetorotational
instabilities -- see e.g. \citet{Kim_WT_Ostriker2003,Piontek_Ostriker2005})
can certainly contribute to
motions within the ISM. The idea that
{self-gravitating instabilities are}
the main source of ISM
turbulence is however challenged by simulations that show gravity-driven
turbulence is transonic only when $Q$ is small and runaway collapse is occurring
\citep[e.g.][]{Kim_Ostriker2007}, and that self-gravitating
ISM models with weak or no feedback have
unrealistically high rates of collapse and 
star formation \citep[e.g.][]{Shetty_Ostriker2008,Hopkins_2011}.
}
However, large-scale gravitational instability is only one among 
several mechanisms that could collect mass into GMCs within a
realistic turbulent, magnetized,
multiphase ISM \citep[see e.g.][]{Dobbs2014}, and there
does not appear to be a link between observed star formation efficiency
and the Toomre parameter \citep{Leroy2008}. In the Milky Way,
observed GMC spin axes
appear to be randomly oriented rather than aligning with the direction of
galactic angular momentum \citep{Koda_2006}, which would argue against GMC
formation being driven solely by large-scale gravitational instability.

Whether disks are susceptible to large-scale (essentially two-dimensional)
instabilities or not, global evolutionary
timescales ($\sim 10^9$yr) are much larger than the turbulent crossing time
(comparable to
the vertical dynamical time, $\sim 10^7$yr)
or the cooling time of warm or cold ISM gas ($\sim 10^6$ or $10^4$yr,
respectively).
It is therefore imperative  to understand what
sustains the turbulent and thermal pressure in the face of rapid dissipation
and cooling, and what implications this may have for star formation.
Since a local deficit
of thermal and turbulent pressure in a region smaller than the disk scale
height $h_{\rm gas}$
could lead to gravitationally-driven contraction  even if the disk is gravitationally stable on scales $h_{\rm gas}$,
regulation of the ISM pressure and star formation in disks may be primarily a
local process that depends on energy inputs from recently-formed (massive)
stars, regardless of larger-scale galactic dynamics.

In the two-phase ISM model of \citet{FGH_1969} as updated by
\citet{Wolfire_1995}, most of the ISM mass is divided between cold
($T\sim 10^2\K$) and
warm ($T\sim 10^4\K$)
phases at the same pressure \citep[see][for \ion{H}{1} observations]{Heiles_Troland2003,Murray_2018}, with photoelectric heating by stellar
FUV incident on small grains responsible for maintaining the thermal pressure.
Based on empirical estimates of the heating rate in
the Milky Way, models show that the thermal
pressure of two atomic phases could be sustained at all radii
(\citealt{Wolfire_2003}; see also \citealt{Dickey_2009} for observations).  The three-phase ISM model of \citet[][]{McKee_Ostriker1977}
argues that for the observed SN rate in the Milky Way,  hot ISM gas with
pressure consistent with observations could be maintained \citep[see also][]{Cox_Smith1974,McCray_1979}. \citet{McKee_Ostriker1977} also note that
balancing collisional losses in a warm-cold cloud distribution
with  kinetic energy inputs from SNe,
the velocity dispersion would be roughly  in agreement with
observations.  \citet{TQM_2005} argued that 
both the radiation pressure and the 
turbulent pressure in cold gas driven by SNe should increase linearly
with the star formation rate per unit area, and that if disks
are self-regulated such that $Q\sim 1$, the star formation rate in normal disks
would scale as the square of the gas surface density
\citep[see also][for related work]{Faucher_2013,Hayward_Hopkins2017,Orr_2018}.
In an approach to large-scale self-regulation that does not presuppose $Q\sim 1$, 
\citet{Franco_Shore1984} argued that
the star formation rate can be estimated by assuming stellar winds and
SNe supply the ISM's energy, with shells expanding until
their kinetic and self-gravitational energies are comparable; under
this hypothesis they found that the star formation rate per unit gas mass
would be proportional to the square root of the gas density. Other
star formation self-regulation models consider the porosity of hot gas produced
by SN feedback as the controlling 
lever \citep[e.g.][]{Silk_2001,Dekel_2019}.

\subsection{Thermal and Dynamical Equilibrium Model
of Star Formation/ISM Co-Regulation
}\label{sec:intro_TDE}

Adopting the assumptions that the main source of energy in the ISM disk is
young stars and that the conversion from gas to stars is slow,
to obtain a general theoretical model 
it is conceptually useful to think of a local patch in
the star-forming ISM as a quasi-equilibrium ecosystem that has
approximately plane-parallel vertical structure (in a
time-averaged sense).
Predictions for the star formation rate as well as the volumetric ISM
quantities (pressure and density) then emerge from considering the
requirements for equilibrium to be maintained self-consistently
\citep[][hereafter \citetalias{OML_10} and \citetalias{OS_11}]{OML_10,OS_11}.
In particular, treating the ISM disk as a fluid system with source terms and
spatially averaging horizontally,
time-independent versions of the energy equation and vertical momentum
equation should be satisfied.  The pressure plays a special role as it
appears in the energy density, energy flux, and momentum flux.  As we
shall describe in more detail in \autoref{sec:theory}, from the
vertical momentum equation the total pressure at the
midplane 
\citep[see e.g.][for consideration of magnetic terms]{Boulares_Cox1990,Piontek_Ostriker2007}
must balance the weight of the ISM, which depends on gas
surface density and the stellar and
dark matter potentials.   The thermal portion of the pressure must also
satisfy balance between heating and cooling, where the heating is
proportional to the star formation rate if the main source is 
stellar radiation.
At the same time, the midplane turbulent pressure
is sourced by the vertical momentum flux from expanding individual
SN remnants or superbubbles, which is proportional to the star formation
rate.  Satisfying
the energy equation and vertical momentum equation simultaneously
then requires the
star formation rate per unit area in the disk, $\SSFR$, to
be proportional to the weight of the ISM per unit area, $\cal W$.
The latter is often referred to as the dynamical equilibrium pressure
since these  must balance in equilibrium.  Here, we shall use the expression
``$\PDE$'' to refer to a commonly-adopted estimate for the weight $\cal W$ (see
\autoref{sec:weight}).

Based on the above physical considerations,
\citetalias{OML_10} and \citetalias{OS_11} formulated the theory of
pressure-regulated, feedback-modulated (PRFM)\footnote{We here introduce the ``PRFM'' appellation to reflect the fact that pressure is regulated -- i.e. dictated -- by the laws of momentum conservation and gravity in the vertical direction, while feedback from star formation modulates -- i.e. tunes -- the component energy densities of the ISM gas to match the required total pressure.}
star formation, with
\citetalias{OML_10}
addressing heating-cooling balance and comparing the prediction of
$\SSFR$ to observations of atomic-dominated regions of galaxies (where
thermal pressure is significant), and \citetalias{OS_11} focusing on turbulent
pressure driving and comparing the prediction of $\SSFR$ to observations of
molecule-dominated galactic centers and starburst galaxies (where
thermal pressure is negligible).
\citet[][hereafter \citetalias{KKO_2011}]{KKO_2011} and
\citet[][hereafter \citetalias{KOK_2013}]{KOK_2013}
developed hydrodynamic simulations within the shearing-box local disk  
framework to test the PRFM theory, with star formation (following
self-gravitating collapse) setting the rate of photoelectric heating
and SN momentum injection to warm-cold ISM gas.  These simulations
covered a range of gas surface density and
stellar potential such that
the resulting $\SSFR$ varied over two orders of magnitude, and 
successfully validated the hypothesis that a quasi-steady state
is reached with pressure simultaneously satisfying the 
equilibrium requirements from the energy and vertical
momentum  equations.  \citetalias{KKO_2011} and \citetalias{KOK_2013}
demonstrated that $\SSFR$ is approximately proportional to either the ISM
weight or midplane total pressure (also showing these are equal)
and computed the feedback yields  (ratios of pressure components 
to $\SSFR$; see \autoref{sec:yield_theory}), showing that the thermal yield is
consistent with the prediction of \citetalias{OML_10} and the turbulent
yield is consistent with the prediction of \citetalias{OS_11}.
Magnetic fields were included in the
numerical simulations of \citet{Kim_Ostriker2015b}, validating 
the magnetohydrodynamic (MHD) generalization  of the PRFM
theory and demonstrating
that magnetic pressure and tension terms are comparable in magnitude
to thermal and turbulent ones.

\subsection{Observational Tests}\label{sec:intro_obs}

Several observational studies have combined tracers
of molecular and/or atomic gas with tracers of star formation to test
several aspects of the model predictions from \citetalias{OML_10,OS_11,KOK_2013}
for the relationships among pressure, weight, and star formation rate.
In \citetalias{OML_10}, the theoretical prediction for $\SSFR$ was compared
to observations of a set of 11 nearby spiral galaxies, using  
azimuthally averaged radial profiles of molecular and atomic gas
(based on CO $J = 2-1$ and $1-0$ and \ion{H}{1} 21 cm maps, respectively),
the stellar surface density and dark matter density (based on 3.6$\mu$m
maps and gas rotation curves, respectively), and star formation surface
densities (based on FUV  and 24$\mu$m maps), all drawn from  \citet{Leroy2008}.
In \citetalias{OS_11}, the theoretical prediction for $\SSFR$
was compared to observations for a set of starbursts (both local
and higher redshift) collected by \citet{Genzel_2010}.
Good agreement was demonstrated for
both ``normal'' and ``starburst'' conditions.

\citet{Herrera-Camus_2017} used \textit{Herschel} observations of the \ion{C}{2}
158$\mu$m line in a set of 31 galaxies to confirm that the thermal pressure is consistent with the requirement for 
two-phase equilibrium of atomic gas
\citep[][see also \autoref{sec:theory}]{Wolfire_2003} for the expected
radiation field, given the star formation rate as measured from a combination
of 24$\mu$m and H$\alpha$ observations.   \citet{Herrera-Camus_2017} also
showed that the relationship between $\PDE$ and $\SSFR$ is consistent
with the numerical results of \citetalias{KOK_2013}.  Using data from 28 PHANGS
galaxies, \citet{Sun_2020a} showed
that the turbulent pressure in molecular gas, as estimated at $\sim 100$pc
scale from ALMA observations of CO $J=2-1$ \citep{Leroy_2021}, increases
approximately linearly with the $\sim$kpc-scale $\SSFR$ as
estimated from a combination of
near-UV and $12\mu$m data \citep{Leroy_2019}, consistent with
\citetalias{OS_11}.  
\citet{Sun_2020a} also found a relationship between $\PDE$ and $\SSFR$
quantitatively similar to the PRFM theory prediction, but with slightly
shallower slope
than found in the simulations of \citetalias{KOK_2013} (0.8 \textit{vs.} 1.1).
\citet{Barrera-Ballesteros_2021}, using observations of 96 galaxies
mapped at $\sim$ kpc scale in CO for the EDGE survey
\citep{Bolatto_2017} and in optical emission for the CALIFA IFU survey
\citep{Sanchez_2012}, found good agreement with both the slope and
coefficient of the $\PDE$ vs. $\SSFR$ relation from \citetalias{KOK_2013} when using
an $\alpha_\mathrm{CO}(Z,\Sigma_*)$ relation  based on \citet{Bolatto_2013},
or a slightly shallower slope ($0.96$) when using constant  $\alpha_\mathrm{CO}$.

The above studies of normal galaxies cover the range $\SSFR \sim 10^{-3} -10^{-1} \sfrunit$, but similar results have also been found in the regime of
higher $\SSFR$. 
Using ALMA and SINFONI observations with $\sim$ kpc resolution of two disk-like
starbursts at $z\sim 0.1$, \citet{Molina_2020} found that the $\SSFR$
and estimated $\PDE$ are intermediate between those of PHANGS and a
set of local
ULIRGs with high-resolution ALMA observations of CO (1-0) and radio continuum
(tracing free-free emission) as collected by \citet{Wilson_2019};  they
noted however that the $z\sim 0.1$ starburst $\PDE$ may be slightly
overestimated because they assumed the stellar and gas scale heights are the
same (rather than the former being larger).  \citet{Molina_2020} found that
combining all data (over 6 orders of magnitude in pressure), $\SSFR$ and
$\PDE$ are related by a power law with slope 0.8.  The true slope  may be
closer to unity, however, because they adopted an assumption of constant 
$\alpha_\mathrm{CO}$ rather than $\alpha_\mathrm{CO}$ decreasing at higher
$W_\mathrm{CO}$ \citep[cf.][]{OML_10,Narayanan_2012,Gong_2020}; constant
$\alpha_\mathrm{CO}$  tends to overestimate $\Sigma_{\rm mol}$ and $\PDE$ at
high $W_\mathrm{CO}$. From the $z\sim 0.1$
DYNAMO sample in a similar regime of $\SSFR$ and $\PDE$ to 
the \citet{Molina_2020} sample, \citet{Fisher_2019} and \citet{Girard_2021}
also found results consistent with a power-law slope 0.8.  

\subsection{Testing Theory with  TIGRESS and other Star-Forming,
  Multiphase ISM Simulations}\label{sec:intro_sim}

The computational studies of \citetalias{KKO_2011,KOK_2013}, \citet{Kim_Ostriker2015b} (and \citetalias{OS_11},
\citet{Shetty_Ostriker2012} for the starburst regime) validated the
basic principles of the PRFM theory for the warm and cold gas that represents
the majority of the ISM mass. However, these were not complete models of the
three-phase ISM
since momentum from SNe was injected ``by hand'' in warm-cold gas  (a similar
approach has been adopted in many galaxy formation simulations).
In reality, in a given SN remnant the momentum deposited in ambient gas
increases in time as the remnant
initially expands adiabatically, and after the onset of cooling the 
leftover hot gas joins the hot phase of the ISM. The combined hot gas from many SN remnants fills a significant fraction of the ISM volume.
\citet[][hereafter \citetalias{Kim_Ostriker2017}]{Kim_Ostriker2017}
introduced the TIGRESS\footnote{{\bf T}hree-phase {\bf I}nterstellar medium in
  {\bf G}alaxies {\bf R}esolving {\bf E}volution with {\bf S}tar formation and
  {\bf S}upernova
  feedback}  computational 
framework 
built on the {\it Athena} MHD code \citep{Stone_2008,Stone_Gardiner2009},
in which
warm and cold gas are treated similarly to our previous work, but 
SNe are modeled via energy rather than momentum  injection and expanding
individual remnants form cool outer shells
when the leading shock front drops below $\sim 200\kms$
\citep{Kim_Ostriker2015a}.  With this approach, the 
Sedov-Taylor stage of evolution is resolved, thereby producing a hot ISM while
also allowing momentum injection to adapt to the local environment.

In a fiducial TIGRESS MHD simulation with background conditions similar to the
solar neighborhood,
\citetalias{Kim_Ostriker2017} showed that a self-regulated, quasi-equilibrium
state is reached with mean thermal, turbulent, and magnetic pressure
as well as $\SSFR\approx 0.005 \sfrunit$ 
in agreement with observed values\footnote{Observational estimates employing a variety of methods indicate a mean value $\SSFR\sim 0.003-0.005 \sfrunit$ for the solar neighborhood, but there is also evidence of significant bursts    \citep{Bertelli_Nasi_2001,Vergely_2002,Fuchs_2009,Tremblay_2014,Mor_2019,
    Ruiz-Lara_2020,Zari_2022}.}, and comparable warm
and hot volumes and pressures in the three-phase ISM
near the midplane.  In addition, the solar neighborhood TIGRESS simulation
demonstrated that correlated SNe produce superbubbles
\citep[see also][]{Kim_Ostriker_Raileanu2017}. These superbubbles
drive a warm-cold  fountain  flow reaching several kpc from the
midplane, and their breakout leads to a hot, fast galactic wind
\citep{Kim_Ostriker2018,Vijayan_2020}.  Confirming the previous
finding by \citet{Kim_Ostriker2015b} that weight and total pressure balance
as a function of height $z$ in warm-cold ISM MHD simulations,
\citet{Vijayan_2020} found that the same balance
is satisfied for the three-phase TIGRESS solar neighborhood simulation.
At the midplane, the turbulent, thermal, and magnetic terms are all comparable
in the warm gas, while thermal pressure dominates in the hot phase; total
pressures are comparable across thermal phases. The majority of
the volume is filled with warm gas below $|z| \sim 1 \kpc$, while hot
wind occupies the majority of the volume at higher
altitudes where warm fountain flow clouds
turn around \citep[see also][for analysis of the
  distribution and properties of photoionized gas]{Kado-Fong2020}.

Using the same TIGRESS computational framework, we have recently
explored a range of
conditions in galactic disks.  As we shall describe further in
\autoref{sec:theory} and \autoref{sec:numer},
the two essential ``background''  galactic disk
properties that may be varied as independent parameters are the total
surface  density of gas, $\Sgas$ and the midplane density of stars
{plus dark matter, $\rho_\mathrm{sd}$}. The star formation
rate per unit area and the distributions of ISM density, pressure,
velocity, and magnetic fields then emerge self-consistently.  
Based on seven different TIGRESS disk models with emergent
$\SSFR$ varying by four orders of magnitude,
\citet{Kim_Ostriker_etal2020a} presented a detailed investigation of the
dependence of multiphase outflow  properties on the star formation
rate and ISM properties.  There, the outflow mass, momentum, energy, and
metal loading factors, as well as velocities,
were separately measured for warm-cold gas and
hot gas, with scaling relations calculated as a function of $\SSFR$ and
midplane pressure or ISM weight.  In \citet{Kim_Ostriker_etal2020b},
these overall loading factors were combined
with measurements of the mass-loading PDFs to derive analytic joint probability
distributions (in velocity and  sound  speed) of mass, momentum, energy, and
metal loading
as a function of $\SSFR$.   

The TIGRESS implementation has also been
used to study effects of spiral arms on star formation rates and ISM dynamical
equilibrium \citep{Kim_WT_etal2020}, demonstrating the local validity of the
PRFM model, and showing that spurs can form downstream from arms due to
correlated feedback. Extending to galactic centers,
\citet{Moon_etal2021,Moon_2022}
applied the TIGRESS computational
framework to model star-forming nuclear rings created by bar-driven
inflows and showed that the PRFM theory is also satisfied in these  more
extreme environments, with the interesting twist that the ring mass adjusts as
needed for the star formation rate to match the inflow rate.

Simulations of the large-scale star-forming ISM with feedback have
also been conducted by a number of other groups.  Those most similar
to the TIGRESS models focus on a local ISM patch at $\sim$ kpc scale
\citep[see][and references
  therein]{Kannan_2020,Brucy_2020,Rathjen_2021}, as this affords
sufficient resolution to follow both gravitational collapse in cold
gas and the evolution and interaction of the hot gas
(produced by SN shocks) with other thermal phases.  One difference from the TIGRESS models
is that other simulations have generally been run for shorter timescales,
typically $\sim 100$ Myr; this may be compared to 700 Myr for the
TIGRESS solar neighborhood model presented in \citetalias{Kim_Ostriker2017}.
Because there may be a strong early
burst of star formation (depending on initial conditions and feedback
ingredients), simulations run for relatively short periods may not
have reached a quasi-steady equilibrium state.  For solar neighborhood
(unmagnetized) simulations with {\sc AREPO-RT} 
\citep{2010MNRAS.401..791S,2019MNRAS.485..117K}
that include just
SNe in comparison to simulations with SNe plus radiation,
\citet{Kannan_2020} found a factor 3-10 higher $\SSFR$ than in
\citetalias{Kim_Ostriker2017}.
The initial set of solar neighborhood SILCC MHD simulations with {\sc
  FLASH}
\citep{2000ApJS..131..273F}  
reported in \citet{Gatto_2017} had similarly high $\SSFR$,
while SILCC simulations that included radiation and stellar wind
feedback as well as SNe reported by \citet{Peters_2017} found
comparable $\SSFR$ to that of \citetalias{Kim_Ostriker2017} (and
observations), and the additional inclusion of cosmic ray feedback in
\citet{Rathjen_2021} reduced $\SSFR$ even below this level. In an
initial set of RAMSES MHD
\citep{2002A&A...385..337T}
simulations including SNe plus
constant background heating, \citet{Hennebelle_Iffrig2014} found
$\SSFR\sim 0.01 -0.1 \sfrunit$, significantly higher than in
\citetalias{Kim_Ostriker2017}.  The simulations from other groups
mentioned above did
not include background shear, and as a result magnetic fields (if they
were included) decayed over time.  In RAMSES MHD simulations with
initial $\Sgas=19 \surfunit$ that also included ionizing
radiation, \citet{Colling_2018} studied the effects of shear, finding
$\SSFR \sim 0.03 - 0.01 \sfrunit$ for shear varying between the solar
neighborhood value and twice that.  \citet{Brucy_2020} considered a
wider range of initial gas surface density,
$\Sgas=13-160 \surfunit$, finding $\SSFR \sim 0.003 - 1
\sfrunit$
with SN and radiative heating 
feedback, and somewhat lower
values at high $\Sgas$ when additional large-scale turbulent driving
is included.

While none of the above kpc-scale ``ISM patch'' studies
have directly investigated whether the PRFM theory predictions
hold in their simulations, there have been a few studies using 
lower resolution simulations that have explored the issues
of the pressure-weight balance and the pressure-$\SSFR$ relation.
Because these global galaxy and cosmological zoom simulations are not able
to follow the Sedov-Taylor expansion and subsequent radiative
stages of individual
SN remnants or resolve the hot ISM more generally, subgrid models
are adopted for the SN feedback that drives turbulence.  
\citet{Benincasa_2016}
employed isolated global-galaxy simulations with a Milky-Way like model to
study pressure balance and the role of pressure in star formation
regulation.  They found that the total midplane pressure (azimuthally averaged
and averaged over 100 Myr) agrees well
with the total vertical weight of the ISM, and this is insensitive to
the adopted star formation efficiency per dynamical time or density
threshold for star formation.  
Using FIRE-2 cosmological zoom simulations of disk galaxies,
\citet{Gurvich_2020} conducted a detailed analysis of the pressure and
weight in vertical columns decomposed in $z$ into slabs, separately
measuring thermal and turbulent pressures and the partial
contributions within a slab from distinct temperature bins.  They found
that the median over time of the ratio between total pressure and vertical
weight is generally within $\sim 20\%$ of unity, with departures attributed
to the approximate treatment of ``long-range'' radiation and to the fact that
the spatial region for SN momentum deposition 
(at their numerical resolution)
is an appreciable
fraction 
of the gas scale height.  They also found that the total pressures in different
temperature bins agree within a factor of $\sim 2$, and that near
the midplane the turbulent pressure is the largest contribution for
$T<10^5\K$ gas while the thermal pressure dominates for hotter gas.
\citet{Gurvich_2020} confirmed as well that the total pressure is  roughly
linearly proportional to $\SSFR$ with a coefficient compatible with theoretical
predictions.

In this paper, we return to the PRFM theory, comparing predictions with
numerical results based on the same set of TIGRESS
MHD simulations used to analyze outflow properties in
\citet{Kim_Ostriker_etal2020a}.  We shall show that the key elements
of the theory
are validated: namely, a state of quasi-equilibrium is reached in which
(a) there is both warm and cold gas at the midplane with thermal pressure
at a level predicted by the photoelectric heating rate, and turbulent pressure
at a level predicted by the SN rate; (b) the total
pressures of hot and warm-cold gas are comparable, and this matches the
vertical weight of the ISM.  Consistent with our
previous  theoretical and computational  results, we shall show
that {as a consequence of (a) and (b)}, $\SSFR$ has a 
nearly  linear dependence on $\cal W$ (or $\PDE$)  over four decades.
We shall also provide quantitative analysis of other measures related
to pressure, including the relative proportions of different components
(thermal, turbulent, magnetic), the component and total
feedback yields (ratios of pressures to $\SSFR$),
and the effective equation of state (pressure-density relation). 

The plan of the paper is as follows:  In \autoref{sec:theory} we review and update key
elements of
the PRFM theory and define necessary terminology.  \autoref{sec:numer}
briefly summarizes the numerical methods and model inputs used in our
TIGRESS simulations, while \autoref{sec:results} presents results from
analysis of our simulation suite.  In \autoref{sec:summary} we
summarize and discuss our conclusions.

\vfil

\section{Pressure-Regulated, Feedback-Modulated Theory of the Equilibrium
Star-Forming ISM }\label{sec:theory}

\subsection{
Pressure Requirement: Dynamical Equilibrium ISM Weight and Gas Scale Height
}\label{sec:weight}

In a disk system, the weight (per unit area) of the ISM at a given midplane
location is obtained by a vertical integral of the product of the gas
density $\rho$ and the vertical component of the combined gravitational
field,
\begin{equation}\label{eq:weight}
  \begin{split}
    {\cal W} &\equiv \int_0^{z_{\rm max}} dz\ \rho (g_{\rm gas}  + g_{\rm ext}) =
    \frac{\Sgas}{2}
    \langle g_z \rangle \\
&= {\cal W}_{\rm gas} + {\cal W}_{\rm ext},
\end{split}
  \end{equation}
where $g_{\rm ext}=\partial \Phi_{\rm ext}/\partial z$ and the
external gravitational potential $\Phi_{\rm ext}$ includes that of the
old stellar disk as well as dark matter (the vertical gravity from the
former is dominant within the actively star-forming disk of normal
galaxies).  For plane-parallel (slab) geometry (i.e.~if the density
and the gravitational field are functions only of $z$, e.g. based on a
horizontal average), $\rho = (4\pi G)^{-1} \partial g_{\rm
  gas}/\partial z$, and we can define $\tilde{g} \equiv g(z)/g(z_{\rm
  max})$ for either the gas or external component of the potential.
In slab geometry $g_{\rm gas}(z_{\rm max}) = 2 \pi G \Sgas$
for the total gas surface density $\Sgas$ and similarly for
an external potential component with $\Sigma_{\rm ext}$ the 
total equivalent surface density.\footnote{For a stellar disk, the contribution
is $g_*(z_{\rm max})=2\pi G \Sigma_*$ for $\Sigma_*$ the stellar
  surface density. The dark matter potential is presumably
  approximately spherical rather than disk-like, and the same would be
  true for a stellar bulge potential.  In the (typical) case that the
  gas scale height is smaller than the gradient scale of the spherical
  dark matter or bulge potential, however, the local vertical gravity
  is still just a linear function of height, $g_{\rm dm} \approx
  4\pi G\rho_{\rm dm} z$ for a dark matter potential with a flat
  rotation curve and local density $\rho_{\rm dm}$, or $g_b
  \approx 4\pi G \rho_b z/3$ for a stellar bulge with uniform density
  $\rho_b$ \citepalias[see respectively][]{OML_10,OS_11}.  The total
  equivalent ``external'' surface density is then $\Sigma_{\rm ext}
  \equiv \Sigma_* + 2 z_{\rm max} (\rho_{\rm dm} + \rho_b/3)$.}
We therefore have the contribution to the weight from the gas gravity
\begin{equation}\label{eq:W_gas}
  {\cal W}_{\rm gas} = \pi G \Sgas^2 \int_0^{z_{\rm max}} dz
  \pderiv{\tilde{g}_{\rm gas}}{z} \tilde{g}_{\rm gas}  
  = \frac{\pi G \Sgas^2}{2}
\end{equation}
and from the external gravity
\begin{equation}\label{eq:W_ext}
  {\cal W}_{\rm ext} = \pi G \Sgas\Sigma_{\rm ext}
  \int_0^{z_{\rm max}}dz
  \pderiv{\tilde{g}_{\rm gas}}{z} \tilde{g}_{\rm ext} .
\end{equation}
Since $\tilde{g}_{\rm gas}(z_{\rm max})=1=\tilde{g}_{\rm ext}(z_{\rm max})$,
the dimensionless integral in \autoref{eq:W_ext} has an upper limit unity;
the value depends on the vertical profile shapes of $\tilde{g}_{\rm gas}$ and
$\tilde{g}_{\rm ext}$ (with a value 1/2 when the profiles are the same).
In the case that both density
profiles follow exponentials $\rho \propto \exp(-z/h)$,
$\tilde{g} =1-\exp(-z/h)$, the integral is $h_{\rm gas}/(h_{\rm gas} + h_{\rm ext})$.

In most circumstances the thickness of the mass-containing portion of the
gas disk is smaller than that of the stellar disk and dark matter,
so that within the gas layer 
$\Sigma_{\rm ext} \tilde{g}_{\rm ext} \approx 2 \rho_{\rm sd} z$ for
$ \rho_{\rm sd}= \rho_* + \rho_{\rm dm}$
the midplane density of stars plus dark matter.\footnote{In galactic
  center regions the external
  potential is dominated by a stellar bulge rather than the combination
  of a stellar disk plus dark matter halo, and the appropriate substitution
  in all of the following formulae  is $\rho_{\rm sd} \rightarrow \rho_b/3$.}
\autoref{eq:W_ext} can then be expressed as 
${\cal W}_{\rm ext} = 2 \pi \zeta_d G \rho_{\rm sd} \Sgas^2/\rho_0$ for $\rho_0$
the midplane gas density  and $\zeta_d \approx 1/3$
(see Equation 6 of \citetalias{OML_10}). The exact value
of $\zeta_d$ depends on the functional form of the vertical density profile
$\rho \propto \partial \tilde{g}_{\rm gas}/\partial z$.  If gas gravity dominates
the  potential, the resulting ${\rm sech}^2$ density profile yields
$\zeta_d =\ln(2)/2=0.35$, while a potential dominated by  external gravity
leads to a Gaussian density profile and $\zeta_d =1/\pi=0.32$.

We can define the half-thickness of the gas disk as
$h_{\rm gas} \equiv \Sgas/(2 \rho_0)$;
with this definition,
\begin{equation}\label{eq:W_from_h}
  {\cal W} =  \frac{\pi G \Sgas^2}{2} +
  4 \pi \zeta_d G \Sgas \rho_{\rm sd}  h_{\rm gas}.
\end{equation}
We note that \autoref{eq:W_from_h} holds quite generally, given a value of $h_{\rm gas}$.  Proceeding further to obtain an estimate of $h_{\rm gas}$ in terms of large-scale disk properties, one must assume that the disk structure represents a (quasi) equilibrium state, in the sense that an average over a few vertical dynamical times (typically a few tens  of Myr) is well defined, evolving only over  a longer timescale.   

In vertical dynamical equilibrium, the ISM weight $\cal W$ must be equal to
$\Delta \Ptot$, the difference in the total vertical
momentum flux across the ISM layer.
Here we will consider the terms in $\Ptot$ associated with thermal, turbulent,
and magnetic stresses, although in principle $\Ptot$ may also contain terms
associated with radiation and cosmic ray pressure \citepalias[see][]{OS_11}.
Writing
$\Delta \Ptot \approx \Ptot(z=0)
\equiv \sigma_{\rm eff}^2 \rho_0=\sigma_{\rm eff}^2 \Sgas/(2h_{\rm gas})$
for a total effective velocity
dispersion $\sigma_{\rm eff}$ (assumed constant in $z$) and equating $\Delta \Ptot = {\cal W}$, we can
solve a quadratic to obtain
\begin{equation}\label{eq:hgas}
  h_{\rm gas} = \frac{2 \sigma_{\rm eff}^2}{ \pi G \Sgas + \left[(\pi G \Sgas )^2
      + 32 \pi \zeta_d G \rho_{\rm sd}  \sigma_{\rm eff}^2\right]^{1/2}}.
\end{equation}
and
\begin{equation}\label{eq:W_terms}
  {\cal W} = \frac{\pi G \Sgas^2}{4}
  \left\{1+  \left[1+\frac{32  \zeta_d  \rho_{\rm sd}\sigma_{\rm eff}^2}{\pi G \Sgas^2}  \right]^{1/2}\right\}.
\end{equation}  
\autoref{eq:hgas} reduces to the familiar limits
$h_{\rm gas} \rightarrow \sigma_{\rm eff}^2/(\pi G \Sgas )$ or
$h_{\rm gas} \rightarrow \sigma_{\rm eff}/(8 G \rho_{\rm sd} )^{1/2}$ in the
gas-gravity and external-gravity dominated limits, respectively, where for
the latter case we also take $\zeta_d \rightarrow 1/\pi$.  
If we consider limiting forms for the square root in
\autoref{eq:W_terms}, we obtain
a simplified (good within 20\%) expression for the weight,
\begin{equation}\label{eq:PDEdef}
{\cal W}\approx \PDE \equiv \frac{\pi G \Sgas^2}{2} + \Sgas(2 G \rho_{\rm sd})^{1/2} \sigma_{\rm eff},
\end{equation}
which has been adopted in many observational studies
\citep[e.g.][]{Blitz_Rosolowsky_2004,Herrera-Camus_2017,Sun_2020a,Barrera-Ballesteros_2021}.\footnote{We note that the expression in \autoref{eq:PDEdef}
  and the related form $(\pi G \Sgas/2)(\Sgas + \Sigma_*\sigma_{\rm eff}/\sigma_*)$  from 
  \citet[][{based on the assumption of vertical equilibrium in the stellar component}]{Elmegreen_1989}, with  $\Sigma_*$ and $\sigma_*$ the surface density and velocity dispersion of the old stellar disk, 
  are sometimes referred to as the ``hydrostatic
  pressure.''  However, we deprecate this term in favor of
  ``dynamical equilibrium pressure'' as the gas in the ISM is nonstatic.}
  Comparing \autoref{eq:PDEdef} to
\autoref{eq:weight}, this also implies that under equilibrium conditions, 
the average gravity
in the vertical direction is given by
$\langle g_z \rangle = \pi G \Sgas + 2(2 G \rho_{\rm sd})^{1/2} \sigma_{\rm eff}$.   We note that $\sigma_{\rm eff}$ in \autoref{eq:hgas} - \autoref{eq:PDEdef}  represents the square root of the ratio of total vertical stresses to density, and therefore should include a magnetic contribution (see \autoref{sec:pred_SFR}  and \autoref{sec:evolution}).

\autoref{eq:hgas}  and \autoref{eq:W_terms} give the ISM scale height and
weight under the
assumption that it consists of a volume-filling medium with an effective
velocity dispersion $\sigma_{\rm eff}$.  \citetalias{OML_10} provided
generalized expressions in the case that the gas is divided into
a diffuse component of surface density  $\Sigma_{\rm diff}$
and velocity dispersion $\sigma_{\rm eff}$ that is volume-filling, 
and a gravitationally bound cloud (GBC)
component  of surface density  $\Sigma_{\rm GBC} = \Sgas - \Sigma_{\rm diff}$
with negligible volume and scale height.  The resulting
expression for the midplane weight or pressure in equilibrium is 
\begin{multline}
  {\cal W} = \frac{\pi G \Sigma_{\rm diff}^2}{4}
  \Bigg\{1+  2\frac{\Sigma_{\rm GBC}}{\Sigma_{\rm diff}} +  \\
    \left[\left(1+ 2\frac{\Sigma_{\rm GBC}}{\Sigma_{\rm diff}}\right)^2 +
    \frac{32  \zeta_d  \rho_{\rm sd}\sigma_{\rm eff}^2}{\pi G \Sigma_{\rm diff}^2}
    \right]^{1/2}\Bigg\} 
\end{multline}
\citepalias[see Eq. 11 of][]{OML_10}, with diffuse-gas scale height
$h_{\rm diff} = \sigma_{\rm eff}^2\Sigma_{\rm diff}  /(2{\cal W})$.
These expressions reduce to \autoref{eq:hgas} and \autoref{eq:W_terms} 
when $\Sigma_{\rm GBC}\ll \Sigma_{\rm diff} \approx \Sgas$.  

Although the traditional view has been that GMCs are gravitationally bound
entities and therefore might be considered  as collectively comprising a
``GBC'' component of the ISM, with the atomic gas comprising the ``diffuse''
component, this has been called into question by recent work.  On the
theoretical side, numerical simulations of star formation in turbulent clouds
(with feedback) show that both the star formation rate per free-fall time
and the lifetime SFE exceed observational estimates unless the cloud-scale
virial parameter is at least $\sim 2-4$ 
\citep[e.g.][]{padoan12,Kim_JG_Ostriker_Filippova2021,Evans_2022}.  The analysis of dense
structures in TIGRESS simulations also shows that only a small fraction
of the mass is gravitationally bound,
and that many structures that would be classified
as gravitationally bound based on typical observational estimates
in fact contain only a small fraction of bound gas when detailed internal
structure and tidal effects are taken into account \citep{Mao_2020}.  On
the observational side, meta-analysis of several Galactic and extragalactic
surveys suggests that the fraction of molecular gas that is in bound structures
may be well below unity \citep{Evans_2021}, and the mean value of the virial
parameter for molecular gas at $\sim 100\pc$ scale
in a large sample of PHANGS galaxies is closer to 4 than to 1
\citep{Sun_2020b}.  GMCs are more likely to be gravitationally bound in
atomic-dominated (rather than molecular-dominated)
regions  of galaxies because they correspond to higher
overdensities in those environments, but since  by  definition 
$\Sigma_{\rm H_2} \ll \Sigma_{\rm HI}$ when atomic gas dominates it would correspond
to the limit $\Sigma_{\rm GBC} \ll \Sigma_{\rm diff}$ so that \autoref{eq:W_terms}
holds.  In any case, the short
lifetime and transient nature of GMCs
\citep[e.g.][]{Leiszwitz_1989,Kawamura_2009,Kruijssen_2019} implies that
regardless of the exact values of their virial parameters 
it may be most appropriate to consider GMCs as temporary
condensations within the dynamic ISM.  

Thus, we adopt the assumption that regardless of its chemical state, the
ISM may be treated as a volume-filling diffuse medium so that in
statistical equilibrium the 
weight and total midplane pressure are equal and may be expressed in terms
of the disk parameters by \autoref{eq:W_terms} or \autoref{eq:PDEdef}.  

We emphasize that because $\sigma_{\rm eff}$
enters the expressions (\autoref{eq:W_from_h} and \autoref{eq:W_terms}) for ISM weight (or equilibrium midplane pressure) by way of representing
the vertical extent $h_{\rm gas}$ of the mass-containing component of  the ISM,
any observational estimate of  $\sigma_{\rm eff}$  must be 
\textit{mass-weighted} and correspond to an average over
the vertical direction. 
In face-on galaxies, the thermal and turbulent motions in the vertical direction combine in quadrature to produce the observed linewidth; the mass-weighted large-scale average of this (based on proxies such as \ion{H}{1} 21 cm and CO)
is the thermal and turbulent contribution to $\sigma_{\rm eff}^2$.
The azimuthal component of the magnetic field is usually the largest, so 
that the magnetic contribution to $\sigma_{\rm eff}^2$ is dominated by the mass-weighted large-scale average of $ B_\phi^2/(8\pi\rho)$.

\subsection{Pressure Response: Feedback Modulation and Yields}\label{sec:yield_theory}

We next consider how pressures are modulated by feedback  from young,
massive  stars.  The two most important direct feedback mechanisms for
setting the large-scale pressure in the warm and cold gas that makes up the
ISM's mass reservoir are the driving of kinetic turbulence by expanding SN
remnants and
the photoelectric heating of atomic and low-A$_V$ molecular gas induced
by far-UV stellar photons impinging on small grains.  The resulting
turbulent pressure
scales with the momentum injection rate from SNe, while the thermal
pressure scales with the FUV intensity (see below).

We note that within young
($\lesssim 10 \Myr$) star-forming molecular clouds, photoionization produces
high-pressure gas that, in combination with radiation pressure on dust,
is quite important for dispersing dense gas  clouds. However, 
the momentum injection from dusty \ion{H}{2} region expansion over the lifetime of the massive-star population (averaging over the IMF) is much less than 
the momentum injection produced when the same stars die as SNe
\citep[e.g.][]{Kim_JG2018}.  {While idealized spherical solutions for
  \ion{H}{2} region expansion around star clusters can produce momentum per
  unit stellar mass approaching that produced by SNe 
  \citep[see e.g. Eq.~14 of][]{PPVII_GMC2022}, the actual momentum 
  in simulations of cluster-forming turbulent clouds that use ray tracing
  to follow radiation is lower by a factor $\sim 5-10$.  This reduction is  in
  part because distributed star formation and fractal ionization fronts lead
  to force cancellation, and
  in part because a significant fraction of photons are absorbed by  dust or
  escape from inhomogeneous clouds
through low-density channels \citep{Kim_JG2019}.}
Stellar winds also inject considerable energy
in their environments, but most is radiated away and the momentum
injection is comparable to that from radiation pressure
\citep{Lancaster_2021,Lancaster_2021b}.
{In the extremely high-column environments of nascent super star clusters,
  the pressure of infrared radiation reprocessed by dust can exceed the direct
  radiation pressure and even the pressure of photoionized gas
  \citep[e.g.][]{MQT_2010}.  However,
  the gravitational force in these environments is also extremely high.
Gravity exceeds the reprocessed radiation  force unless the IMF is top-heavy
  or the dust abundance is significantly enhanced; numerical radiation
  hydrodynamic simulations show that reprocessed radiation
  is therefore ineffective at limiting star formation unless these exceptions
  apply \citep{Skinner_Ostriker2015}.
}

In high-A$_V$ regions where FUV photons are
shielded, low-energy
CRs are the main mechanism for heating gas and therefore setting
the thermal pressure.  However, the low temperature produced by CR heating
\citep[e.g.][]{Gong_2017} generally implies the thermal pressure is lower
than kinetic turbulent pressure (as well as magnetic pressure) in molecular gas
\citep[e.g.][]{Heyer_Dame2015}.   

In the Milky Way, the midplane
energy density of $\sim$GeV CRs is $\sim 1{\rm \ eV \pcc}$
\citep[e.g.][]{Grenier_15},
comparable to the thermal, turbulent, and magnetic energy densities.  However,
the ion-neutral damping of Alfv\'en waves in the primarily-neutral gas
leads to strong CR diffusion and is expected to limit
the support against gravity by CRs, as confirmed by 
low CR pressure gradients near the midplane
in recent numerical simulations of CR transport in TIGRESS simulations
\citep{Armillotta_etal2021,Armillotta_2022}.

We shall define $m_*$ as the total mass in stars formed for every high mass
star that undergoes a SN; $m_* \approx 100  \Msun$ for a \citet{Kroupa_2001}
IMF.  In quasi steady state, the rate per unit area in the disk
of core collapse SNe from high mass stellar death
will then be given by $\SSFR/m_*$.  Taking the spherical momentum injected per
SN as $p_*$, and assuming that the turbulent pressure is equal to
the vertical momentum flux injected to each side of the disk,
\citetalias{OS_11} predicted that the turbulent pressure will follow
\begin{equation}\label{eq:Pturb_theory}
P_{\rm turb} = \frac{1}{4} \frac{p_*}{m_*} \SSFR.
\end{equation}
\citetalias{OS_11} allowed for an additional multiplicative
factor $f_p$ between unity (for strong dissipation)
and 2 (for weak dissipation); since the numerical simulations of
\citetalias{KOK_2013} found $f_p = 1.2 (\SSFR/0.001 \sfrunit)^{-0.11}$,
here we adopt $f_p=1$ as a good approximation.  
The scaling of turbulent pressure with $\SSFR$
can also be obtained by equating
the turbulent energy driving rate with the turbulent energy
dissipation rate \citep{TQM_2005,OS_11,Hennebelle_Iffrig2014}.
Per unit area in the disk, the kinetic energy input rate from SNe is
an order-unity factor times $\sigma_v(p_*/m_*)\SSFR$ for $\sigma_v$ the
velocity dispersion on the energy-containing scale.  Assuming the dissipation
timescale is comparable to the vertical crossing time 
$h_{\rm gas}/\sigma_v$
\citep{Stone_1998,MacLow_1998}, the kinetic energy dissipation rate
per unit area is then equal to
$(1/2)\sigma_v^3 \Sgas/h_{\rm gas} = \sigma_v P_{\rm turb} $ times an order-unity
factor.  Equating leads to $P_{\rm turb} \sim (p_*/m_*) \SSFR$, but without
a specific prediction for the numerical coefficient.  
  
The momentum injection per SN  $p_*$  is insensitive to the ambient environment
because it is primarily set by the condition for the expanding blast
wave to become radiative when the shock velocity drops to
$v_{\rm cool} \sim 200 \kms$
(for solar metallicity; see Eq.~9 of \citealt{Kim_Ostriker2015a} or
Eq.~39.22 of \citealt{Draine_2011}).  This 
yields
momentum  $\sim 0.6 E_{\rm SN}/v_{\rm cool}$ at the shell formation time
since the kinetic energy in
the Sedov-Taylor  stage is $E_\mathrm{kin}= 0.283 E_\mathrm{SN}$.
The numerical simulations of
\citet{Kim_Ostriker2015a} for single SN explosions at a range of
ambient (uniform) density of hydrogen nuclei
$n_H = 0.1 - 100 \pcc$ found that the momentum
increases by $\approx 50\%$ after cooling and shell formation, reaching
a level $p_* \approx 2.95 \times  10^{5}\Msun \kms (n_H/1 \pcc)^{-0.16}$,
consistent with theoretical expectations (these and other simulations
assume $E_\mathrm{SN} =10^{51}\erg$; e.g., \citealt{Cioffi_1988,Thornton_1998}).
For inhomogeneous conditions corresponding to a two-phase ISM the
result is quite similar,
$p_* \approx 2.8 \times  10^{5}\Msun \kms (n_H/1 \pcc)^{-0.17}$
\citep[see also][]{Iffrig_Hennebelle2015,Martizzi_2015,Walch_Naab2015};
here $n_H$ is the density of hydrogen nuclei in the ambient medium
averaged over both phases. 
These results are also consistent with the radial shell momentum of seven radiative SN remnants 
inferred from \ion{H}{1} observations \citep{Koo2020}.
The momentum at shell formation also shows a weak dependence on the metallicity $\propto Z^{-0.17}$ at $Z>0.01Z_\odot$ -- and nearly constant otherwise
(J.-G. Kim et al. 2022, {submitted}; see also \citealt{Thornton_1998, Oku_2022}).

While correlation of SN originating from a star cluster (or neighboring
clusters) could in principle
enhance the momentum per event if most of the injected SN energy is retained
by the superbubble 
\citep{McCray_1987,Gentry_2017}, this
requires a contact discontinuity to be maintained at the interface
between the hot bubble interior and the surrounding gas.  However, this
is unlikely in a realistic, clumpy ambient medium.  Instead,
Kelvin-Helmholz and other instabilities at interfaces excite turbulence that drives mixing between hot gas and
surrounding higher-density (warm or cold) gas with subsequent rapid cooling
\citep{Fielding_2020,Lancaster_2021}, so that the energy in the interior
of the hot bubble drops in between SN events.
\citet{Kim_Ostriker_Raileanu2017} reported on numerical simulations
of multiple SN events in a cloudy ambient
medium, considering a range of ambient mean
density ($n_H =0.1-10\pcc$) and SN intervals ($0.01-1 \Myr$), finding quite
similar momentum per event to the single-SN case,
$p_* \sim 0.7 - 3 \times  10^{5}\Msun \kms$.  The lower (upper) value is for
an interval between SNe of 0.01 (1)$\Myr$, corresponding to a total
clustered stellar mass of $4\times 10^5$ ($4\times 10^3)\Msun$.
Realistic clustered stellar masses (which may include several individual
star clusters that are born within a few tens of $\Myr$ of each other)
are likely in the middle of this range, implying $p_* \sim 2 \times
10^5 \Msun \kms$.

The turbulent feedback yield is given by the ratio between $P_{\rm turb}$ and
$\SSFR$, which from \autoref{eq:Pturb_theory} and the idealized
supernova simulations described above is expected to be 
\begin{equation}\label{eq:Upsturb_theory}
\begin{split}
  \Upsilon_{\rm turb} \equiv \frac{P_{\rm turb}}{\SSFR}
   & =
   250 \kms \frac{p_*/m_*}{1000 \kms } \\
   &\sim 500 \kms.
\end{split}
\end{equation}
The SN momentum injection is relatively insensitive to ambient
conditions so we expect $\Upsilon_{\rm turb}$ to decrease only weakly in
the higher-density, inner regions of galaxies.

For a given FUV radiation field, a wide range of thermal pressures are
possible in the atomic ISM, with the lowest pressures associated with
warm, low-density gas (cooled mostly by Ly$\alpha$ and recombination
on grains), and the highest pressures associated with cold, high-density gas
(cooled mostly by \ion{C}{2} and \ion{O}{1} fine structure lines).
Which among these possible pressures does the ISM select?
\citetalias{OML_10} hypothesized that the equilibrium midplane
pressure should be in the ``two-phase''
range where both cold and warm phases are available, adopting the geometric mean
$P_{\rm two} \equiv (P_{\rm max,warm} P_{\rm min,cold})^{1/2}$
between the largest possible
pressure for a purely warm medium and the smallest possible pressure
at which a cold medium becomes possible.  The resulting thermal
pressure is given in Equation 15 of \citetalias{OML_10}, which uses
the \citet{Wolfire_2003} analytic fit to their thermal equilibrium
curve (their Eq. 33), also assuming that the mean pressure is a factor 1.4 above
the minimum possible pressure of the cold medium.

The primary scaling of the equilibrium thermal pressure is with
the FUV radiation field mean intensity, $J_{\rm FUV}$. Since the FUV
originates from young stars, we expect $J_{\rm FUV} \propto \SSFR$,
with an additional attenuation factor that depends on radiative
transfer in the ISM.  For the simplest possible case of a slab
containing a uniform distribution of dusty gas producing extinction
and uniformly-distributed
stars producing emission, the radiation field will
be {$J_{\rm FUV} = \Sigma_{\rm FUV}f_\tau/(4\pi)$ where $\Sigma_{\rm FUV} \propto \SSFR$ is the energy
  per unit area per unit time in FUV produced by recently formed stars, and} 
\begin{equation}
  f_\tau \equiv \frac{1-E_2(\tau_\perp/2)}{\tau_\perp}
\label{eq:shielding}
\end{equation}
\citepalias{OML_10}. Here,
$E_2$ is the second exponential integral and $\tau_\perp= \kappa_{\rm FUV}
\Sgas$ is the mean optical depth to FUV vertically through the disk;
$E_2(x)/x$ is logarithmic at small argument and decreases exponentially
at large argument.
Taking $\kappa_{\rm FUV} = 10^3  \cm^2\ \g^{-1}$
and total surface density $\Sgas=10 \surfunit$ in the solar
neighborhood, the local value would be $f_{\tau,\odot}\approx 0.41$.
From Equation (15) of \citetalias{OML_10},
we can then write the predicted yield for thermal pressure as
\begin{equation}\label{eq:Upsth_theory}
\begin{split}  
  \Upsilon_{\rm th} &\equiv \frac{P_{\rm th}}{\SSFR} \\
  &= 240 \kms
  \frac{4.1 f_\tau/f_{\tau,\odot}}
       {1 + 3.1 \left(\frac{\Sgas Z_d^{'} f_\tau/f_{\rm \tau,\odot}
         }{10 \surfunit}  \right)^{0.4}}
\end{split}       
\end{equation}
where $Z_d^{'}$ is the dust abundance relative to the solar neighborhood value
and the dust abundance is assumed to scale with gas metallicity
($Z_d^{'}= Z_g^{'}$).  In \autoref{eq:Upsth_theory}
we explicitly show the dependence on $f_\tau$, which was omitted for
simplicity in Equation 18 of \citetalias{OML_10}
for the relationship between $P_{\rm th}$ and $\SSFR$. 
We note that in contrast to the relatively weak variation of
$\Upsilon_{\rm turb}$ with environment,
attenuation of FUV radiation is expected to strongly
reduce $\Upsilon_{\rm th}$ under higher-density galactic conditions.
In particular, $f_\tau \sim 1/\tau_\perp$ at high
optical depth, with $\tau_\perp \approx \Sgas/5 \surfunit$.

In addition to turbulent kinetic and thermal pressure, magnetic pressure also
helps to support against gravity.  Magnetic fields are driven by dynamo
activity, with kinetic turbulence and large-scale galactic shear both
contributing to amplification via folding and stretching of field lines,
and buoyancy and superbubble expansion combining with Coriolis forces
to create poloidal from toroidal fields
\citep[e.g.][]{Kulsrud2005}.  Many aspects of dynamo theory -- including
how mean fields are generated -- remain however poorly understood,
even in idealized
situations that lack the complexity of the multiphase ISM.  Nevertheless,
the previous numerical simulations
of \citet{Kim_Ostriker2015b}  do show amplification of both the turbulent
and mean magnetic field, with the saturation level of turbulent magnetic
pressure about one-third of the saturation level of turbulent kinetic
pressure, and the 
pressure in the mean magnetic field (which is primarily azimuthal) comparable
to that in the turbulent magnetic field.  Since the turbulent magnetic pressure
scales with turbulent kinetic pressure, from \autoref{eq:Pturb_theory} and
\autoref{eq:Upsturb_theory}  we
can expect the turbulent portion of the magnetic pressure to scale nearly
linearly with $\SSFR$.  It is less clear whether the mean component of
the magnetic pressure would also scale with $\SSFR$.  In  summary, 
uncertainties related to dynamo theory preclude making a definitive
prediction for
the magnetic pressure or the magnetic yield at this time, but it is reasonable
to expect $\Upsilon_{\rm mag}$ to be $\sim 0.5-1 \times \Upsilon_{\rm turb}$.  This is consistent with observed estimates of magnetic field strength in the Milky Way and other spiral galaxies, which based on Zeeman splitting, rotation measure, and synchrotron emission find comparable magnetic pressure to the turbulent pressure \citep[e.g.][]{Heiles_Troland_2005,Beck_2019}, but are subject to significant systematic uncertainties. 

Taking the sum of the turbulent kinetic, magnetic, and thermal yield
terms discussed above, the theoretically predicted
value of the feedback yield is $\Upstot \sim 1000 \kms$.  
One of the main goals of the present work is to  evaluate 
$\Upstot$ from realistic MHD simulations with a range of conditions.
This will provide a numerical test of the theory.

Finally, it is worth noting that the drop
in $\Upsilon_{\rm th}$ compared to $\Upsilon_{\rm turb}$ is expected to lead to
a reduction in the fraction of gas mass in the warm phase in high-density
environments.  The warm mass fraction is given by
$f_{\rm m,w} = (P_{\rm th}/P_{\rm tot})(\sigma_{\rm eff}^2/c_w^2)$ where $c_w$ is the thermal speed in warm gas ($\sim 10 \kms$, insensitive to environment).  With
$P_{\rm th}/P_{\rm tot} = \Upsilon_{\rm th}/\Upsilon_{\rm tot}$ decreasing
$\propto 1/\Sgas$  at high surface density due to the $1/\tau_\perp$
dependence of $f_\tau$ in \autoref{eq:Upsth_theory}, $f_{\rm m,w}$ is expected to
become small in these dense environments.  While an increase in
$\sigma_{\rm eff}$ (dominated by  turbulence) could offset this to some extent,
empirically the effective velocity dispersion
does not increase as rapidly as the gas surface density
\citep[e.g.][]{Sun_2018,Wilson_2019}.

\subsection{Predicting Large-Scale Galactic SFRs}\label{sec:pred_SFR}

From \autoref{sec:yield_theory}, if the energy  and pressure
in the ISM are sustained
by  feedback from star formation, we can relate the pressure and
star formation rate via $\Ptot = \Upstot \SSFR$,
where $\Upstot$ is the total feedback yield, including thermal,
turbulent (kinetic), and magnetic terms.
For current purposes, the $\Ptot$ we are interested
in is the total midplane pressure as it would appear in the vertical component of
the vertical momentum equation, consisting of the sum $\Ptot = \Pth + \Pturb + \Pimag$ for 
$\Pth\equiv \langle \rho c_s^2\rangle $ (thermal pressure),
$\Pturb \equiv \langle \rho v_z^2 \rangle$ (vertical Reynolds stress, or
turbulent pressure),
and $\Pimag \equiv \langle|{\bf B}|^2 - 2B_z^2\rangle/(8\pi) =
\langle B_x^2 + B_y^2 - B_z^2 \rangle/(8\pi)$ (vertical Maxwell stress,
combining magnetic pressure and tension); here angle brackets
denote horizontal averages at the midplane.  In the case of
isotropic velocity  and magnetic  fields, the turbulent and
magnetic terms would be equivalent to arbitrary one-dimensional projections of the vector velocity
and magnetic field.  In reality, however, both the turbulent velocity and magnetic field are generally anisotropic.   

Under the assumption that vertical dynamical equilibrium is satisfied
(at least as a quasi-steady state), and also assuming the midplane pressure
is much larger than the pressure above the mass-containing portion of the
disk, the weight of the ISM calculated in \autoref{sec:weight}
must be balanced by the midplane pressure,
${\cal W}=\Ptot$. The star formation rate per unit area
may then be predicted as a function
of large-scale ISM properties as
\begin{equation}\label{eq:SFR_pred}
\SSFR = \frac{\cal W}{\Upstot} \approx \frac{\PDE}{\Upstot}.
\end{equation}
where \autoref{eq:W_terms} and \autoref{eq:PDEdef} 
express $\cal W$ and $\PDE$,  respectively,  in terms of $\Sgas$, $\rho_{\rm sd}$,
$\sigma_\mathrm{eff}$. 
We expect $\Upstot$ to decrease slightly in regions of high $\Sgas$.
In particular, when $\Sgas$ increases, $\Upsth$ decreases due to
greater radiation extinction, reducing heating; and $\Upsturb$ decreases due to
the reduction in SN momentum injection at higher ambient density.
As a consequence, since
higher $\Sgas$ is associated with higher pressure, this is expected to yield
a dependence of $\SSFR$ on $\PDE$ that is slightly superlinear.
{We note that in the above, we implicitly assume other parameters,
  including metallicity and the IMF, are held fixed; a lower metallicity  or
  top-heavy IMF 
  would tend to increase $\Upstot$.}

With $\Ptot = \seff^2 \Sgas/(2h_{\rm gas})$ for $\seff$ the  effective
total velocity dispersion and $h_{\rm gas}$ the semi-thickness of the
mass-containing disk, \autoref{eq:SFR_pred} can also be expressed as
\begin{equation}\label{eq:tdyn_form}
  \SSFR = \frac{\seff}{\Upstot} \frac{\Sgas}{\tdyn}
  \equiv \varepsilon_{\rm ver} \frac{\Sgas}{\tdyn}.
\end{equation}  
Here, for convenience we have incorporated a factor of two in defining
the vertical
dynamical time $\tdyn \equiv 2 h_\mathrm{gas}/\sigma_{\rm eff}$; an explicit formula for
$\tdyn$ (in terms of $\Sgas$, $\rho_{\rm sd}$,
$\sigma_\mathrm{eff}$)
can be obtained by substituting for
$h_\mathrm{gas}$ from \autoref{eq:hgas}.  In this
formulation, the ratio $\seff/\Upstot \equiv \varepsilon_{\rm ver}$
represents the star formation efficiency per vertical dynamical time;  
this efficiency $\varepsilon_{\rm ver} \sim 1\%$ since
$\sigma_{\rm eff} \sim 10 \kms$ and $\Upstot \sim 1000 \kms$.  
Star formation is also commonly quantified in terms of the gas
depletion time, $\tdep \equiv \Sgas/\SSFR = M_{\rm gas}/\dot M_*$.
Using \autoref{eq:tdyn_form} and \autoref{eq:hgas},
\begin{equation}\label{eq:t_dep}
\begin{split}
\tdep &=\frac{1}{\varepsilon_{\rm ver}}\tdyn = \frac{\Upstot}{\seff}\tdyn \\
& \approx \frac{2\Upstot}{\pi G \Sgas + 2\seff (2 G \rho_{\rm sd})^{1/2}},
\end{split}
\end{equation}
  implying a depletion time 
two orders of magnitude longer than the vertical dynamical time.
Because $\Upstot$ tends to decrease (modestly) and $\sigma_{\rm eff}$ to increase
(modestly) in galactic centers and other high-$\Sgas$ environments, it is
expected that the efficiency of star formation will (modestly) increase
under these conditions.

Given $\Upstot$, quantitative predictions of star formation are obtained through
\autoref{eq:SFR_pred} (or equivalently \autoref{eq:t_dep}).
Theoretical estimates of feedback yields in \autoref{sec:yield_theory}
predict $\Upstot\sim 1000\kms$; our numerical results for
$\Upsilon_\mathrm{tot}$ will be presented in \autoref{sec:yields}.

We point out that \autoref{eq:SFR_pred} is equivalent to Equation (22) of \citetalias{OML_10}.  To make the connection, note that in \citetalias{OML_10} $\alpha \equiv \Ptot/\Pth$.  Taking the ratio of the right-hand side of Equation (18) in \citetalias{OML_10} to $\Pth$ to obtain the adopted value of $1/\Upsth$ in that work, and substituting in  $\alpha \rightarrow \Upstot/\Upsth$ in Equation (22) of \citetalias{OML_10}, one can obtain \autoref{eq:SFR_pred} above.  We further note
that for the reasons discussed at the end of \autoref{sec:weight} (see also \autoref{sec:small_scale}, where we argue that the star formation timescale in gravitationally bound gas is likely to  be quite small), Equation (22) of \citetalias{OML_10} (rather than Equation 23) is expected to apply in general. Since the ratio $\Upstot/\Upsth$ is not constant, \autoref{eq:SFR_pred} combined with values of $\Upstot$ that calibrate for varying ISM conditions is preferred over Equation (22) of \citetalias{OML_10}.

\subsection{Connection to Small-Scale Star Formation}\label{sec:small_scale}

It is worth remarking on the connection between the PRFM theory
of star formation regulation that is motivated by maintaining the
average conditions in the ISM,
and the localized process of star formation
in highly overdense structures.
Since star formation involves collapse, it presumably
takes place within individual gravitationally bound structures, which
comprise a (small) fraction
$f_{\rm gb}$
of the total ISM mass.  If the
free-fall time for these structures is $t_{\rm ff,gb}$ and their star formation
efficiency per free-fall time is
$\varepsilon_{\rm ff,gb}$, we can set  \autoref{eq:tdyn_form} equal to
$\varepsilon_{\rm ff,gb} \langle f_{\rm gb}\rangle \Sgas/t_{\rm ff,gb}$ to obtain 
\begin{equation}\label{eq:f_gb}
\langle f_{\rm gb} \rangle = \frac{\varepsilon_{\rm ver}}{\varepsilon_{\rm ff,gb}}\frac{t_{\rm ff,gb}}{\tdyn},
\end{equation}
where the angle brackets denote a time average.  
The value of $\varepsilon_{\rm ff,gb}$ is set by 
small-scale gravoturbulent fragmentation processes \citep[e.g.][]{Dobbs2014,Padoan_2014}.
Physically, we can  understand \autoref{eq:f_gb} as saying
that given the small-scale $\varepsilon_{\rm ff,gb}$
together with the large-scale $\varepsilon_{\rm ver}$
(as derived from PRFM theory), the fraction of material that
is contained in bound clouds will adjust to satisfy both constraints
(in a time-averaged sense).
Quantitatively, numerical
simulations show that for gravitationally-bound systems,
$\varepsilon_{\rm ff,gb} \gtrsim 0.1$ 
\citep[e.g.][]{padoan12,Raskutti_2016,Kim_JG_Ostriker_Filippova2021}. Then
with $\varepsilon_{\rm ver} \sim 0.01$ and $t_{\rm ff,gb} < t_{\rm ver}$,
the expected $\langle f_{\rm gb}\rangle \lesssim 0.1$, i.e. only a small fraction of the
ISM will be in bound structures.  The \citet{Mao_2020}
analysis of the TIGRESS solar neighborhood simulation indeed shows only
a very  small fraction of gas ($0.01-0.1$) is gravitationally bound. 

In both the real ISM and in numerical
simulations, star formation takes place within massive clouds that are
overdense and overpressured with respect to average conditions at the
ISM midplane. These massive clouds are typically observed as GMCs in
CO lines, although at low metallicity CO emission may be weak, and if
self-shielding is low enough they would primarily consist of
\ion{H}{1} rather than H$_2$
\citep[e.g.][]{Bialy_Sternberg2016,Gong_2017}.  Analogously to
\autoref{eq:f_gb}, we may write the expected time-averaged
fraction of ISM material in GMCs as
$\langle f_\mathrm{GMC}\rangle =(\varepsilon_{\rm ver}/\varepsilon_{\rm ff,GMC})
(t_\mathrm{ff,GMC}/\tdyn)$.
While GMCs are self-gravitating, recent
work suggests that especially in molecule-dominated regions of galaxies
their typical virial parameters are closer to $\sim 4$ rather than 1
\citep{Sun_2020b}, and they would
therefore have $\varepsilon_{\rm ff,GMC}\lesssim 0.01$
since the efficiency decreases exponentially with increasing
virial parameter
\citep[e.g.][]{Krumholz_McKee2005,padoan12,Federrath_Klessen2012,Kim_JG_Ostriker_Filippova2021}.  The fraction of the ISM's mass in GMCs can thus
be significant.

We emphasize that the above $\langle f_\mathrm{GMC}\rangle$
is what must hold in a time-averaged sense, while individual GMCs continually
form and disperse.  Formation is subject to the level of turbulence
in the diffuse ISM (primarily driven by SNe), while dispersal of
the denser clouds that have formed massive stars is likely due to
\ion{H}{2} regions (see \autoref{sec:intro}).
The instantaneous $f_\mathrm{GMC}$ may be above or below
the equilibrium value.  If $f_\mathrm{GMC}$ and $f_\mathrm{gb}$ are
above the equilibrium level, the ``excess'' star formation and
feedback that ensue will drive greater-than-equilibrium heating and
momentum injection on large scales (after dispersing existing GMCs),
temporarily limiting contraction of diffuse gas
into denser, star-forming 
clouds. If $f_\mathrm{GMC}$ and $f_\mathrm{gb}$ are below equilibrium,
the level of feedback will be low enough that new overdense clouds
readily form.  Within these GMCs, gravitationally bound regions will form and
star formation will commence. The cyclic
formation and dispersal of  overdense and star-forming structures is evident
in the time series correlation
analysis of the TIGRESS solar neighborhood simulation
by \citet{Mao_2020} \citep[see also][]{Semenov_2017,Orr_2019,Moon_2022}.

\subsection{Testing the theory}

The remainder of this paper will largely focus on testing the key elements of the theory laid out in this section.  The tests will make use of a set of numerical simulations that 
sample a range of parameters representing
normal star-forming disk galaxies.  We first describe the numerical methods employed in these simulations and the model parameter set (\autoref{sec:numer}). Then we proceed to present results from our simulation analysis that (1) confirm the prediction of vertical dynamical equilibrium from \autoref{sec:weight}, i.e. that midplane $\Ptot$ values (in both hot and warm/cold phases) agree with $\cal W$ and $\PDE$   (see \autoref{sec:equil}); (2) characterize the individual pressure components, and compare to the theoretical predictions of \autoref{sec:yield_theory} for thermal and turbulent  feedback yields  (see \autoref{sec:press_frac}, \autoref{sec:yields}); (3)  for comparison to the prediction for large-scale star formation described in \autoref{sec:pred_SFR}, quantify 
the relationship between midplane weight 
($\cal W$ or $\PDE$) or pressure ($\Ptot$) and $\SSFR$ as measured in the simulations, and compare the simulation
and theory results to observations 
(\autoref{sec:SFR}, \autoref{sec:obs}).

\section{Numerical Methods and Models}\label{sec:numer}

\begin{deluxetable*}{lCCCCCCCCC}
\tablecaption{TIGRESS Simulation Model Parameters \label{tbl:model}}
\tablehead{
\colhead{Model} &
\colhead{$\Sgas$} &  
\colhead{$\Sigma_\mathrm{gas,0}$} &
\colhead{$\Sigma_*$} &
\colhead{$\rho_*$} &
\colhead{$\rho_{\rm dm}$} &
\colhead{$t_{\rm orb}$} &
\colhead{$L_x,L_y$} &
\colhead{$L_z$} &
\colhead{$\Delta x$} 
\\
\colhead{} &
\colhead{$M_\odot\pc^{-2}$} &
\colhead{$M_\odot\pc^{-2}$} &
\colhead{$M_\odot\pc^{-2}$} &
\colhead{$M_\odot\pc^{-3}$} &
\colhead{$M_\odot\pc^{-3}$} &
\colhead{${\rm Myr}$} &
\colhead{pc} &
\colhead{pc} &
\colhead{pc}
}
\colnumbers
\startdata
R2    & 70 & 150 &  450 & 0.92 & 0.08   &61.4 & 512  &  3584 & 2\\
R4    & 30 &  50 &  208 & 0.42 & 0.024  & 114 & 512  &  3584 & 2\\
R8    & 10 &  12 &   42 & 0.086& 0.0064 & 219 & 1024 &  7168 & 4\\
R16   & 2.5& 2.5 & 1.71 & 0.0035& 0.0014& 518 & 2048 &  14336& 8 \\
LGR2  & 70 & 150 &  110 & 0.11 & 0.015  & 123 & 512  &  3584 & 2\\
LGR4  & 40 &  60 &   50 & 0.05 & 0.005  & 205 & 512  &  3584 & 2\\
LGR8  & 11 &  12 &   10 & 0.01 & 0.0016 & 410 & 1024 &  7168 & 4\\
\enddata
\tablecomments{
Model parameters listed are: (1) model name, (2) target gas surface density, (3) initial gas surface density,
(4) stellar surface density, (5) stellar midplane volume density, 
(6) dark matter midplane volume density,
(7) galactocentric orbit time, (8) horizontal box dimensions,
(9) vertical box dimension,
(10) spatial resolution of simulation.
}
\end{deluxetable*}

\subsection{TIGRESS Implementation}\label{sec:TIGRESS_method}

We use the TIGRESS numerical framework to simulate the three-phase ISM with
self-consistent star formation and feedback \citepalias{Kim_Ostriker2017};
TIGRESS is built on the
{\it Athena} finite-volume code for magnetohydrodynamics
\citep{Stone_2008,Stone_Gardiner2009}.  To focus on local patches within
a differentially rotating galactic disk, we employ shearing-periodic boundary
conditions in the local radial direction ($\hat x$), periodic boundary
conditions in the local 
azimuthal direction ($\hat y$) \citep{Stone_Gardiner2010}, and
open boundary conditions in the vertical direction ($\hat z$). We
use piecewise-linear reconstruction with the Roe Riemann solver.  
The gravity
of the gas and of star particles (representing stellar clusters)
is obtained via a fast Fourier
transform solution of the Poisson equation
\citep{Gammie_2001,Koyama_Ostriker2009}, with the mass of each particle
mapped onto the grid using a triangle-shaped cloud kernel
\citep{Hockney_Eastwood1981}.

Star particles are initially created
as sinks for mass and momentum
on the grid when gravitational collapse causes the numerical
solution to be unresolved, adopting the criteria and methods of
\citet{Gong_Ostriker2013} with modifications as described in
\citetalias{Kim_Ostriker2017}, \citet{Kim_Ostriker_etal2020a}. 
Star particle positions and velocities
are advanced using a symplectic kick-drift-kick leapfrog integrator for Hill's equations \citep{Quinn2010}. 
Star particles may accrete further gas over time and merge with other sinks
up until the point when the first SN occurs  (typically after 3 - 4 Myr).
Each star particle
is surrounded by a $3^3$-cell control volume which is treated as ghost zones for
actively-accreting particles, with the accretion rate  determined
by fluxes of mass and momentum returned by the Riemann solver at
the surfaces of the control volume.  The mass and momentum accreted in
this way is shared between the sink particle and the cells in the
control volume.    
After its first SN event, and up until its
lifetime of 40 Myr, a star particle will no longer accrete or merge.
Throughout their lifetimes, star particles are sources of FUV radiation.

We employ simple cooling functions
suitable for the warm-cold ISM 
at $T<10^{4.2}\K$ \citep[][see \citet{KKO_2008} for form with
correction of a typographical error]{Koyama_Inutsuka2002}, 
and for the ionized and hot ISM at $T>10^{4.2}\K$ \citep{Sutherland_Dopita1993}.
The adopted cooling functions are appropriate for ISM gas at solar
neighborhood abundances
(we do not follow changes in metallicity in the current simulations).

The star cluster particle attributes that lead to feedback are assigned
based on their mass and age using the STARBURST99 population synthesis
package \citep{Leitherer_1999}, assuming a  \citet{Kroupa_2001} IMF and the Geneva evolutionary tracks for non-rotating, solar metallicity stars.  The FUV intensity is taken to be
\begin{equation}\label{eq:JFUV}
J_{\rm FUV}(t)= \frac{\Sigma_{\rm FUV}(t)}{4\pi} f_\tau,
  \end{equation}
with the age-dependent luminosity-to-mass ratio in FUV from STARBURST99
used to obtain the FUV luminosity per unit area $\Sigma_{\rm FUV}$
(averaged over the whole domain) by summing over star particles. The
(time-dependent) factor $f_\tau$ given in \autoref{eq:shielding}  
takes into account attenuation in an  approximate manner, based on the
solution of the equation of radiation transfer in a slab for uniform
emissivity \citepalias{OML_10}, with
$\tau_\perp (t)= \kappa_{\rm FUV} \Sgas(t)$
the mean optical depth to FUV in the direction perpendicular to the disk.
The heating rate coefficient from the photoelectric effect in cells containing
warm or cold gas is set to 
\begin{equation}\label{eq:Gamma}
\Gamma = \Gamma_0\left(\frac{J_{\rm FUV}}{J_{\rm FUV,0}} + 0.0024 \right)
  \end{equation}
where $\Gamma_0=2\times 10^{-26} \ergpers$ and
$J_{\rm FUV,0}= 6.8 L_\odot \pc^{-2} /(4\pi)=2.2\times 10^{-4} \ergpers {\rm cm}^{-2} {\rm sr}^{-1}$
are adopted as reference
values for the heating rate coefficient and mean FUV intensity
in the solar neighborhood.
Photoelectric heating is not applied to gas at $T>10^5\K$.  
As we do not explicitly track ionization in warm
and cold gas in the
present simulations, our photoelectric heating efficiency is effectively
constant, rather than depending on a  grain charging parameter that is
sensitive to $n_e$
\citep{Wolfire_1995,Wolfire_2003}.  
We note that with our adopted heating and cooling functions, the geometric
mean  $P_{\rm two} \equiv (P_{\rm max,warm} P_{\rm min,cold})^{1/2}$
between the maximum warm and minimum cold pressure in thermal equilibrium
is given by 
\begin{equation}\label{eq:Ptwo}
P_{\rm two}/k_B = 3.1 \times 10^3 \pcc \K \left(\frac{J_{\rm FUV}}{J_{\rm FUV,0}} + 0.0024  \right).
\end{equation}
Thus,
if $\Pth = P_{\rm two}$ the value of $\Upsth$ would be
$172 f_\tau/f_{\tau,\odot}  \kms$ with
our adopted heating and cooling functions, which is slightly lower than
it would be with the \citet{Wolfire_2003} heating and cooling functions
(see \autoref{eq:Upsth_theory}).  The above assumes
${J_{\rm FUV}}/{J_{\rm FUV,0} }\gg 1$, as is generally the case.

The treatment of SNe is as described in \citetalias{Kim_Ostriker2017}, where
full details and tests of the method are presented. The SN
event rate from any given star particle is set by its mass and age, and
we allow for runaways by ejecting a massless test particle with 50\%
probability for each event.
We turn off runaways in the R2 model, however,
for the sake of computational efficiency.  Different treatments of
SN events are applied depending on the density in ambient gas, which is used to
compute the ratio ${\cal R}_M$ between the mass in the feedback region
and the mass that the remnant would have (at that density) when it becomes
radiative, calibrated by \citet{Kim_Ostriker2015a} to be
$M_{\rm sf}= 1540 M_\odot (n_{\rm H}/ \pcc)^{-0.33}$.
For the majority of cases, where a feedback region of radius
at least $R_{\rm min}=3\Delta x$  has ${\cal R}_M <1$,
the Sedov-Taylor stage is resolved
and we deposit $10^{51}$ ergs of energy within the feedback region
(72\% thermal and 28\% kinetic).
For a minority of cases, the ambient 
density may be high enough so that ${\cal R}_M >1$
even for the smallest allowed feedback region radius, in which case
the Sedov-Taylor stage is unresolved and
we deposit momentum equal to $p_{\rm final}= 2.8 \times 10^5 M_\odot \kms
(n_{\rm H}/\pcc)^{-0.17}$  \citep[calibrated by][]{Kim_Ostriker2015a} within
the feedback region.  For the simulations presented here, more 
than 90\% of the SN are well resolved, with ${\cal R}_M <0.1$, and
are treated with energy rather than momentum deposition.

\subsection{Model Parameters}\label{sec:models}

The MHD simulations we analyze here are the same as those used for analysis
of outflow properties in \citet{Kim_Ostriker_etal2020a,Kim_Ostriker_etal2020b}.
Physical and numerical parameters for the seven simulations are listed in
\autoref{tbl:model}.  These simulations cover a range of background states
for the gas and gravitational potential (stellar and dark matter) that
would be encountered in nearby disk galaxies.  Although galactocentric
radius does not directly enter in the equations for a local shearing-box
model, galaxies generally have gas and stellar surface densities
that decline with increasing radius, so we label our models based on a nominal
galactocentric radius (in kpc).  There are two sets of models, R2 to R16 (nominal radius $R=$~2 to 16 kpc with
higher external gravity), and LGR2 to LGR8 ($R=$~2 to 8 kpc with  lower external
gravity).

\autoref{tbl:model} lists the (constant in time) parameters for
the stellar disk and dark matter halo.  The scale height of the stellar
disk is $z_*=245\pc$ (R2, R4, R8, R16) or
$z_*=500\pc$ (LGR2, LGR4, LGR8), with stellar surface density $\Sigma_*$
related to midplane stellar volume density by $\rho_*  = \Sigma_*/(2 z_*)$ based
on Equation 6 of \citetalias{Kim_Ostriker2017}.  The Table lists both the initial
gas surface density $\Sigma_\mathrm{gas,0}$ at the time the simulation is begun,
and the target surface density $\Sgas$
after a transient stage of evolution;
ISM and star formation properties are measured when the
mean gas surface density  of  the
disk is near this target value.  The orbit time for the simulation
domain about the galactic center is listed, together with the physical
box size.  The number of zones in the
domain is $N_x \times N_y \times N_z = 256 \times 256 \times 1792$ for
all models.

All models are initiated with a horizontal magnetic field
aligned in the
$\hat y$ (i.e.~azimuthal) direction, with plasma $\beta \equiv 8 \pi \Pth/B^2$
everywhere equal to $\beta=10$ for all models except R2, which has initial
$\beta =2$.  We set the initial magnetic field higher (lower $\beta$)
for this model because the
orbital time is significantly shorter for  model R2 than other models, so
that the background magnetic field would not have time to grow.

The R8 model is the most similar in its parameters to conditions in
the solar neighborhood. A simulation with the same parameters (but slightly
different treatment of star particles and initial turbulence) was previously
presented as the fiducial TIGRESS model in \citetalias{Kim_Ostriker2017},
with outflow properties analyzed in \citet{Kim_Ostriker2018,Vijayan_2020}.  
The LGR4 model is the most similar to the mean (weighted by molecular mass)
properties in the PHANGS survey of nearby star-forming galaxies.

We note that in addition to the standard runs listed in \autoref{tbl:model},
additional
simulations with the same physical parameters but
different numerical resolution and computational domain size 
have been run to confirm convergence; see Section 4 in \citetalias{Kim_Ostriker2017}
and Appendix A in \citet{Kim_Ostriker_etal2020a}.

\begin{deluxetable*}{lCCCCCCCCCCCC}
\tabletypesize{\scriptsize}
\tablecaption{Measured Star Formation and ISM Properties \label{tbl:properties}}
\tablehead{
\colhead{Model} &
\colhead{$\Sigma_\mathrm{gas}$} &
\colhead{$\Sigma_\mathrm{SFR}$} &
\colhead{$\tdep$} &
\colhead{$P_\mathrm{DE}$} &
\colhead{$P_\mathrm{tot,hot}$} &
\colhead{$P_\mathrm{tot,2p}$} &
\colhead{$P_\mathrm{th,2p}$} &
\colhead{$P_\mathrm{turb,2p}$} &
\colhead{$\Pi_\mathrm{mag,2p}$} &
\colhead{$n_\mathrm{H,2p}$} &
\colhead{$\overline{\sigma}_{{\rm eff,2p}}$} &
\colhead{$H_\mathrm{2p}$} 
\\
\colhead{} &
\colhead{$(M_\odot/{\rm pc}^{2})$} &
\colhead{$(M_\odot/{\rm pc}^{2}/{\rm Myr})$} &
\colhead{(Myr)} &
\colhead{(${\rm cm}^{-3} \, {\rm K}$)} &
\colhead{(${\rm cm}^{-3} \, {\rm K}$)} &
\colhead{(${\rm cm}^{-3} \, {\rm K}$)} &
\colhead{(${\rm cm}^{-3} \, {\rm K}$)} &
\colhead{(${\rm cm}^{-3} \, {\rm K}$)} &
\colhead{(${\rm cm}^{-3} \, {\rm K}$)} &
\colhead{$({\rm cm^{-3}})$} &
\colhead{$({\rm km\, s^{-1}})$} &
\colhead{$({\rm pc})$}
}
\colnumbers
\startdata
R2   & 70.9   &  1.10               & 6.45\times\!10^{1} & 2.00\times\!10^{6}& 1.36\times\!10^{6} & 1.92\times\!10^{6} & 1.13\times\!10^5 & 1.26\times\!10^6 &  5.37\times\!10^5 & 26.1 & 43.4 & 282\\
R4   & 30.2   &  5.37\times\!10^{-2} & 5.63\times\!10^{2} &3.57\times\!10^{5} & 1.44\times\!10^{5} & 2.41\times\!10^{5} & 1.76\times\!10^4 & 1.95\times\!10^5 & 2.22\times\!10^4 & 2.52 & 30.3 & 294\\
R8   & 9.87   &  2.67\times\!10^{-3} & 3.70\times\!10^{3} &2.34\times\!10^{4} & 1.70\times\!10^{4} & 1.91\times\!10^{4} & 5.02\times\!10^3 & 5.71\times\!10^3 & 7.86\times\!10^3 & 1.18 & 13.9 & 351\\
R16  & 2.45   &  6.21\times\!10^{-5} & 3.94\times\!10^{4} &1.03\times\!10^{3} & 1.03\times\!10^{3} & 7.77\times\!10^{2} & 3.39\times\!10^2 & 1.88\times\!10^2 & 1.67\times\!10^2 & 0.0919 & 10.7 & 679\\
LGR2 & 65.9   &  1.17\times\!10^{-1} & 5.65\times\!10^{2} &9.14\times\!10^{5}& 6.66\times\!10^{5} & 1.01\times\!10^{6} & 6.60\times\!10^4 & 6.41\times\!10^5 & 2.77\times\!10^5 & 15.3 & 34.3 & 460\\
LGR4 & 39.6   &  5.41\times\!10^{-2} & 7.32\times\!10^{2} &1.80\times\!10^{5} & 1.01\times\!10^{5} & 1.38\times\!10^{5} & 1.34\times\!10^4 & 1.01\times\!10^5 & 1.92\times\!10^4 & 3.49 & 19.4 & 307\\
LGR8 & 11.2   &  2.16\times\!10^{-3} & 5.17\times\!10^{3} &1.09\times\!10^{4} & 7.39\times\!10^{3} & 9.65\times\!10^{3} & 2.36\times\!10^3 & 4.78\times\!10^3 & 1.91\times\!10^3 & 0.738 & 11.3 & 438\\

\enddata
\tablecomments{Numerically measured quantities in each model.  Reported
  values are medians of the distribution for the sampling 
  period indicated in \autoref{fig:Sigma_hist}. Pressures and density are
  based on horizontal averages at the midplane.  The ``2p'' subscript
  indicates that only warm-cold gas ($T<2\times10^4\K$)  
  is included in the measurement.  Effective velocity
  dispersion  includes all pressure components (see text).}
\end{deluxetable*}

\vfil

\section{Numerical Results}\label{sec:results}

\begin{figure}
\centering\includegraphics[width=0.45\textwidth]{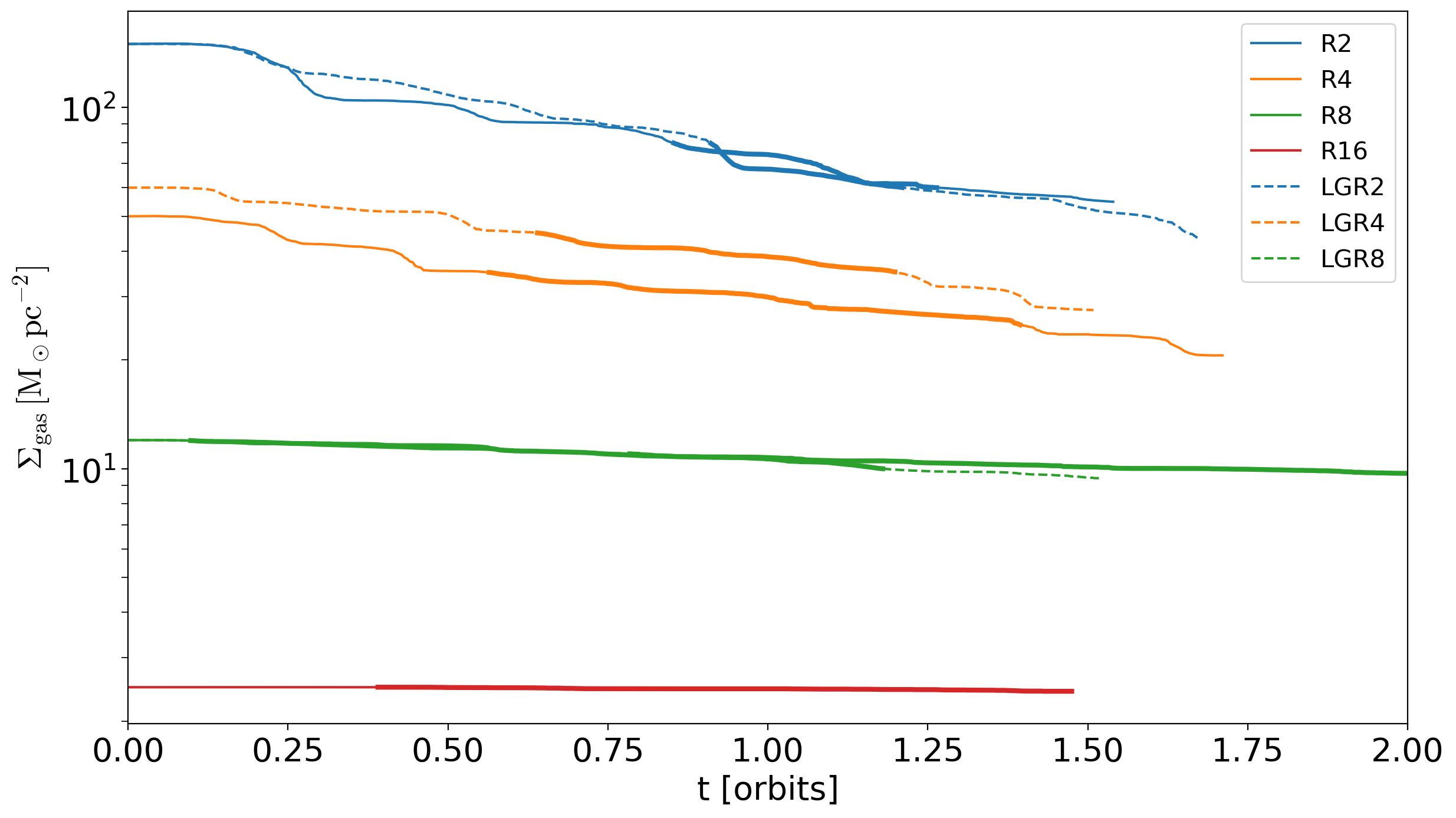}
\caption{Evolution of gas surface density in all models.  The interval selected
  for computing averaged {tabulated} quantities in each model is marked with a heavy curve.
\label{fig:Sigma_hist}
}
\end{figure}

\begin{figure}
\centering\includegraphics[width=0.45\textwidth]{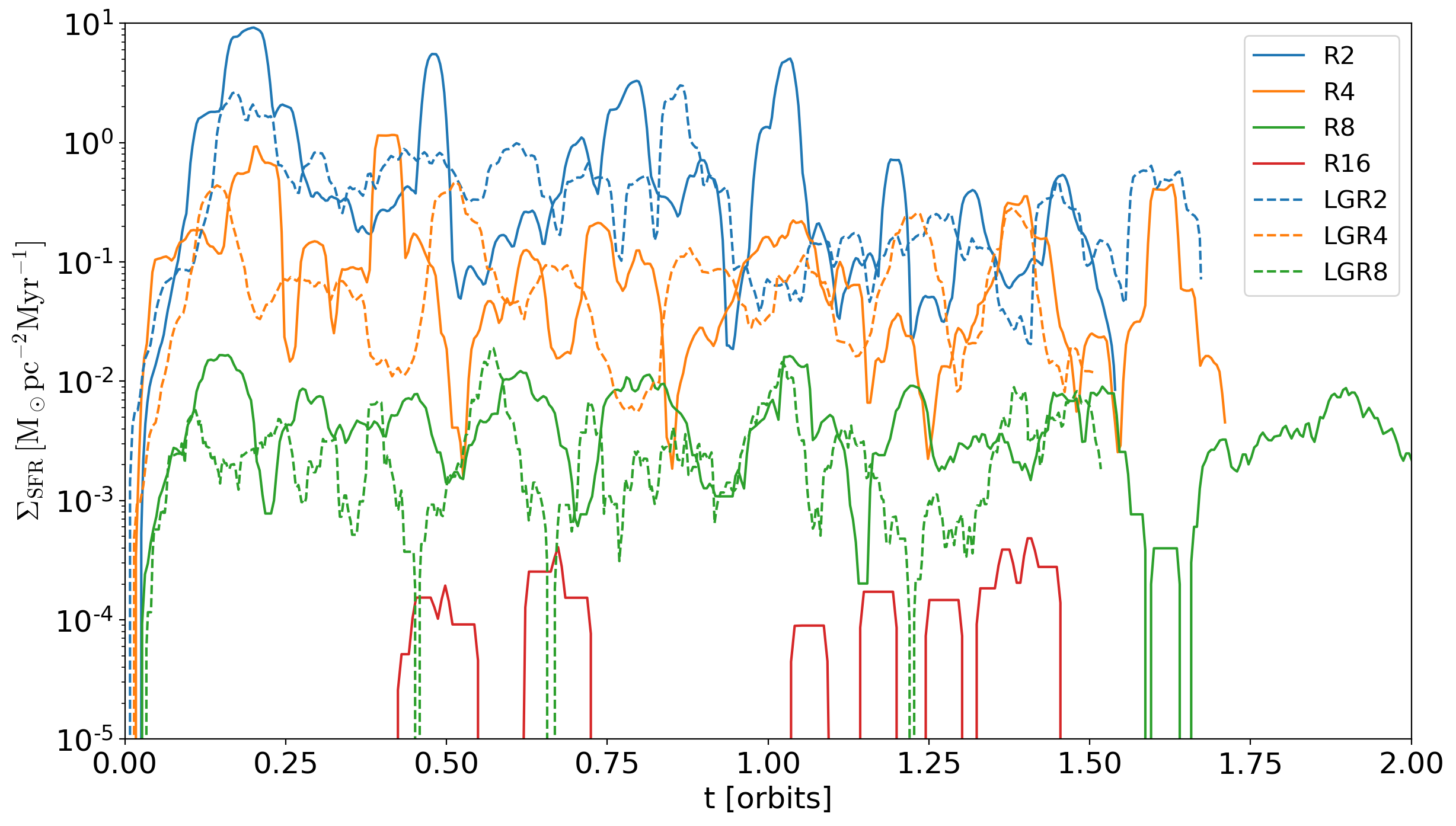}
\caption{Evolution of the star formation rate surface density in all models, calculated based on the mass in new stars formed within the previous 1 Myr, and then
  smoothed over 10  Myr.
\label{fig:Sigma_SFR_hist}
}
\end{figure}

\subsection{Temporal evolution and ISM structure}\label{sec:evolution}

Evolution of the gas surface density $\Sgas\equiv M_\mathrm{gas}/(L_x
L_y)$
for all TIGRESS models is shown in \autoref{fig:Sigma_hist}.  For each
model, the temporal range used for computing the ISM and star
formation properties {we tabulate} is marked as a heavy curve.
{This interval is chosen to have a range within $\sim 10\%$ of  the target
  $\Sgas$, so that tabulated results
  represent well-defined conditions in the disk.}
The median value of
$\Sgas$ over this sampling interval for each model is listed in
\autoref{tbl:properties}.  Evolution of the star formation rate surface
density $\SSFR \equiv \dot M_*/(L_x L_y)$ for all models is shown in
\autoref{fig:Sigma_SFR_hist}, with the median $\SSFR$ over the
sampling interval as well as the depletion time defined by
$\tdep=\Sgas/\SSFR$ listed in \autoref{tbl:properties}.

As previously
discussed in \citet{Mao_2020,Vijayan_2020,Kim_Ostriker_etal2020a}, the values of $\SSFR$
fluctuate in time as the gas cycles between phases where there is
a relatively large quantity of dense gas and star formation, and
phases where feedback has dispersed much of the dense gas, reducing
$\SSFR$. Feedback also causes the gas scale height $H$ to oscillate in time
\citep{Kim_Ostriker_etal2020a}, but for both $\SSFR$ and $H$ (and other box-averaged
variables)
there are well-defined quasi-steady mean values subsequent to the initial
transients.  \autoref{tbl:properties} includes the median value of the
root mean square scale height $H\equiv (\sum_{ijk} \rho z^2  /\sum_{ijk} \rho )^{1/2}$
for the warm-cold ($T< 2\times 10^4\K$) ``two-phase'' gas. Hereafter we shall
use the subscript ``2p'' to denote quantities computed based on selecting
only zones with gas in this warm-cold range.

\begin{figure*}
\centering
\includegraphics[width=\textwidth]{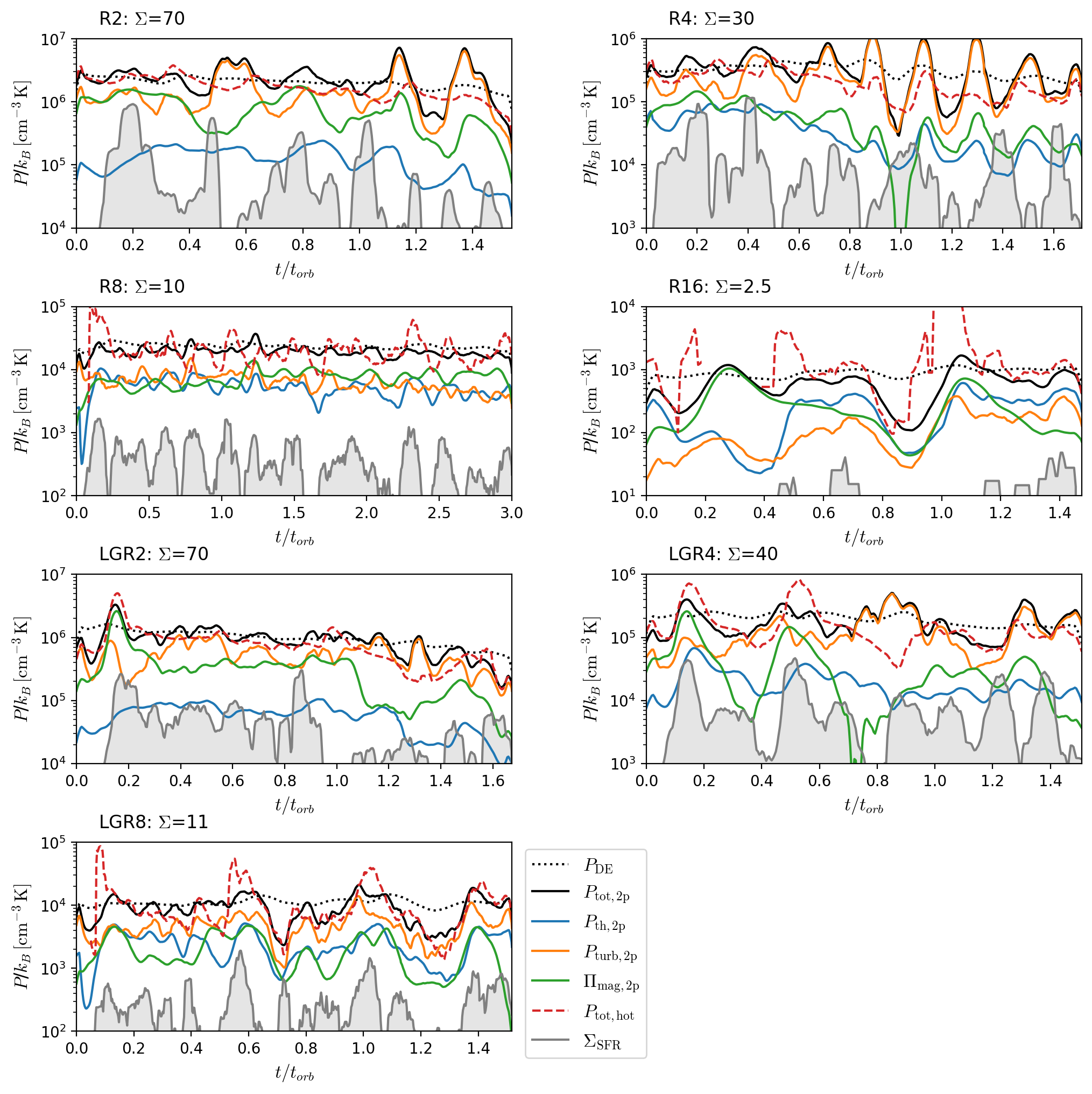}
\caption{Evolution of midplane pressures in all models. Solid curves show total pressure
  and component pressures (thermal, turbulent, magnetic) for the
  two-phase (warm-cold) gas, dashed curves show total pressure for hot gas, and dotted
  curves show estimated dynamical equilibrium pressure.   For reference, $ \SSFR$ is also shown, in units $10^{-5}\sfrunit$ (gray shading).  
\label{fig:pressure_hist}
}
\end{figure*}

\begin{figure*}
  \plottwo{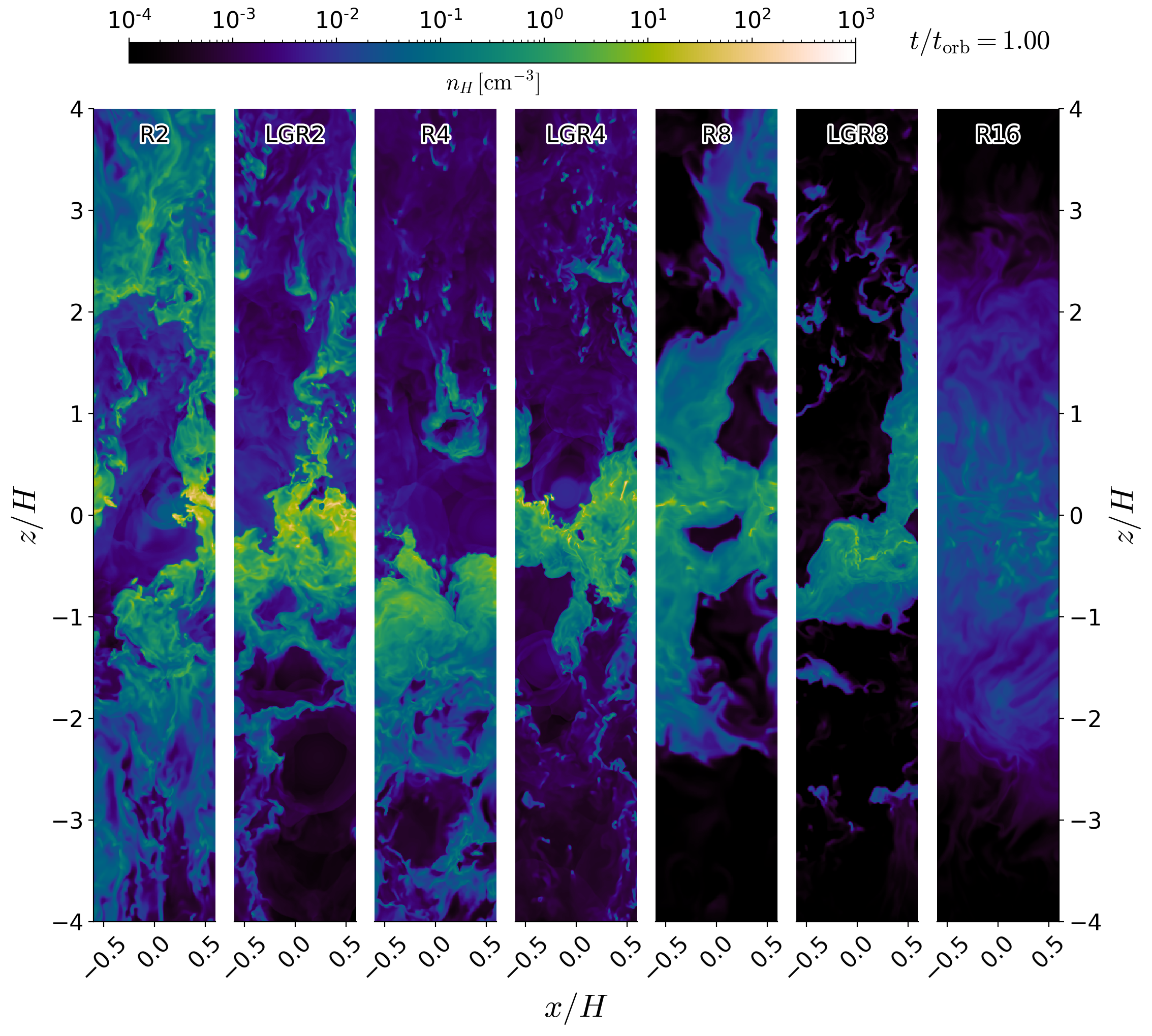}{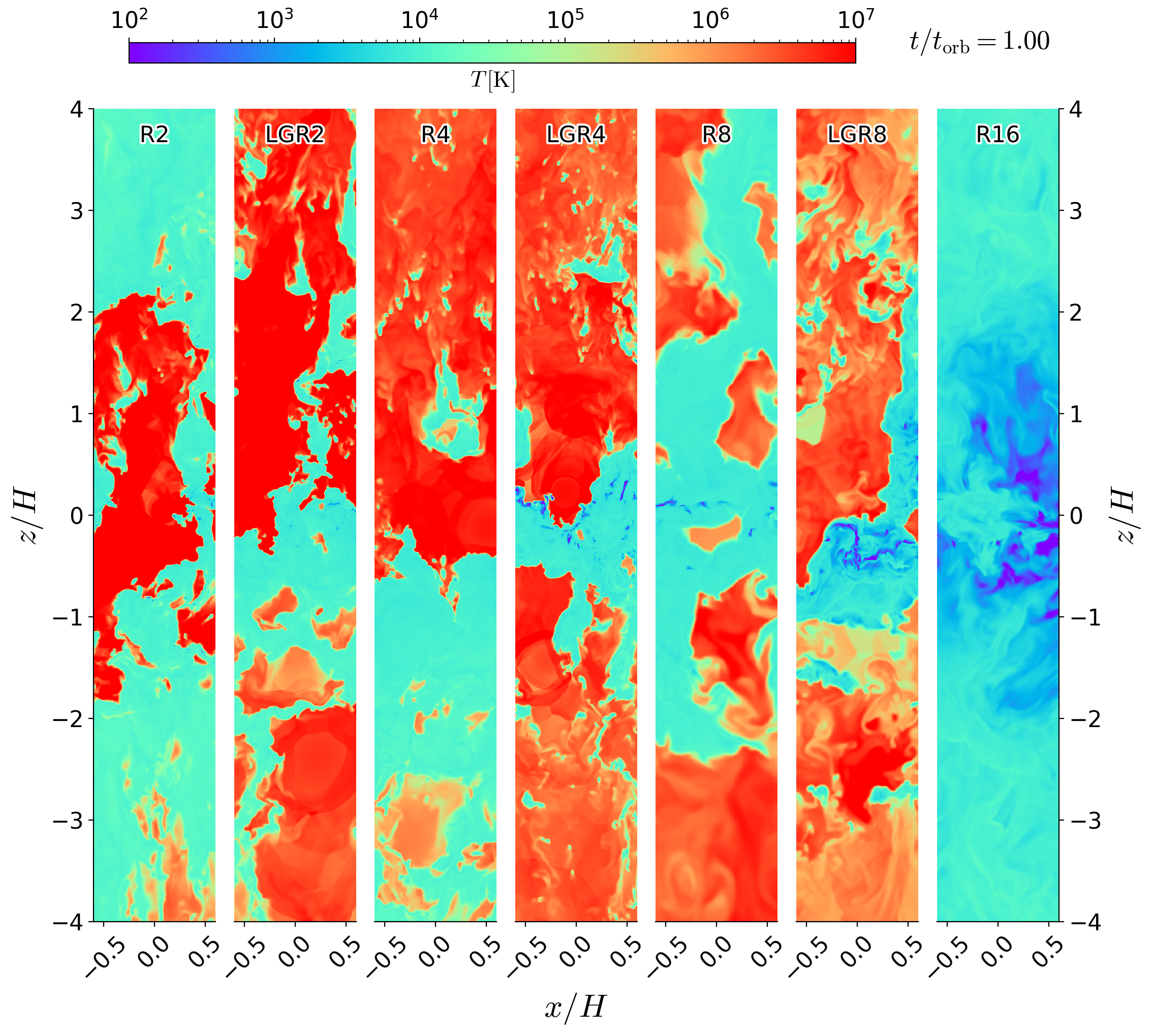}
  \caption{Vertical slice showing hydrogen number density $n_H$ and
    gas temperature $T$ 
    in all models at $t=t_\mathrm{orb}$.
    Axes are in units of the respective scale height $H_{\rm 2p}$ for each  model (see Column (13) in \autoref{tbl:properties}).
\label{fig:den_temp_slice}
  }
  \end{figure*}

Contributions to the total pressure in the TIGRESS simulations include
thermal, turbulent, and magnetic terms.  For the purpose of
considering overall force balance in the direction perpendicular to
the disk, the relevant component pressures are horizontal averages at
the midplane of the corresponding stress
terms in the vertical momentum equation.  The
individual terms are: $\Pth\equiv \rho c_s^2$ (thermal pressure),
$\Pturb \equiv \rho v_z^2$ (vertical Reynolds stress, i.e.~effective
turbulent pressure),
and $\Pimag \equiv (|{\bf B}|^2 - 2B_z^2)/(8\pi) = (B_x^2 + B_y^2
- B_z^2)/(8\pi)$ (vertical Maxwell stress, combining magnetic pressure and
tension).
\autoref{fig:pressure_hist} shows the temporal evolution for all
models of these individual terms in the two-phase gas at the midplane,
as well as the total of these terms,
$P_{\rm tot,2p} =P_{\rm th,2p} + P_{\rm turb,2p} + \Pi_{\rm mag,2p}$.
The evolution of the total pressure $P_{\rm tot,hot}$ for the hot gas ($T> 5\times 10^5\K$)
is also shown, together with the evolution
of the commonly-used estimator for the midplane pressure based on
dynamical equilibrium, $\PDE$ (\autoref{eq:PDEdef}).
For each model,
median values over the sampling interval of individual pressure components
(2p only)
and totals (2p and hot) 
are listed in \autoref{tbl:properties}.
In \autoref{tbl:properties} and elsewhere, pressure values with subscript ``2p'' or ``hot'' denote horizontal averages made at the 
midplane in the simulations.  
As we shall
discuss in more detail below, the total pressure for the warm-cold and
hot phases are comparable, and while these fluctuate in time they remain
close to the expected dynamical equilibrium pressure, $\PDE$.

In any zone, the effective vertical 
velocity dispersion is computed using $\sigma_{\rm eff}^2 \equiv P_{\rm tot}/\rho$, 
and we define the mass-weighted average of the effective vertical velocity dispersion as
$\overline{\sigma}_{\rm eff} \equiv (\sum_{ijk} P_{\rm tot}/\sum_{ijk} \rho)^{1/2}$.
\autoref{tbl:properties} lists the time average of this RMS velocity dispersion computed from all (not just midplane) two-phase gas.  We use this  $\overline{\sigma}_{\rm eff,2p}$ in $\PDE$.
\autoref{tbl:properties} also lists, for the two-phase  gas, 
the median midplane values of the gas hydrogen number density, $n_{\rm H,2p}=\rho_{\rm 2p}/(1.4m_{\rm H})$.

Snapshots of density and temperature slices
(\autoref{fig:den_temp_slice}) show that the gas is highly structured.
The warm and cold gas is concentrated in the midplane, but SNe drive
fountain flows in the warm gas extending to several kpc, together with
hot winds that escape from the disk. Temporally-averaged 
vertical profiles of density (total and individual phases)
as well as pressure (total and individual components) are smooth
\citep[see][]{Kim_Ostriker2018,Vijayan_2020,Kado-Fong2020,Kim_Ostriker_etal2020a},
but the instantaneous
snapshots show a highly inhomogeneous medium, and except for model R16
there is comparable volume near the midplane occupied by hot and warm
gas \citepalias[see][]{Kim_Ostriker2017}.

\begin{figure}
  \centering
  \includegraphics[width=0.45\textwidth]{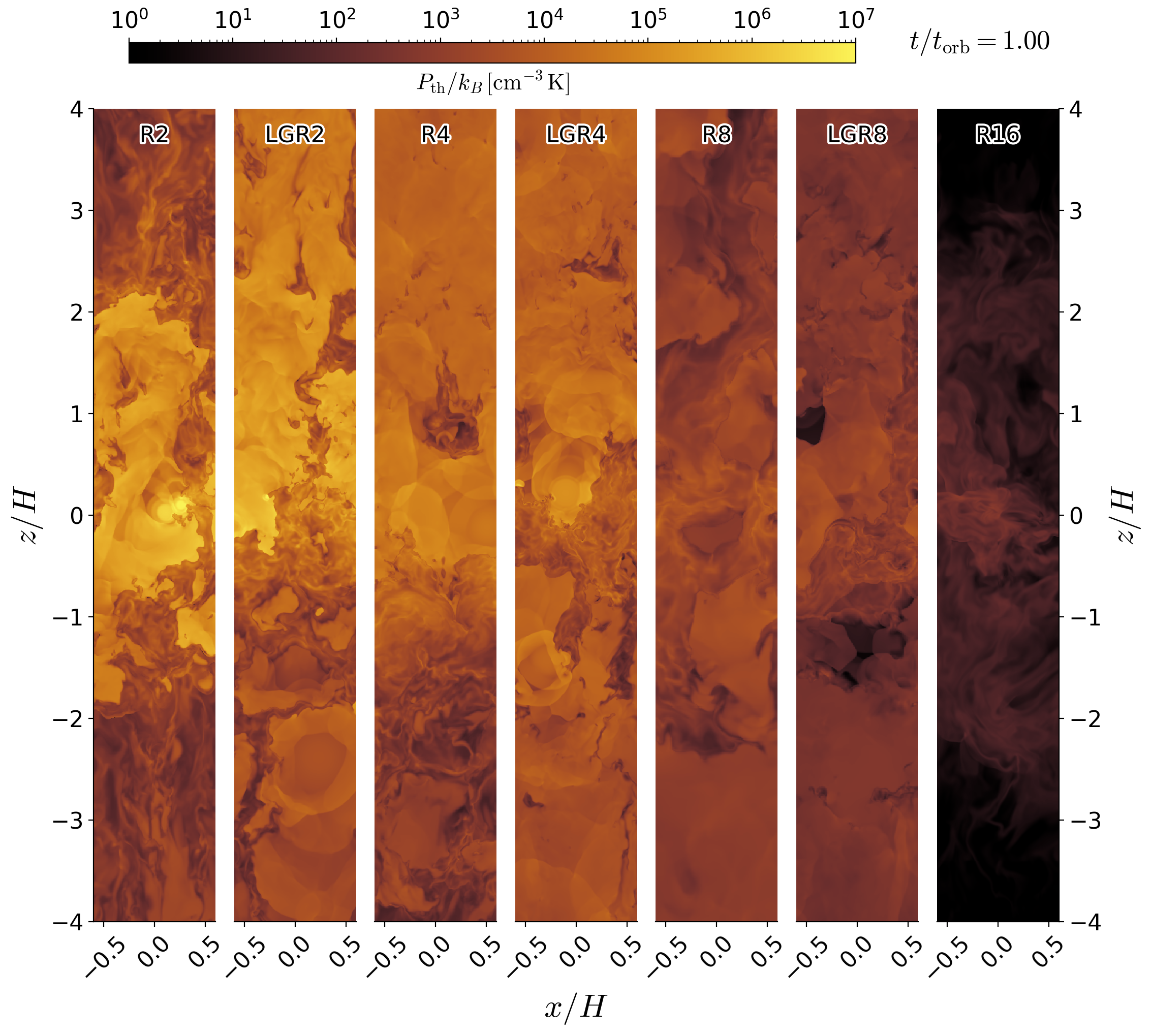}
  \includegraphics[width=0.45\textwidth]{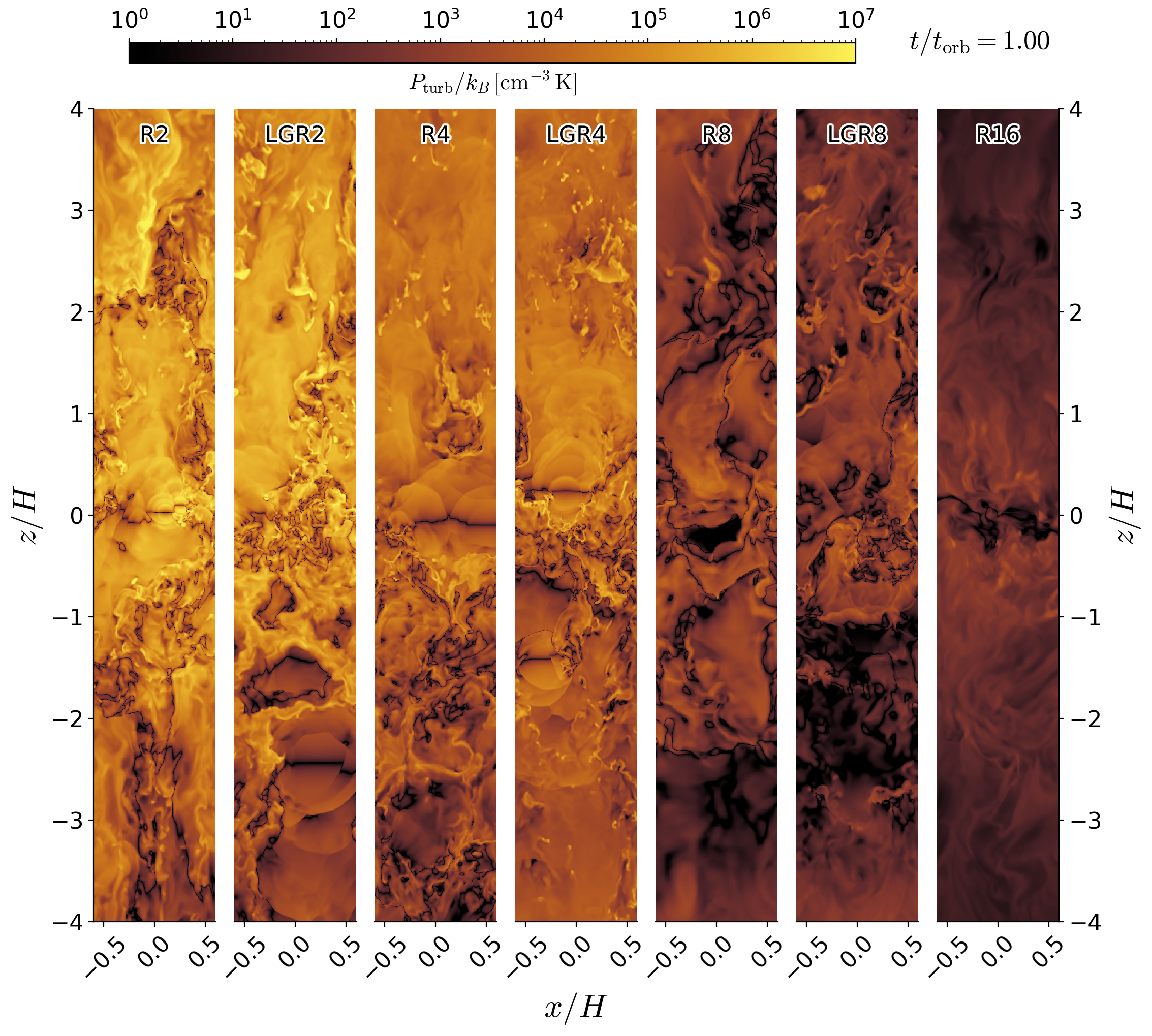}
  \includegraphics[width=0.45\textwidth]{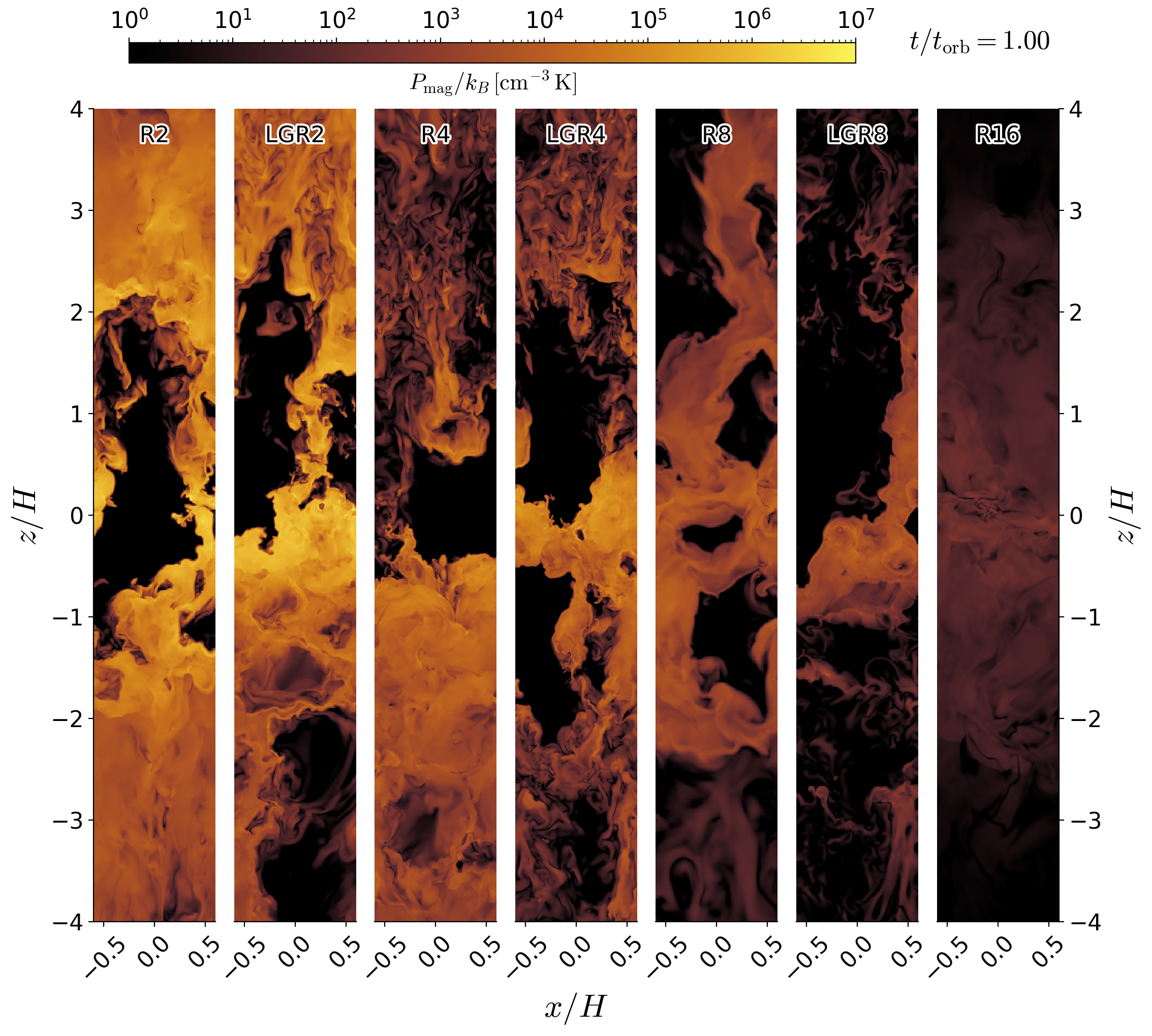}
  \caption{Vertical slices showing thermal pressure $\Pth$ (top),
    turbulent pressure  $\Pturb \equiv \rho v_z^2$  (middle), and
    magnetic pressure  $\Pmag \equiv |{\bf B}|^2/(8\pi)$  (middle)
    in all models (labeled at top) at $t=t_\mathrm{orb}$.
    Axes are in units of the respective scale height for each  model.
\label{fig:Pth_Pturb_Pmag_slice}
  }
\end{figure}

While there are orders of magnitude difference between density and temperature
of the different phases, pressures are much more similar.
\autoref{fig:Pth_Pturb_Pmag_slice} shows that hot bubbles and outflow
``chimneys'' have thermal pressure slightly larger than that of the warm-cold
gas.  Turbulent pressure is similar in magnitude between the hot gas and
warm/cold gas, although the latter has fluctuations on smaller spatial scales.  
The magnetic pressure $P_{\rm mag} \equiv |{\bf B}|^2/(8\pi)$ is, however, quite
  small within the volume occupied by the low-density hot gas.

\begin{figure}
\centering\includegraphics[width=0.48\textwidth]{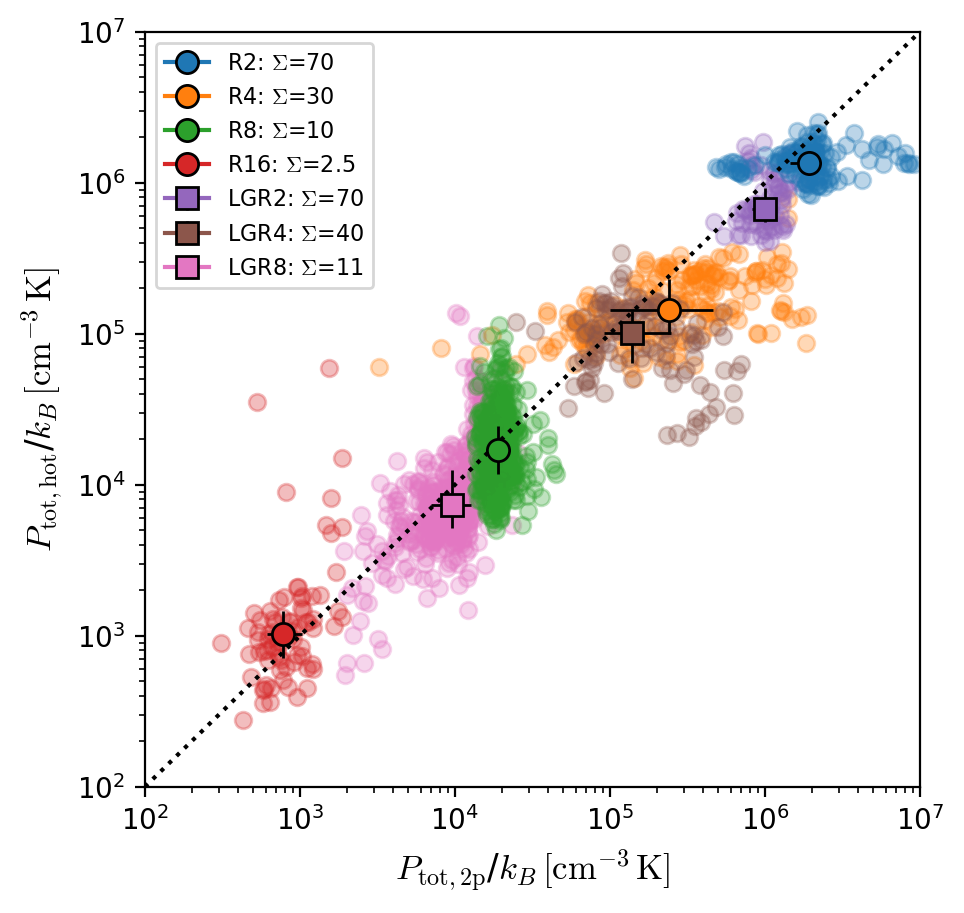}
\caption{Total vertical pressure $P_{\rm tot} \equiv \Pth + \Pturb + \Pimag$ in hot gas vs. two-phase gas for all models.
  Midplane-averaged values at intervals of 1 Myr are shown with individual
  small circles, and medians over the temporal domain shown in \autoref{fig:Sigma_hist}
  are shown as large points with 25$^{th}$ and 75$^{th}$ percentiles indicated.
  For reference the dotted line shows the identity $P_{\rm tot,hot}=P_{\rm tot,2p}$.
\label{fig:P2p_Phot}  
}
\end{figure}

\subsection{Pressure equilibrium}\label{sec:equil}

As noted above, the pressures in warm/cold gas and in hot gas for
individual snapshots are generally similar.  In more detail, from
\autoref{fig:pressure_hist} the hot gas and
two-phase gas do not track each other's fluctuations in the midplane
average pressure (hot gas
pressure is largest during SN feedback episodes; two-phase gas pressure
variations are more complex), but nevertheless the median value of the
total vertical pressure at the midplane is only slightly lower ($\sim
30\%$) for hot gas than for warm/cold gas (see \autoref{tbl:properties}).
\autoref{fig:P2p_Phot} shows
midplane values of the total pressure in the
two-phase and the hot gas at intervals of 1 Myr, along with median values
over the sampling interval.  From a theoretical point of
view, in any quasi-steady non-self-gravitating
system the time-averaged pressures must be similar 
in different thermal phases, because otherwise the component with higher
pressure would have expanded to occupy a larger (average) fraction of the
volume.  Of course, any component that is strongly self-gravitating would
be expected to have significantly larger pressure, but this is not the case
for the two-phase gas overall; only a small fraction of the (cold, dense)
gas in the TIGRESS simulations is gravitationally bound \citep{Mao_2020}.

\begin{figure*}
  \plottwo{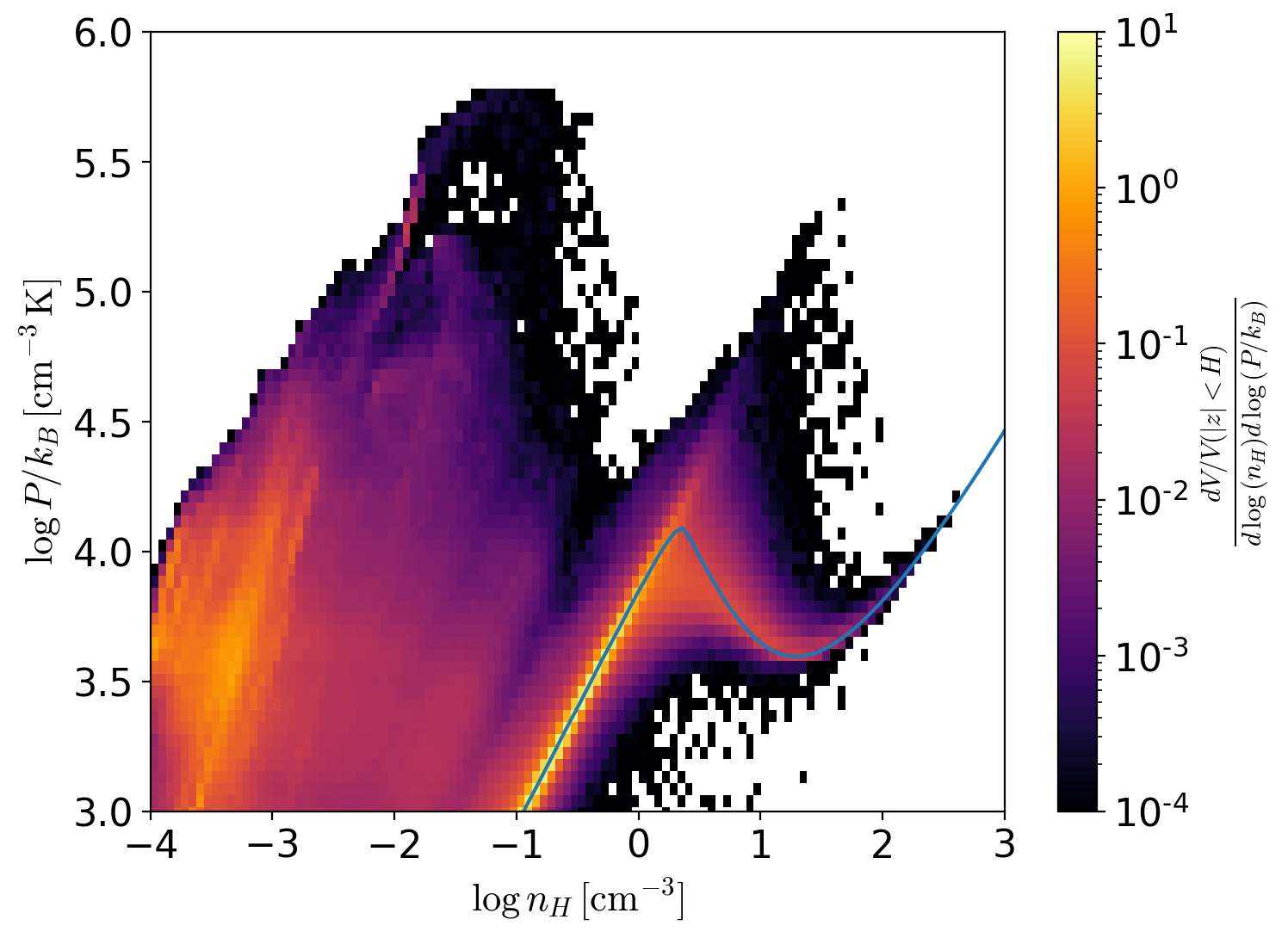}{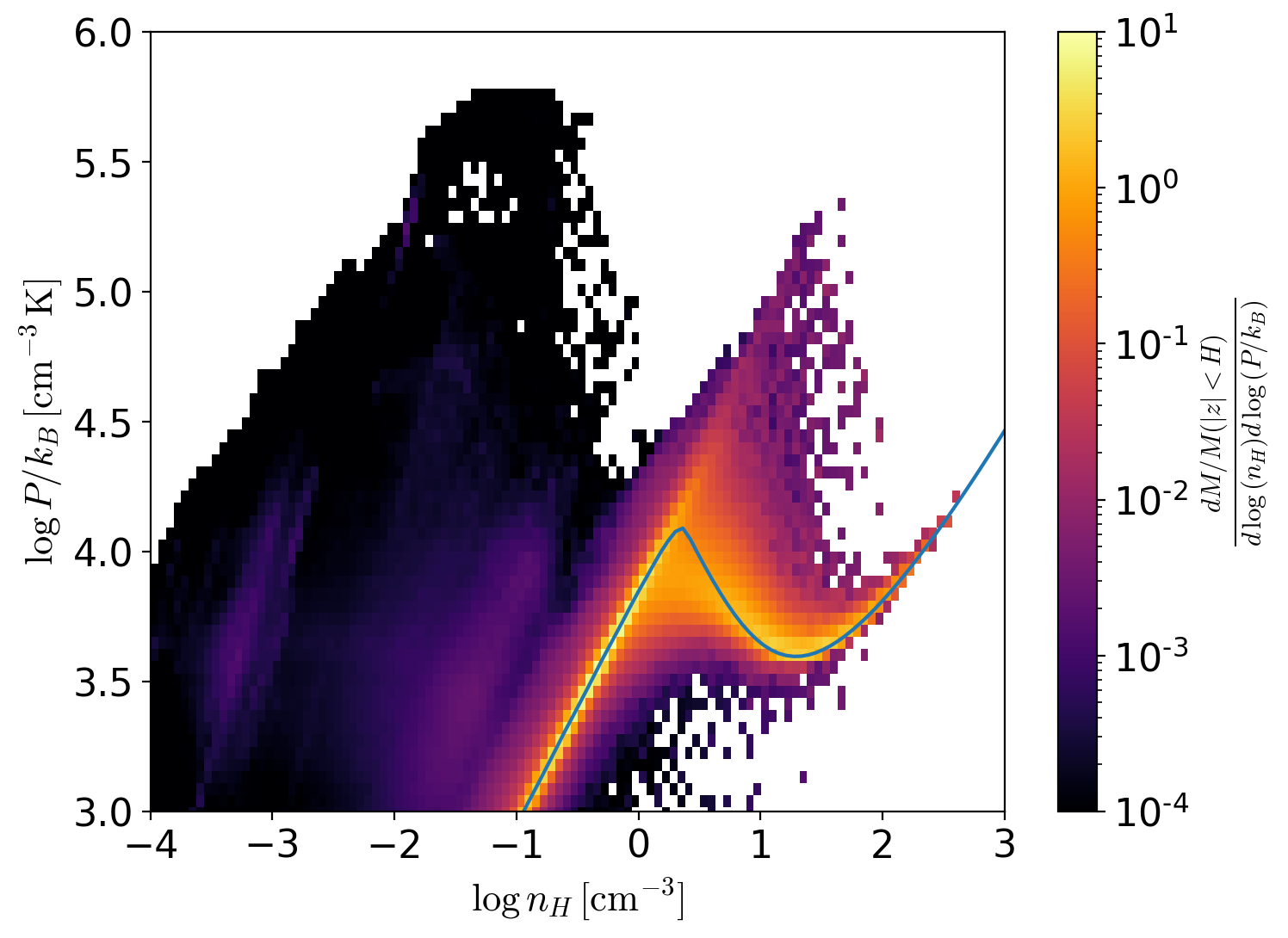}
  \plottwo{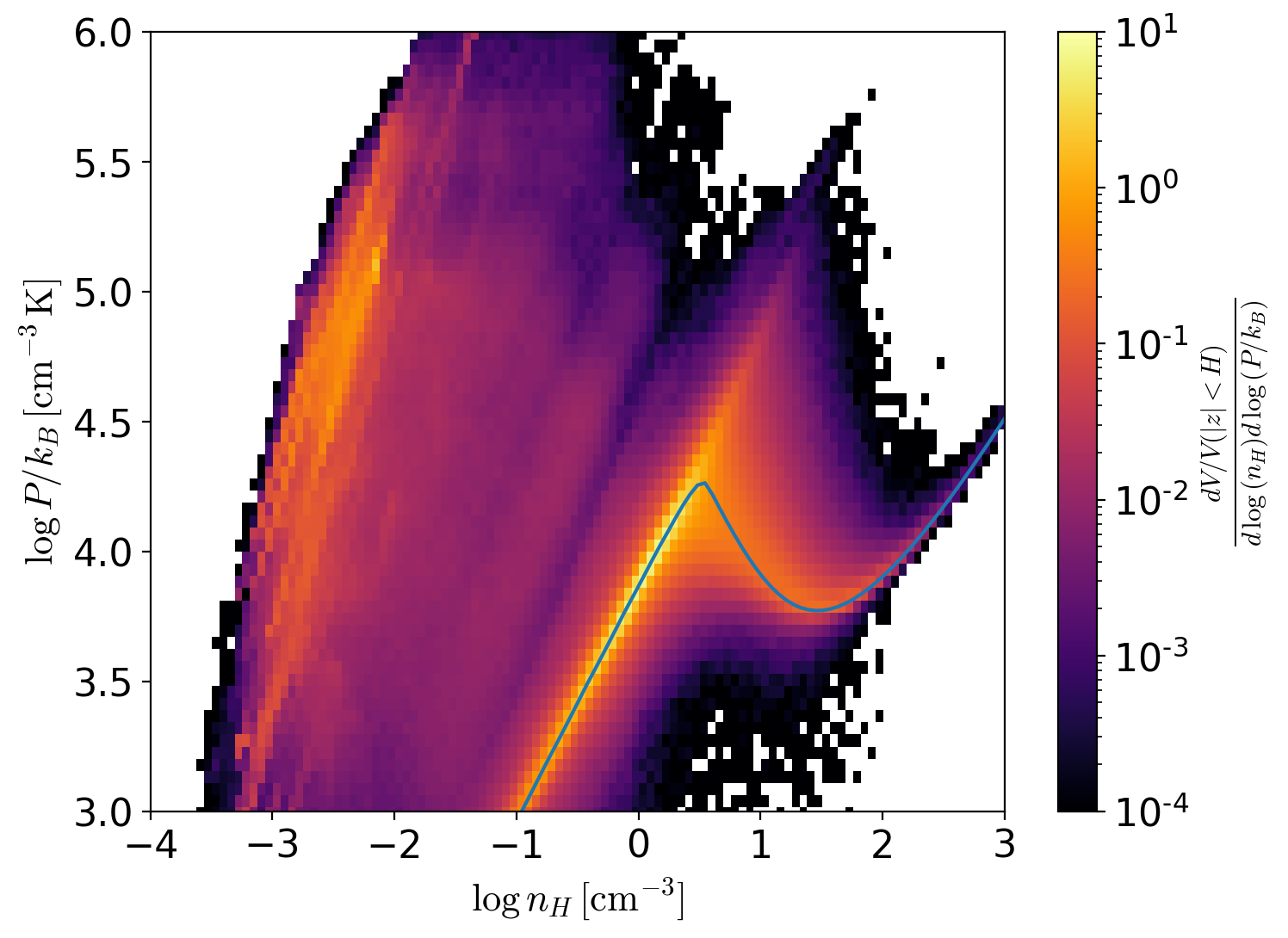}{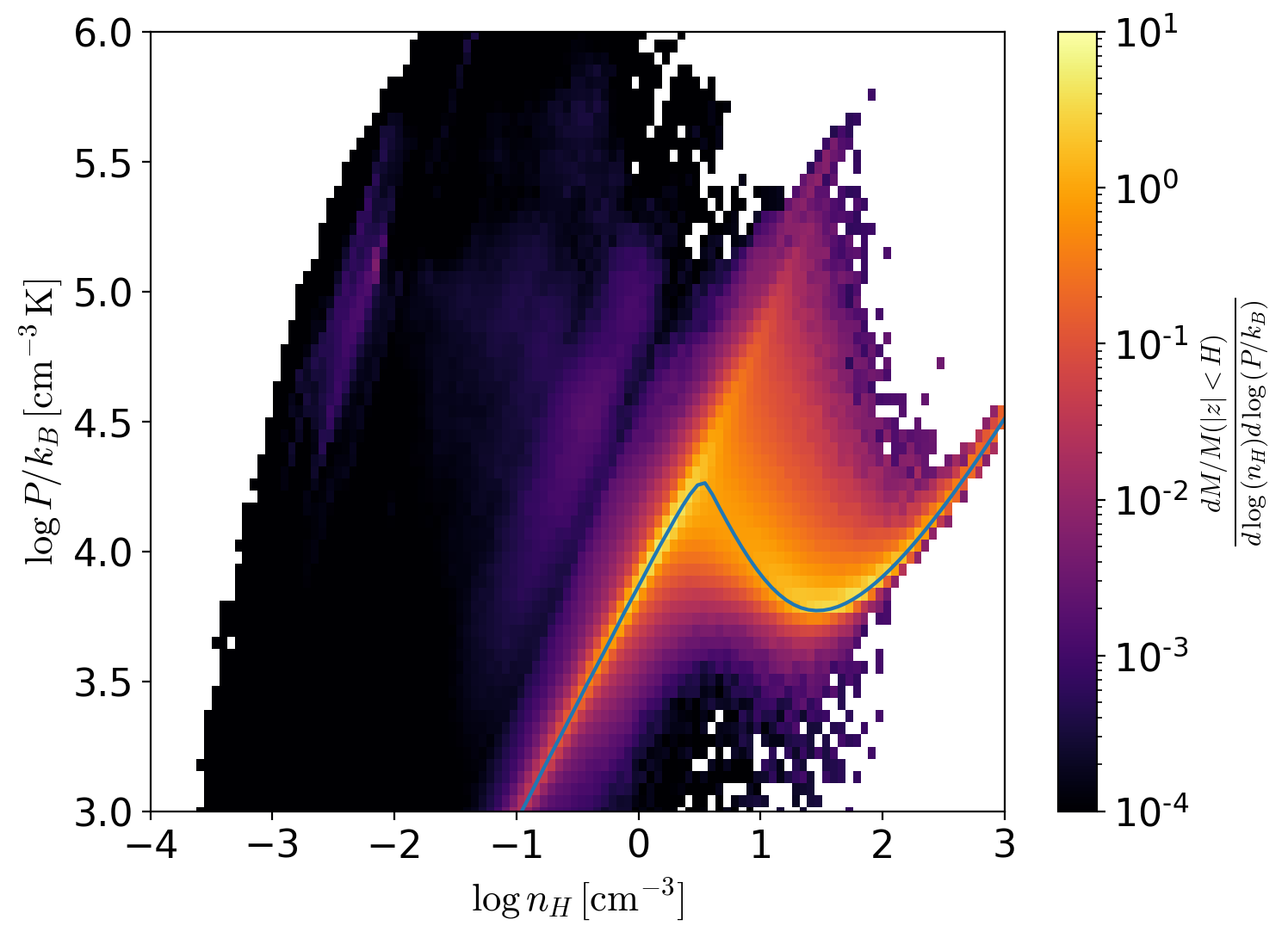}
  \caption{Thermal pressure vs. density for all gas within one scale height of the midplane.  Top row shows model R8, and bottom row shows model LGR4, each at $t=\torb$.  Both volume-weighted (left) and mass-weighted (right) PDFs are shown.  In each panel, an overlay of the equilibrium curve for the instantaneous heating rate is shown.
\label{fig:Pth_vs_n}
  }
  \end{figure*}

In the limit of very short cooling time compared to the dynamical
time, the density and thermal pressure distribution would follow the thermal
equilibrium curve obtained by balancing radiative heating and cooling,
$n \Lambda=\Gamma$, for $\Gamma \propto J_\mathrm{FUV} \propto \SSFR$
when photoelectric heating dominates (see \autoref{eq:Gamma}).
\autoref{fig:Pth_vs_n} shows, for model R8 (top row) and model LGR4
(bottom row) the joint pdfs of density and thermal pressure at $t=\torb$.  The
mass-weighted distributions (right panels) show that the
cold gas (which has the shortest cooling time) is very close to the
thermal equilibrium curve set by the instantaneous heating rate. At
higher (lower) heating rate, the equilibrium curve would shift diagonally
upward/rightward (downward/leftward) with the same characteristic warm
gas temperature, which is set by Ly$\alpha$ cooling.  

For the volume-weighted distributions (left panels), the hot
gas locus is evident at high temperature.  Since the cooling time in the
hot gas is very  long, its pressure adjusts only by dynamical means,
and at any  given time can be either higher or lower than that of the
warm gas (see also \autoref{fig:pressure_hist}); in the  particular snapshots
shown, hot gas is slightly overpressured for LGR4 and slightly
underpressured for R8.  

In addition to thermal equilibrium, a second aspect of equilibrium involving pressure in
galactic disks is dynamical equilibrium.   As described in
\autoref{sec:theory}, vertical dynamical equilibrium is satisfied if
the difference between midplane pressure and the pressure above the
main gas layer is equal to the weight of that gas in the total
gravitational  potential, $\cal W $.

\begin{figure}
\centering\includegraphics[width=0.45\textwidth]{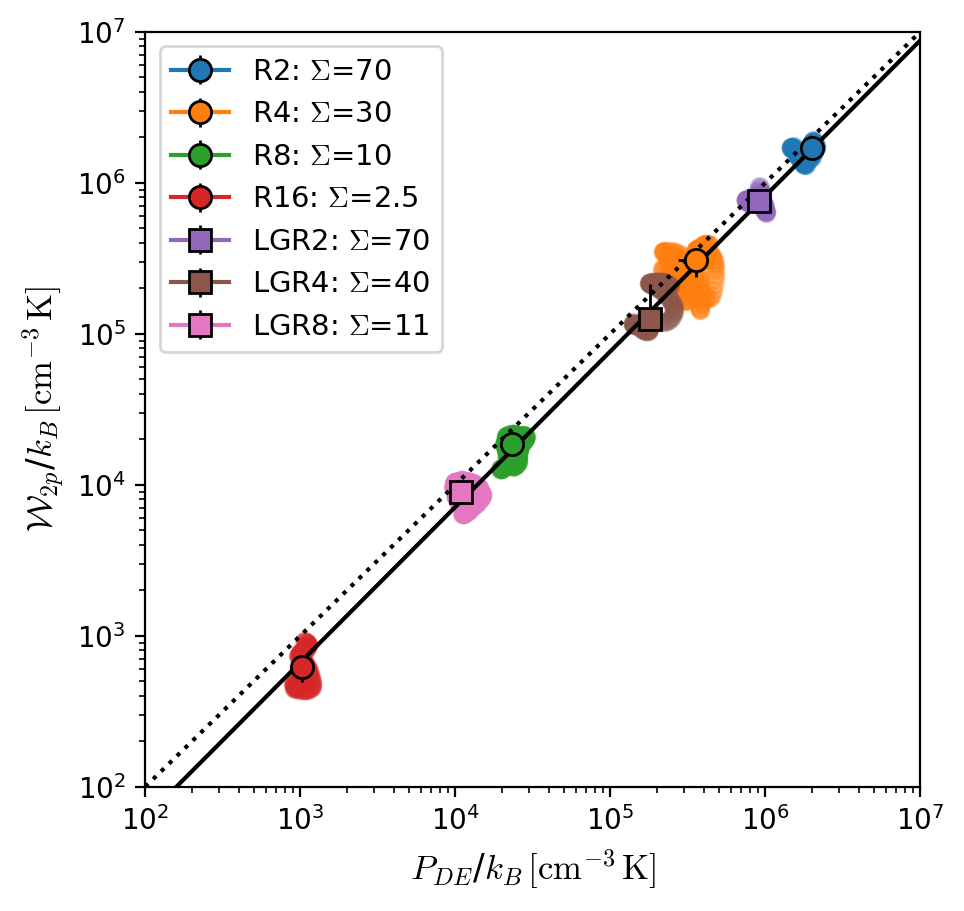}
\caption{Measured weight of warm-cold (two-phase)
  gas, ${\cal W}_{\rm 2p}$, vs. estimated 
  dynamical equilibrium weight, $\PDE$,
  for all models.  Individual points at intervals 1 Myr
  are plotted for each model,
  as well as medians with
  25$^{th}$ and 75$^{th}$ percentiles indicated.    For reference the dotted line shows the identity ${\cal W}_{2p}=P_{DE}$ while the solid line shows the best fit (see text). 
\label{fig:PDE_W2p}
}
\end{figure}

In our simulations, we can measure the weight at any point $(x,y)$ in
the midplane by vertical integration, following the definition
in \autoref{eq:weight}, and average horizontally.  This directly-measured
weight  can be compared to  the commonly-adopted estimator $\PDE$  for
the weight given in
\autoref{eq:PDEdef}, using $\overline{\sigma}_{\rm eff,2p}$.   
When we make this comparison
the weight is for the two-phase gas; in the simulation it is almost
the same for all gas since the hot gas mass is quite low.
\autoref{fig:PDE_W2p} shows that $\PDE$ is indeed an almost linear estimator
for the ISM weight,
\begin{equation}\label{eq:W_PDE_fit}
\log({\cal W}_{\rm 2p}/k_B) = 1.03\log(\PDE/k_B) -0.267,
\end{equation}
although for most models $\PDE$ is $\sim 30\%$ higher than $\cal W$
(the difference
is greater for model R16, driving the larger offset in \autoref{eq:W_PDE_fit}).  Here and in other relations based on fits
to the simulations, pressures and weights
are in units of $k_B \pcc \K$, i.e. we report $P/k_B$ or ${\cal W}/k_B$ in cgs units.  The result of \autoref{eq:W_PDE_fit}
demonstrates the validity of adopting $\PDE$
as a simple estimator, although at the same time shows that it cannot be
expected to recover the true weight to better than a few tens of percent.  

\begin{figure*}
\centering\includegraphics[width=0.8\textwidth]{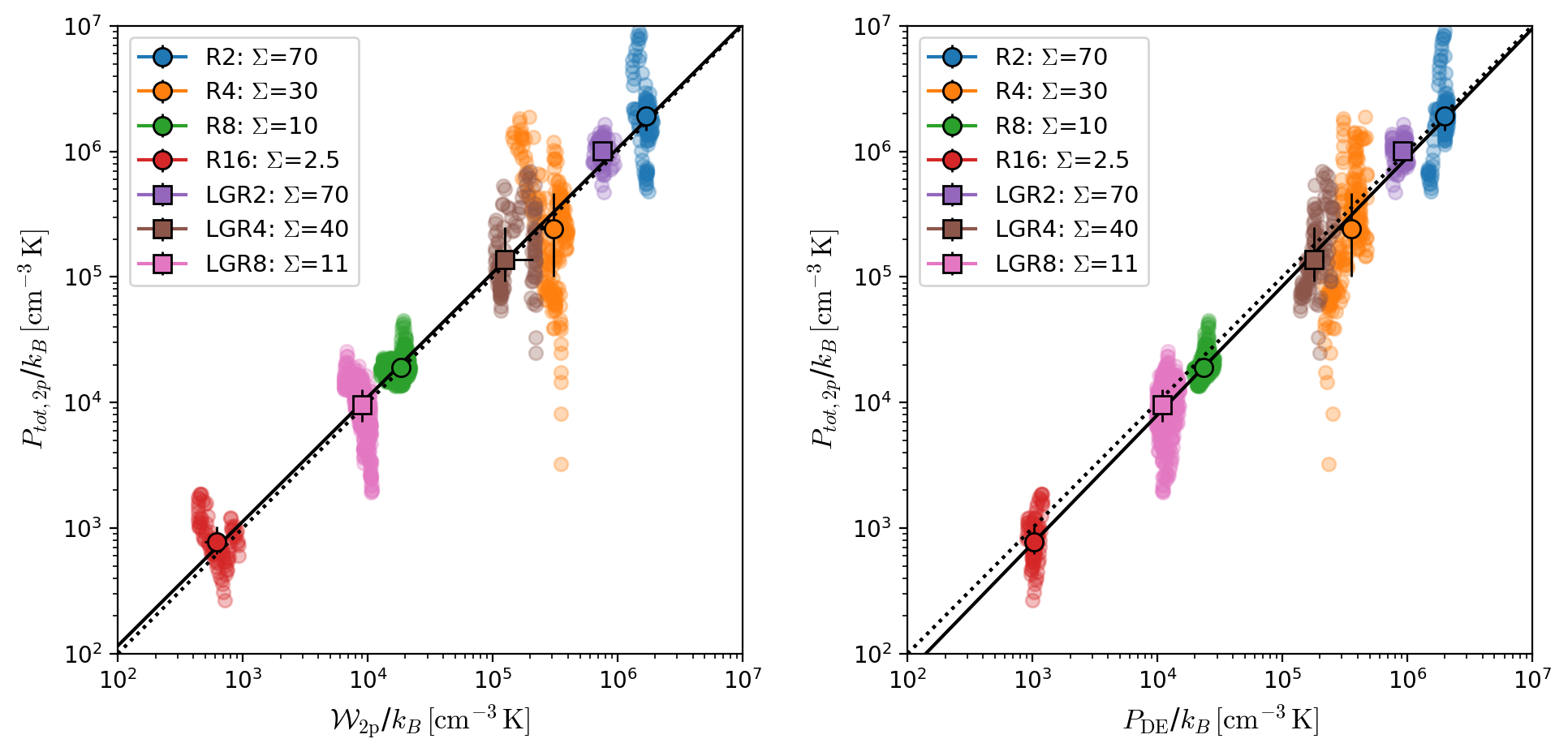}
\caption{Measured total midplane pressure of warm-cold (two-phase) gas, $P_{\rm tot,2p}$, vs.
  measured and estimated weight of gas, ${\cal W}_{\rm 2p}$ and $\PDE$,
  for all models.  Individual points at intervals 1 Myr
  are plotted for each model,
  as well as medians with
  25$^{th}$ and 75$^{th}$ percentiles indicated, with dotted lines showing
  the identity relation in each panel and
  the solid lines showing the best fits (see text). 
\label{fig:weight_pressure}
}
\end{figure*}

With our simulations, we can compare the midplane value of the total
pressure in
the two-phase gas to either the true ISM weight $\cal W$ or the estimator
$\PDE$, with the results shown in \autoref{fig:weight_pressure}.  Best-fit
relations,
\begin{subequations}
\begin{eqnarray}
\label{eq:P_vs_weight_fit}
  \log(P_{\rm tot,2p}/k_B) &=& 0.99\log({\cal W}_{\rm 2p}/k_B) + 0.083\\
\label{eq:P_vs_PDE_fit}
  \log(P_{\rm tot,2p}/k_B) &=& 1.03\log(\PDE/k_B) - 0.199,
\end{eqnarray}
\end{subequations}
show that vertical dynamical equilibrium is satisfied within a few tens of
percent.

\subsection{Pressure components}\label{sec:press_frac}

\begin{figure}
\centering\includegraphics[width=0.48\textwidth]{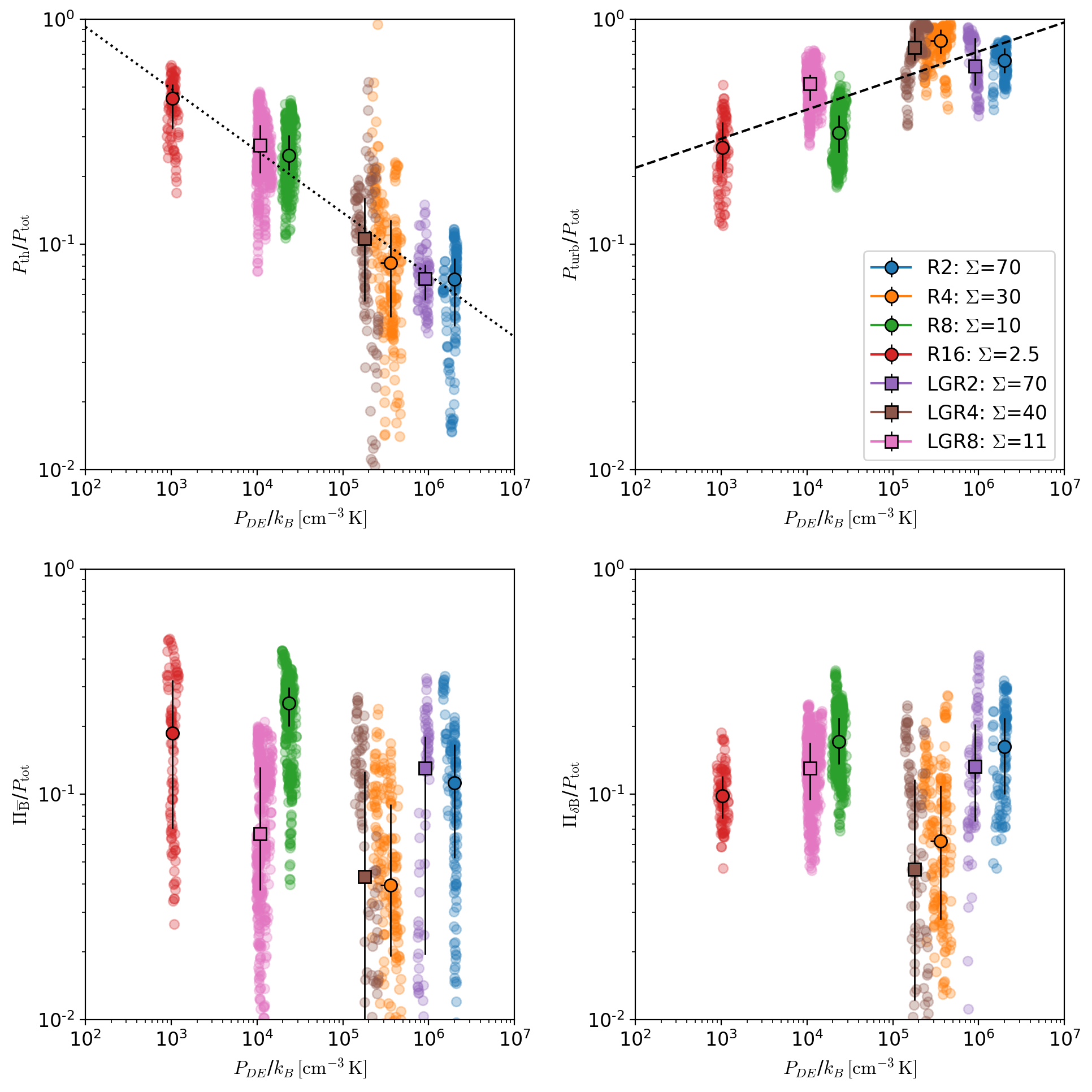}
\caption{Fractional contributions to the total pressure in warm-cold (``2p'') gas
  from thermal, turbulent, and magnetic (mean and perturbation) terms, as a function of $\PDE$.
  Individual points are plotted for each model at intervals 1 Myr,
  as well as medians with
  25$^{th}$ and 75$^{th}$ percentiles indicated.
  Best fit relations are also shown for the thermal and turbulent fractions
  (see text). 
\label{fig:P_components}
}
\end{figure}

In the solar neighborhood of the Milky Way, all pressure components
are observed to be roughly similar \citep[e.g.][]{Boulares_Cox1990}, and \autoref{fig:pressure_hist} shows that
this is also true for our R8 model, which adopts similar background
conditions of gas surface density and external gravitational
potential to the solar neighborhood.
Since the momentum input per SN is relatively
insensitive to the ambient density, we expect from \autoref{eq:Pturb_theory}
that the turbulent pressure
will scale nearly linearly with $\SSFR$. With the higher shielding (lower
$f_\tau$) in
regions of higher surface density (implemented in TIGRESS via
\autoref{eq:shielding}), the radiative heating rate for a
given SFR is reduced, and this is expected
to decrease the thermal
pressure relative to the turbulent pressure.  With magnetic pressure
driven by the dynamo, it is not expected to exceed the turbulent pressure, so
that overall $\Pturbtwo$ will be the largest single contributor to $\Ptottwo$.  
As a consequence, we expect $\Pthtwo/\Ptottwo$ will decrease and
$\Pturbtwo/\Ptottwo$ will increase slightly
for the models of higher surface density, which correspond to
higher pressure.   
In \autoref{fig:P_components} we show for all models the fractional
contributions to the total midplane pressure as a function of $\PDE$, along with
best-fit power law relations for the thermal and turbulent fractions:
\begin{subequations}
\begin{eqnarray}
\label{eq:Pth_frac}
\log\left(\frac{\Pthtwo}{\Ptottwo}\right)   = -0.275\log(\PDE/k_B) + 0.517\quad\quad&&\\
\label{eq:Pturb_frac}
\log\left(\frac{\Pturbtwo}{\Ptottwo}\right) =  0.129\log(\PDE/k_B) - 0.918.\quad\quad&&
\end{eqnarray}
\end{subequations}
The expected trends in the fractional contributions to the median pressure are
evident in these fits.  In addition,
\autoref{fig:pressure_hist} makes clear the overall decrease in
the time-dependent $\Pthtwo/\Ptottwo$ 
from R8 to R4 to R2, or LGR8 to LGR4 to LGR2.

Magnetic fields grow due to the combination of sheared rotation, turbulence, and
buoyancy in our simulations.  \autoref{fig:P_components} 
shows, for all models, the separate contribution to the total pressure from the
mean magnetic component $\Pi_{\overline B, {\rm 2p}}$,
which is computed using the mean magnetic field based on horizontal
averages, and the fluctuation component $\Pi_{\delta B, {\rm 2p}}$, which is computed
using the difference between the horizontally-averaged magnetic field and the
total. As with other quantities, the ``2p'' subscript indicates the average includes only zones containing warm and cold gas. In general, we find that the
magnetic field requires some time to grow, such that it has not necessarily
reached saturation for our simulations.  The exception is model R8,
which has a longer duration (in orbit times) than other models, showing saturation after $t>2\torb$ (see \citealt{Kim_CG2019}).
As previously found in the simulations of \citet{Kim_Ostriker2015b}
with sheared rotation, turbulence, and buoyancy but only two-phase gas,
the level of turbulent magnetic pressure remains
below the level of the turbulent kinetic pressure in our simulations.  
Since there are no clear relationships evident for relative importance
of magnetic pressure in different models (potentially for numerical reasons,
if magnetic growth is not saturated), we do not attempt to obtain fits
for magnetic contributions.

\begin{figure*}
  \centering\includegraphics[width=\textwidth]{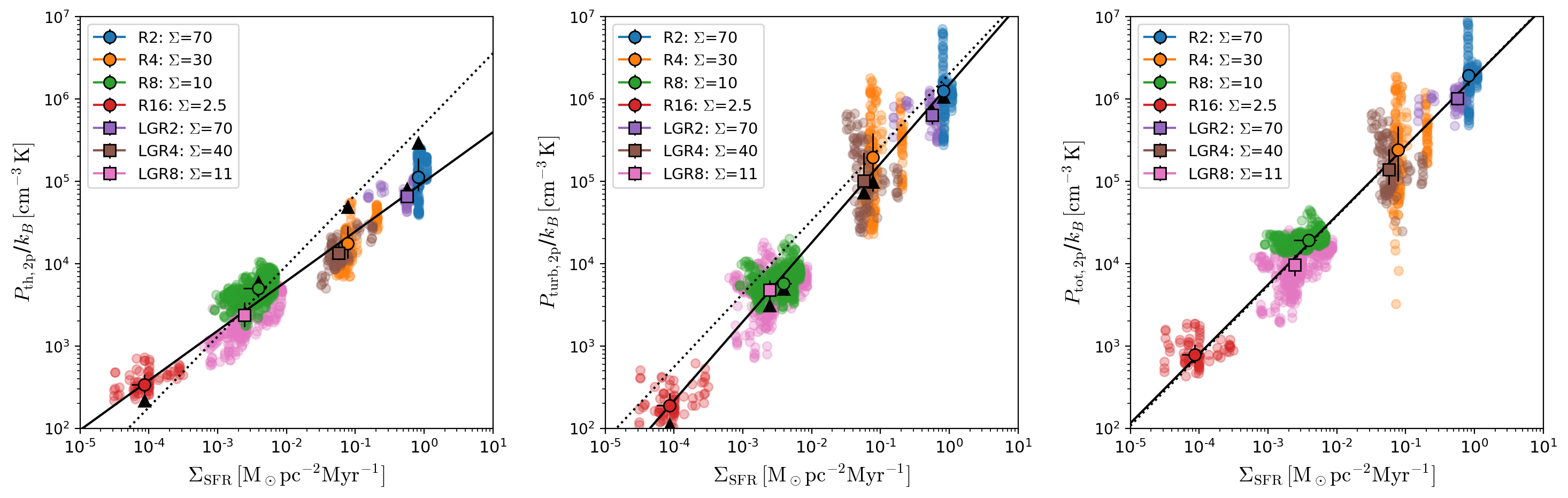}
  \caption{Thermal, turbulent, and total pressure as a function of $\SSFR$.
    Here, $\SSFR$ for each point is based on an average over the past 40 Myr
    in the simulation, and the pressures represent horizontal midplane
    averages.  Median values and 25$^{th}$ and 75$^{th}$ percentiles are
    also shown for each model. Solid lines show best-fit power laws, 
    dotted lines show results from \citetalias{KOK_2013}, and black triangles show the
    analytic predictions for thermal and turbulent pressure (see text).   
\label{fig:P_vs_SFR}
  }
\end{figure*}

We have demonstrated above that the ISM pressure is \textit{regulated} 
in disk systems as it obeys certain ``rules'' that follow
from conservation laws of energy and momentum, in particular the thermal
pressure
must be consistent with a balance between heating and cooling
(for short cooling time), and
the total pressure 
consistent with balancing the weight of the gas (in an average sense), with comparable total midplane pressures in hot and two-phase gas.
At the same time, pressure responds to the star formation rate.
The input FUV flux $\Sigma_{\rm FUV}$ responsible for heating atomic gas scales
linearly with $\SSFR$, but as noted above the
increasing attenuation of FUV in higher surface density regions
(smaller $f_\tau$ in \autoref{eq:shielding})
implies
a relative reduction in the photoelectric heating rate. With the equilibrium
thermal pressure following $P_{\rm th}/k_B = \Gamma T/\Lambda$ for $T/\Lambda$
roughly constant and $\Gamma \propto \SSFR f_\tau$,
the thermal pressure is expected to scale
sublinearly with $\SSFR$.
As discussed in \autoref{sec:theory},
since the momentum injection per SN $p_*$ is relatively insensitive to ambient
conditions, the turbulent pressure driven by SNe is expected to scale 
approximately linearly with the rate of SNe per unit area per unit time
in the disk, leading to an approximately linear relation between $\Pturb$
and $\SSFR$  (\autoref{eq:Pturb_theory}).  
Since the turbulent pressure is the largest single component of $\Ptot$, 
the total pressure is also expected to be roughly linear in $\SSFR$. 

\autoref{fig:P_vs_SFR} shows
the relation between the measured midplane $\Pthtwo$, $\Pturbtwo$, and 
$\Ptottwo$ with $\SSFR$.  The best fit power law relations are:
\begin{subequations}
\begin{eqnarray}
\label{eq:Pth_SSFR}
\log(\Pthtwo/k_B)   &=& 0.603\log(\SSFR) + 4.99\\
\label{eq:Pturb_SSFR}
\log(\Pturbtwo/k_B) &=& 0.960\log(\SSFR) + 6.17\\
\label{eq:Ptot_SSFR}
\log(\Ptottwo/k_B) &=&  0.840\log(\SSFR) + 6.26
\end{eqnarray}
\end{subequations}
with $\SSFR$ in $\sfrunit$; we overlay these fits as solid lines.  

Given the mean value of $\SSFR$ and attenuation factor, we can obtain
$J_{\rm FUV}$, and for our adopted heating and cooling functions
this then leads to a characteristic value for the equilibrium
thermal pressure $P_{\rm two}\equiv (P_{\rm max,warm} P_{\rm min,cold})^{1/2}$ 
given in \autoref{eq:Ptwo}.  For each model, we
show these reference equilibrium values as black triangles in \autoref{fig:P_vs_SFR}. 
The reference
values are slightly above the mean values in the simulation for the
models with higher $\Sgas$ and $\SSFR$, which is not surprising given
the shorter cooling times in these high density models.  Similarly,
the black triangles in the turbulent pressure panel show the prediction
of \autoref{eq:Pturb_theory} assuming a characteristic value
$p_*=10^5 \Msun \kms$ for each SN 
(with the value $m_*=95.5 \Msun$ adopted by TIGRESS).  The numerical
results from TIGRESS follow this prediction quite well overall,
with values consistent with mean $p_* \sim 1.3 \times 10^5 \Msun \kms$.  
The implied values of $p_*$ increase slightly at low $\SSFR$  
presumably due to the slight increase 
in $p_*$ in conditions of lower ambient density (as predicted by theory and 
idealized simulations,
e.g.~\citealt[][and references therein]{Kim_Ostriker2015a,Kim_Ostriker_Raileanu2017}).

Also overlaid in \autoref{fig:P_vs_SFR}
are the corresponding fits from \citetalias{KOK_2013}
(respectively their Eqs. 20 and 22 for $\Pth$, their Eqs. 21 and 23
    for $\Pturb$, and their Eq. 26 for $\Ptot$). 
In \citetalias{KOK_2013}, the heating
rate was taken as simply linear in $\SSFR$, yielding a steeper power law
slope for thermal pressure ($\Pth \propto \SSFR^{0.86}$) than found from
TIGRESS, which takes into account radiation attenuation (albeit in an
approximate fashion).  In \citetalias{KOK_2013}, SN
feedback was realized via direct momentum input to the simulation in
the region surrounding the source, with a constant momentum value
per SN of
$p_* = 3 \times 10^5 \Msun \kms$.  Here, for most SN events we instead
inject energy such that the Sedov-Taylor stage of the SNR remnant
is directly captured and the momentum injection is determined by the
SNR expansion rate when cooling/shell formation occurs
given the conditions in the ambient environment.
While we do not measure the momentum injection directly for each SN
in the TIGRESS simulations, idealized simulations suggest that
the mean value is likely somewhat lower than the value adopted by \citetalias{KOK_2013} (see
also below).
This would explain why the TIGRESS normalization of $\Pturb$ vs $\SSFR$ is
also slightly lower than that in \citetalias{KOK_2013}, although
the slope is quite similar {($0.96$ here vs.~$0.89$ in
  \citetalias{KOK_2013})}.
The TIGRESS relationship between the total midplane (vertical) pressure and 
$\SSFR$ is almost identical to that found by \citetalias{KOK_2013}.

\begin{figure*}
  \centering\includegraphics[width=\textwidth]{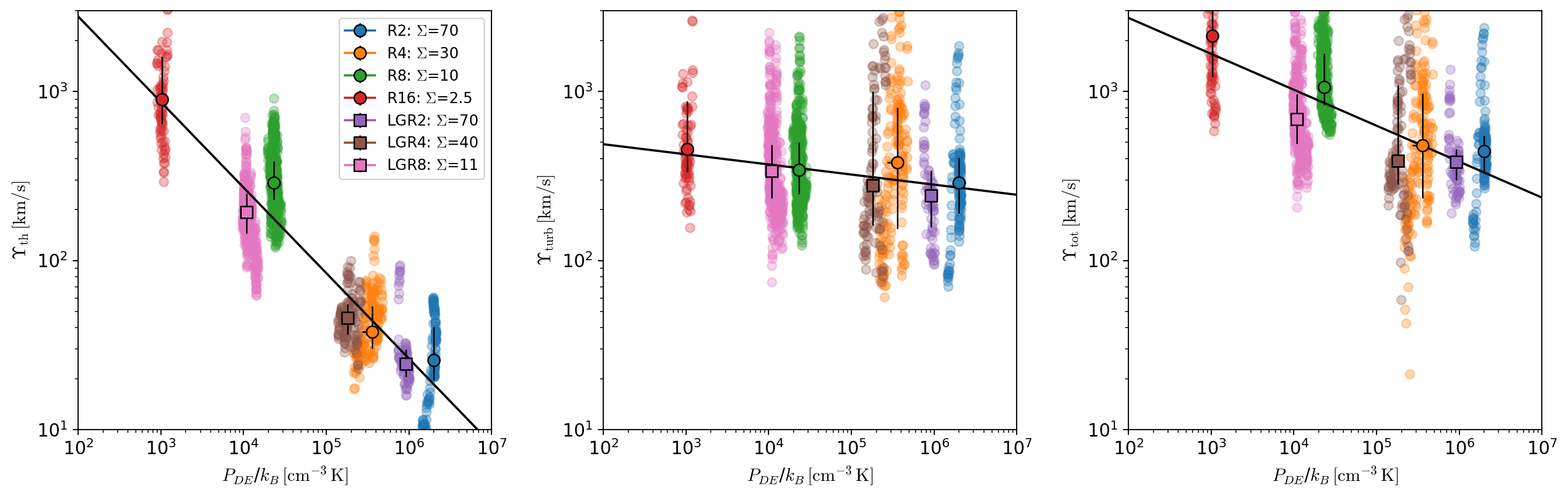}
  \caption{Thermal, turbulent, and total feedback  yield as a function of
    $\PDE$. Median values and 25$^{th}$ and 75$^{th}$ percentiles are
    also shown for each model. Solid lines show best-fit power laws
(see text).   
\label{fig:Ups_vs_PDE}
  }
\end{figure*}

\subsection{Feedback yields}\label{sec:yields}

The quantitative \textit{modulation} of individual
pressure components by feedback can be characterized by the \textit{yield}
parameters $\Upsilon \equiv P/\SSFR$. Since pressure has units of
momentum/time/area and $\SSFR$ has units of mass/time/area, the natural unit
for the feedback yield is a velocity.  Due to the shielding of radiation in
regions of high surface density, $\Upsilon_{\rm th}$ decreases with
increasing $\SSFR$, whereas
$\Upsilon_{\rm turb}$  is relatively flat because SN momentum input is
insensitive to environment. Since turbulent and magnetic terms
are at least as large as
thermal terms  and are relatively insensitive to ambient conditions, 
the total yield only decreases slightly
at higher star formation rate.
Fits to the TIGRESS simulations give
\begin{subequations}
\begin{eqnarray}
  \label{eq:Ups_th}
\Upsilon_{\rm th}&=& 110{\rm \, km\, s^{-1}} \left(\frac{\SSFR}{0.01{\sfrunit}}\right)^{-0.4}\\
\label{eq:Ups_turb}
\Upsilon_{\rm turb} &=& 330{\rm \, km\, s^{-1}} \left(\frac{\SSFR}{{0.01\sfrunit}}\right)^{-0.05}\\
\label{eq:Ups_tot}
\Upsilon_{\rm tot} &=& 740{\rm \, km\, s^{-1}} \left(\frac{\SSFR}{0.01{\sfrunit}}\right)^{-0.18}
\end{eqnarray}
\end{subequations}  
where for convenience we normalize here 
relative to typical areal star formation rates in nearby
galaxies. 
Comparing \autoref{eq:Ups_turb}
to \autoref{eq:Upsturb_theory} 
implies that the effective value of momentum/mass from SN injection in 
TIGRESS is $(p_*/m_*)_{\rm eff} = 1300 \kms (\SSFR/0.01\sfrunit)^{-0.05}$.

We may compare the yields for TIGRESS
to the corresponding relations reported in
\citetalias{KOK_2013}: 
$\Upsilon_{\rm th}= 200{\rm \, km\, s^{-1}} \left({\SSFR}/0.01{\sfrunit}\right)^{-0.14}$,
$\Upsilon_{\rm turb} = 700{\rm \, km\, s^{-1}} \left({\SSFR}/{0.01\sfrunit}\right)^{-0.11}$, and
$\Upsilon_{\rm tot} = 770{\rm \, km\, s^{-1}} \left({\SSFR}/{0.01\sfrunit}\right)^{-0.15}$;
we have converted units for most direct comparison (in  addition to
different units for yield, \citetalias{KOK_2013} use the 
notation $\eta$ instead of $\Upsilon$).  The weaker
dependence of $\Upsilon_{\rm th}$ in \citetalias{KOK_2013}
is because shielding of radiation
was not included, while the larger coefficient for $\Upsilon_{\rm turb}$ is
because a (constant) value $p_* = 3 \times 10^5 \Msun \kms$ was adopted,
which is larger than $(p_*/m_*)_{\rm eff}$ in TIGRESS.  
In spite of these differences, the relation between
$\Upsilon_{\rm tot}$ and $\SSFR$ from TIGRESS
is almost the same as that from \citetalias{KOK_2013}, due to the inclusion of magnetic
fields in TIGRESS.  We note that while the \citetalias{KOK_2013} models were unmagnetized,
the \citet{Kim_Ostriker2015b} MHD simulations for a model
representing the solar neighborhood (similar
to R8) found similar thermal and turbulent pressure to those in \citetalias{KOK_2013},
with $\Pi_{\rm mag} \sim 0.7 P_{\rm turb}$.  Thus, inclusion of magnetic fields
happens to compensate for the
lower $(p_*/m_*)_{\rm eff}$  from SNe in TIGRESS
compared to the value imposed in \citetalias{KOK_2013}. 

\begin{figure*}
  \centering\includegraphics[width=\textwidth]{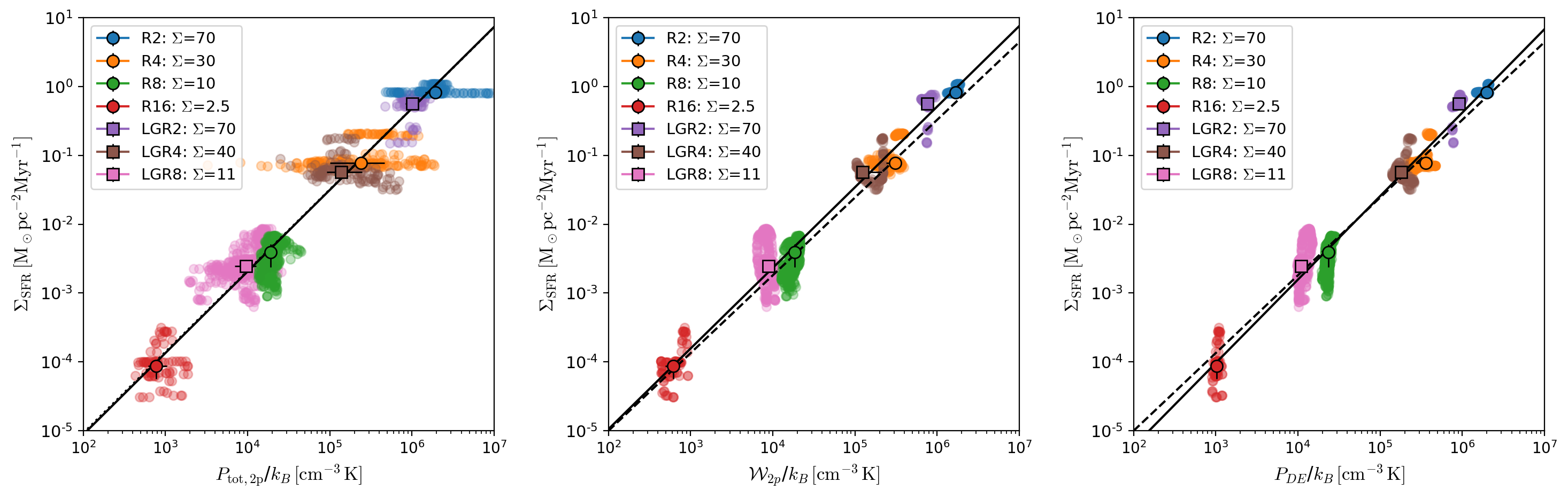}
  \caption{$\SSFR$ (40Myr average)
    as a function of measured total midplane pressure $P_{\rm tot,2p}$,
    measured ISM weight  ${\cal W}_{2p}$, and estimated weight $\PDE$. 
    Individual points from 1 Myr intervals as well as median values and
    25$^{th}$ and 75$^{th}$ percentiles from the sampling interval are
    also shown for each model. Solid lines show best-fit power laws and
    dotted and dashed lines show results from \citetalias{KOK_2013}
(see text).   
\label{fig:SFR_vs_P}
  }
\end{figure*}

Since $\PDE$ can be obtained relatively easily from observable large-scale
parameters in star-forming galaxies (see \autoref{eq:PDEdef}), it is also
useful to see how the feedback yields depend on $\PDE$.
\autoref{fig:Ups_vs_PDE} shows both individual points at 1 Myr intervals, and
medians from the distribution over the sampling interval.  Here, we use
a trailing 40 Myr average for $\SSFR$ in the denominator of each computed
value of $\Upsilon$.  Since midplane pressures have considerable fluctuations,
there is significant scatter in $\Upsilon$ values for each model, but
clear correlations are evident.  The best-fit power law relations shown as
solid lines in  \autoref{fig:Ups_vs_PDE} are:
\begin{subequations}
\begin{eqnarray}
\label{eq:Upsth_PDE}
\log(\Upsilon_{\rm th})  &=& -0.506 \log(\PDE/k_B) + 4.45\\
\label{eq:Upsturb_PDE}
\log(\Upsilon_{\rm turb})  &=& -0.060 \log(\PDE/k_B) + 2.81\\
\label{eq:Ups_PDE}
\log(\Upsilon_{\rm tot}) &=& -0.212 \log(\PDE/k_B) + 3.86
\end{eqnarray}
\end{subequations}
where $\PDE/k_B$ is in $\pcc \K$ as before, and $\Upsilon$ is in $\kms$.  
Evidently, $\Upsilon_{\rm tot}$ is expected to range between a few thousand $\kms$
in low-pressure, far outer-galaxy  regions (or low surface brightness diffuse
disks or dwarfs) to a few hundred $\kms$ in the
high pressure regions surrounding galactic centers (or starburst regions
elsewhere).  This decrease in $\Upsilon$ 
reflects the increasing efficiency of radiative losses in higher pressure
interstellar
environments to the energy that has been deposited in them by stars.

\subsection{Star formation-pressure relations}\label{sec:SFR}

The two main theoretical principles invoked by \citetalias{OML_10} and
\citetalias{OS_11} are (1)
the star-forming ISM constantly adjusts to keep its pressure in equilibrium
with its weight, and (2) in an
equilibrium  state, the supply of energy from massive, luminous stars must
match the demand for energy to resupply the ISM's continual losses.  Under
the simplifying assumption of constant feedback yield this principle
leads to a prediction that $\SSFR$ varies linearly with the ISM weight
${\cal W}\approx \PDE$.  The TIGRESS simulations have validated the
theoretical principles of pressure balancing weight and energy
supply matching demand, but have also demonstrated that the feedback
yield is not constant.  We may still express the equilibrium
star formation rate in terms of the equilibrium weight using the feedback
yield as $\SSFR = \PDE/\Upsilon_{\rm tot}$, but
because $\Upsilon_{\rm tot}$ decreases with increasing $\SSFR$ or $\PDE$
(see \autoref{eq:Ups_tot} or \autoref{eq:Ups_PDE}), the prediction for pressure-regulated,
feedback-modulated star formation is that $\SSFR$ will increase superlinearly
with pressure.

\autoref{fig:SFR_vs_P} shows $\SSFR$ vs. the measured midplane pressure in
the warm-cold gas ($P_{\rm tot,2p}$), the measured ISM weight (${\cal W}_{2p}$),
and the estimated ISM weight ($\PDE$).  The best-fit power law relations
overlaid in the figure are
\begin{subequations}
\begin{eqnarray}
\label{eq:SFR_P}
\log(\SSFR)  &=&  1.18 \log(P_{\rm tot,2p}/k_B) -7.43 \\
\label{eq:SFR_W}
\log(\SSFR)  &=&  1.17 \log({\cal W}_{2p}/k_B) -7.32 \\
\label{eq:SFR_PDE}
\log(\SSFR) &=&   1.21 \log(\PDE/k_B) -7.66.
\end{eqnarray}
\end{subequations}
As expected, these relations are slightly superlinear, and almost the
same for the three versions of the pressure.  In the figure,  we also show
for reference the fits reported in \citetalias{KOK_2013}, respectively
$\log(\SSFR)  =  1.18 \log(P_{\rm tot,2p}/k_B) -7.4$ (left panel) and 
$\log(\SSFR)  =  1.13 \log(\PDE/k_B) -7.3$ (center and right panels).  In
spite of having a far more complex model of the ISM than in \citetalias{KOK_2013},
the TIGRESS simulations
show quite similar results for the relationship between star formation
and pressure as the earlier simulations.  {This validates the conceptual
  and practical simplifications in modeling the 
  ISM  adopted in  \citetalias{KOK_2013}, in particular the assumption
  that the main role of SNe in modulating star formation is driving turbulence
  in the warm-cold gas via momentum injection  during radiative SNR stages.}

\subsection{Effective equation of state and velocity dispersion}\label{ssec:EOS}

\begin{figure}
\includegraphics[width=0.49\textwidth]{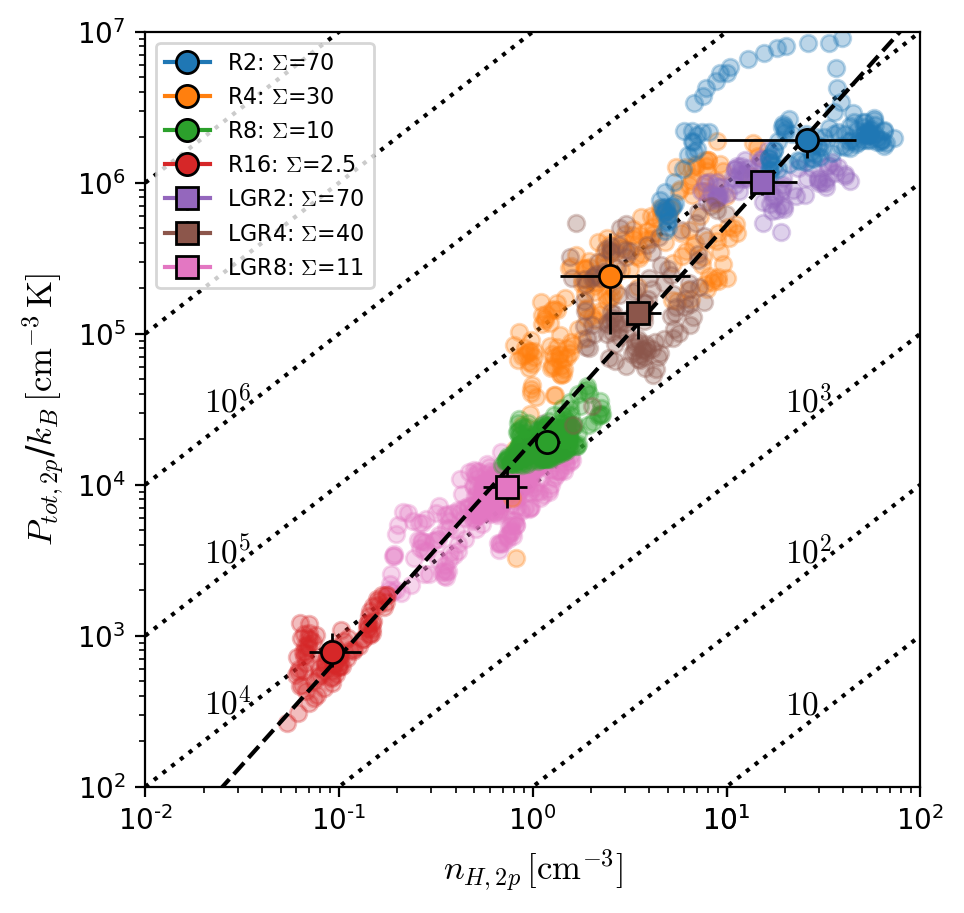}
\caption{Total pressure $P_{\rm tot} \equiv \Pth + \Pturb + \Pimag$ vs. hydrogen density $n_{\rm H}$ in two-phase gas for all models.
  Midplane-averaged values at intervals of 1 Myr are shown with individual
  small circles, together with medians and 25$^{th}$ and 75$^{th}$ percentiles
  from the sampling interval shown as large points. The best-fit power law
  with slope 1.43 is
  shown as a dashed line.  Dotted lines indicate isotherms
  of $T_{\rm eff} = P_{\rm tot,2p}/(n_{\rm H,2p} k_B) = 
  1.7  \times 10^4\K (\sigma_{\rm eff, 2p}/10\kms)^{2}$. 
\label{fig:EOS}  
}
\end{figure}

In situations where it is either  not necessary or not possible to
follow the detailed thermodynamics of a gaseous system, an effective equation
of state is often adopted that relates the gas pressure and density.  We
can use the results from our TIGRESS simulations to propose an effective
equation of state for star-forming interstellar
gas.  We base this effective equation of state on the fitted relationship between midplane averages of the pressure and density in the warm-cold ISM, which represents the majority of the mass; 
this gas is subject to its own internal  thermodynamics and MHD, and to intricate interactions with hot gas. Both pressure and density are responsive to  inhomogeneous and intermittent energy injection by feedback.
\autoref{fig:EOS} shows pressure and density from individual snapshots,
median values for each model, and the best-fit power-law relationship
\begin{equation}\label{eq:EOS_fit}
\log(P_{\rm tot,2p}/k_B) = 1.43\log(n_{\rm H,2p}) + 4.30.
\end{equation}  
Interestingly, the exponent 1.43 in this pressure-density relation
exceeds the minimal value (4/3) required for spherical polytropes to be stable
\citep[e.g.][]{Bonnor_1958}.
{Isotherms
  of $T_{\rm eff} = P_{\rm tot,2p}/(n_{\rm H,2p} k_B)$ are also shown in the figure;
  $T_{\rm eff}\sim 10^4 - 10^5\K$ for midplane density between
  $n_{\rm H,2p} \sim 0.1 -10 \pcc$ for our suite of TIGRESS simulations.}

As discussed in \autoref{sec:evolution}, the 
ISM gas is characterized by an effective vertical velocity dispersion $\seff^2=\Ptot/\rho$, which takes into account turbulent, thermal, and magnetic stress terms.  If we consider just the two-phase gas at the midplane,  the fit in \autoref{eq:EOS_fit} 
corresponds to an effective midplane velocity dispersion 
$\sigma_{\rm eff,2p}\equiv[P_{\rm tot,2p}/(1.4 m_H n_{\rm H,2p})]^{1/2} = 9.8  \kms [P_{\rm tot,2p}/(10^4 k_B \pcc \K)]^{0.15}$, ranging over
$\sim 7-20 \kms$ for our set of models.   
If instead we consider contributions from warm-cold gas over the whole volume, 
the result is $\sim 30\%$ higher.
These values range
from $\sim 10-40 \kms$ for individual models, listed as $\overline{\sigma}_{\rm eff, 2p}$ in \autoref{tbl:properties}.  
This mass-weighted mean effective velocity dispersion increases with higher pressure 
following 
$\overline{\sigma}_{\rm eff,2p} = 12 \kms [\PDE/(10^4 k_B \pcc \K)]^{0.22}$ for $\PDE/k_B > 10^4 k_B \pcc \K$.

\subsection{Observational comparisons}\label{sec:obs}

\begin{figure}
\centering\includegraphics[width=0.48\textwidth]{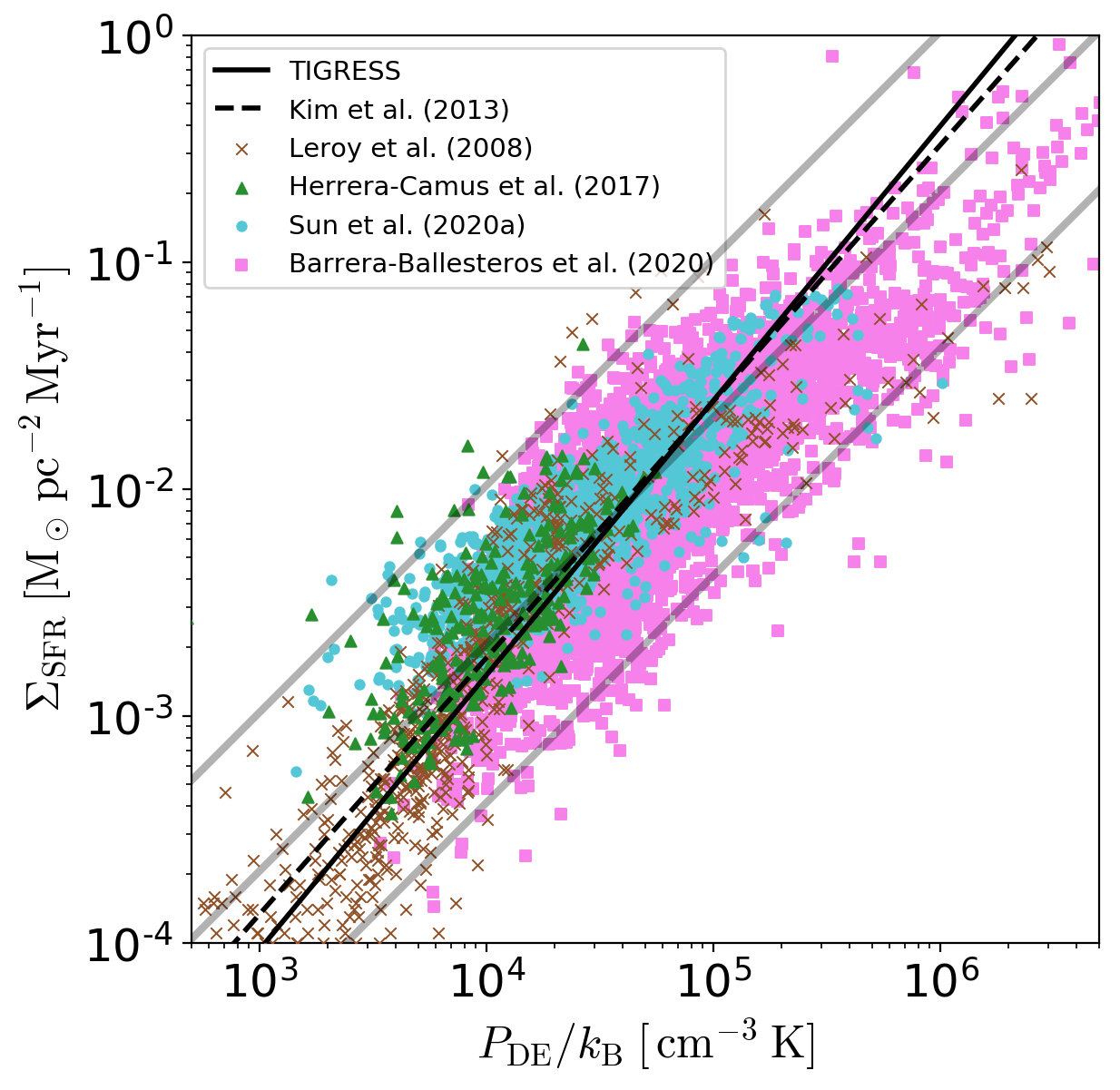}
\caption{ $\SSFR$ as a function of estimated weight $P_{\rm DE}$, comparing
  TIGRESS numerical results (\autoref{eq:SFR_PDE}, solid, as shown in
  \autoref{fig:SFR_vs_P}) to observations from several recent surveys of
  galaxies resolved at $\sim \kpc$ scale.  Observational results shown are from
  \citet{Leroy2008,Herrera-Camus_2017,Sun_2020a,Barrera-Ballesteros_2021}.
  The previous
  numerical result from \citetalias[][Eq. 27]{KOK_2013} is also
  shown (dashed). 
  Also overlaid for reference is \autoref{eq:obs_unit} with 
  constant $\Upstot=200,\ 1000,\ 5000\kms$  (light gray lines, top to bottom).
\label{fig:obs}  
}
\end{figure}

As discussed in \autoref{sec:intro_obs}, there have been several previous studies
comparing the predictions of PRFM star formation to observations, starting
with the \citetalias{OML_10} and \citetalias{OS_11} papers.  
In particular, observational
surveys with $\sim{\rm kpc}$-scale resolution  (or in the case of
PHANGS, higher resolution averaged over $\sim{\rm kpc}$ scales) have shown
that there is a near-linear relationship between $\PDE$ and $\SSFR$, with
coefficient consistent with theory and numerical simulations.  A compendium
of observational results based on $\sim{\rm kpc}$ patches from
\citet{Leroy2008,Herrera-Camus_2017,Sun_2020a,Barrera-Ballesteros_2021}
is shown in \autoref{fig:obs}.  For all of these works, $\PDE$ is computed
as in \autoref{eq:PDEdef}; readers are referred to the original
publications for details on the assumptions made in obtaining estimates
of $\Sgas$, $\rho_\mathrm{sd}$, $\sigma_{\rm eff}$, and $\SSFR$  from observables.
Overall, the different surveys show
quite similar results, although there do appear to be some systematic
differences.  

From \autoref{eq:SFR_pred}, the theoretical prediction is that the mean SFR per unit area in the disk will be related to the 
dynamical equilibrium midplane pressure (\autoref{eq:PDEdef}) via the
total feedback yield $\Upsilon_\mathrm{tot}$; this may be written
in commonly adopted units as 
\begin{equation}\label{eq:obs_unit}
 \frac{\SSFR}{\small \sfrunit}
  = 2.07\times 10^{-4}
  \frac{\PDE/k_B \, {\scriptsize [\pcc \K] }}{\Upstot\, {\scriptsize [\kms]}}.
  \end{equation}
The theoretical expectation (see \autoref{sec:yield_theory}) is that
$\Upstot \sim 1000 \kms$ from a combination of thermal, turbulent kinetic,
and magnetic contributions for solar neighborhood conditions, decreasing
a few tens of percent in inner disks where shielding of radiation reduces
the thermal pressure contribution.  In \autoref{fig:obs} we show
\autoref{eq:obs_unit} with $\Upstot = 200,\ 1000,\ 5000 \kms$; evidently,
$\Upstot = 1000 \kms$ characterizes the center of the
observed distribution well, consistent with the theoretical expectation.  

\autoref{fig:obs} overlays on the data
the best-fit power-law relationship between
$\PDE$ and $\SSFR$ from our suite of TIGRESS multiphase ISM simulations
(see \autoref{eq:SFR_PDE} and 
\autoref{fig:SFR_vs_P}).  Additionally, we show the fit
previously obtained by \citet{KOK_2013}, based on simulations of
warm-cold gas in which effects of SNe were treated by fixed momentum
injection.  Since the two sets of simulations have only slight differences
in $\Upstot$, the resulting predicted $\SSFR$ is also quite similar, and
both are in good agreement with the observations.  The decrease in
the feedback yield in higher-density, higher-pressure conditions makes
the numerical relations superlinear, with the numerical fits crossing
the $\Upstot=1000\kms$ line near the center of the observed distribution.
This corresponds to slightly higher-pressure, higher-$\SSFR$ conditions
than the solar neighborhood.  

\section{Summary and Discussion}\label{sec:summary}

In this paper, we investigate the co-regulation of star formation and
ISM properties in disk galaxies for a range of conditions
representative of the nearby Universe.  Our goal is to test the PRFM
theory first enunciated in \citetalias{OML_10} and \citetalias{OS_11}, which
posits that the star formation rate and mean pressure in the
multiphase ISM are intimately linked through the energetic feedback
provided by high mass stars, and both can be quantitatively predicted
via simple considerations of thermal and dynamical equilibrium.
The key physical concept in the PRFM theory is that the same midplane pressure
that balances the vertical weight of the ISM ``atmosphere'' must
be equal to the sum of individual pressures
derived from
considerations of balance between energy and/or momentum injection by
stellar feedback into the ISM, and losses from the ISM.  Previously, in
\citetalias{KKO_2011,KOK_2013}; and \citet{Kim_Ostriker2015b}, we used numerical
simulations focused just on the two-phase ISM to test these ideas and
to compute feedback yields, defined as $\Upsilon_i = P_i/\SSFR$ for $P_i$
representing thermal, turbulent, or magnetic pressure.

In the
present work, we use a set of seven TIGRESS MHD simulations to
provide further numerical tests of the PRFM principles -- now with numerical
models that
include a hot ISM component produced by correlated SNe, 
and to  measure the pressure components and feedback yields.  The simulations we
employ represent horizontal patches ranging in size from $(512\pc)^2$ to
$(2048\pc)^2$, with a vertical dimension 7 times as large, and minimum
physical resolution in the range $2-8\pc$. Each simulation is run  for at
least 1.5 orbits at the corresponding galactic radius.

\subsection{Summary of key numerical results}

The main conclusions from analysis of our simulations are as follows:

\paragraph{Quasi-steady state}
In all simulations, a quasi-steady state is reached after a few tenths of a galactic orbital time (\autoref{fig:pressure_hist}).
In this quasi-steady state, the SFR fluctuates
temporally, leading to fluctuations of feedback and pressure. Because feedback
is extended in time over the lifetimes of massive stars
(several tens of $\Myr$), temporal pressure fluctuations  
have lower amplitudes compared to those in $\SSFR$. 
 In general, the ISM includes all three phases
of gas (hot, warm, and cold).   The hot gas is produced by repeated
shocks from correlated SN, with the resulting
superbubbles expanding preferentially
in the vertical direction to create chimneys where hot galactic winds
are vented.  In the warm and cold phases, the loci
of highest occupation in the pressure-density phase plane
(\autoref{fig:Pth_vs_n}) follow the
thermal equilibrium curve set by the instantaneous heating rate, but there is
non-negligible occupation of the out-of-thermal-equilibrium regime due to
dynamical effects.

\paragraph{Multiphase pressure equilibrium}
Hot gas and two-phase (warm+cold) gas reach a state of
approximate mutual pressure equilibrium.
Medians of $P_{\rm tot,hot}$ and
$P_{\rm tot,2p}$ are within 50\% of each other at the midplane of the disk, in
all models   (\autoref{fig:P2p_Phot}).  The models with ``inner-galaxy''
conditions (higher $\Sgas$ and $\rho_*$) have systematically
higher emergent $\Ptot$ and $\SSFR$.   
For the two-phase gas, the ratio $\Pth/\Ptot$ declines and $\Pturb/\Ptot$
increases in the higher-pressure, inner galaxy models
(\autoref{eq:Pth_frac}, \autoref{eq:Pturb_frac}),
with the former due to the increased shielding applied for photoelectric
heating at higher $\Sgas$.  Magnetic pressure
can become comparable to 
turbulent pressure in the warm-cold gas (at the midplane),
but is negligible in the hot gas.
In the hot phase, thermal pressure and Reynolds stress are comparable overall,
although thermal exceeds turbulent pressure at the midplane.

\paragraph{Vertical force balance}
The midplane total pressure is in vertical dynamical
equilibrium with the weight of the ISM (\autoref{fig:weight_pressure}), as
an immediate consequence of quasi-steady state for the vertical component of
the momentum equation.
While the total pressure $\Ptot \equiv \Pth + \Pturb + \Pimag$
has larger variations in time than the weight $\cal W$ (defined in
\autoref{eq:weight}), the mean values are within 15\% of each other (except for model R16, where the difference in means is within 30\%).  The simple 
form $\PDE$ given in \autoref{eq:PDEdef}, commonly used in observations,
provides an excellent estimate of $\cal W$. Although $\PDE$ is
consistently larger than
$\cal W$, the difference is within $20-30\%$, except for model R16.   Thus,
the mean value of $\PDE$ is within $10-40\%$ of the mean midplane total
pressure  $\Ptot$.  It is important to recognize that $\PDE$ should not be thought of as 
an ``external'' force/area that acts on clouds.
Rather, $\PDE$ is equal to the statistical average of the total pressure (the
vertical component of the momentum flux)
over the gas at the midplane.  Since the nonthermal (turbulent, magnetic)
stresses are dominated by large scales that may exceed the sizes of
individual clouds, these stresses cannot be thought of as ``surface'' terms on individual clouds.

\paragraph{Feedback yields}
Numerical results for the thermal, turbulent, and total feedback
yields {$\Upsth$, $\Upsturb$, and $\Upstot$} are given as a function of $\SSFR$ in
\autoref{eq:Ups_th}-\autoref{eq:Ups_tot}, or as a function of $\PDE$
in \autoref{eq:Upsth_PDE}-\autoref{eq:Ups_PDE}.  Consistent with the
hypothesis of \citetalias{OML_10}, the set of measured thermal yields
$\Upsth \equiv \Pth/\SSFR$ for our simulations is close to expected
values based on thermal equilibrium with pressure equal to the
``two-phase'' value ($\Pth = P_{\rm two}\propto \SSFR f_\tau$ from
\autoref{eq:Ptwo} and \autoref{eq:JFUV})
for our adopted photoelectric heating rate and
cooling functions; this is equivalent to \autoref{eq:Upsth_theory} for
the \citet{Wolfire_2003} heating and cooling functions for atomic gas.
The turbulent yield $\Upsturb \equiv \Pturb/\SSFR$ measured in the
simulations is consistent with the prediction of \citetalias{OS_11} as
given in \autoref{eq:Upsturb_theory} with $p_*/m_* \sim 1300 \kms$
averaged over models, and less than a factor $2$ decrease in $\Upsturb$ from
outer-to-inner disk models.  The weak variation in
$\Upstot  \equiv \Ptot/\SSFR$ with galactic conditions
($\Upstot \propto \SSFR^{-0.2} \propto\PDE^{-0.2}$) reflects the fact that
the largest component of the pressure
in the ISM is $\Pturb$, and the net momentum injection per SN event is
insensitive to environment (as previously demonstrated in idealized
simulations).  We note that the total yield $\Upstot$ agrees to better than 10\% with that previously reported
in \citetalias{KOK_2013}.

\paragraph{$\SSFR$ - pressure relation}
There is a nearly linear relationship between $\SSFR$ and midplane
pressure:
\autoref{eq:SFR_P}, \autoref{eq:SFR_W}, \autoref{eq:SFR_PDE} respectively
give the best-fit power law relations between $\SSFR$ and $\Ptot$, $\cal W$,
and $\PDE$, which follow the same scalings 
($\SSFR \propto P^{1.2}$) and have very similar coefficients.  The relationship
between $\SSFR$ and $\Ptot$ 
reflects the role of feedback in setting physical
ISM pressures  (through the yields),
while the relationship between $\SSFR$ and $\cal W$ (or $\PDE$)
reflects both the role of feedback and vertical dynamical equilibrium.
The weak decrease of total feedback yield in higher-pressure (higher
density) environments
explains why  
$\SSFR = \PDE/\Upstot$ is slightly superlinear.  Quantitatively, we find
essentially the same relation between $\SSFR$ and $\PDE$ as previously
reported in \citetalias{KOK_2013} (see comparison in \autoref{fig:SFR_vs_P}).
The relation between $\SSFR$ and $\PDE$ is both the most important
physical concept and the most useful practical result of the PRFM theory,
because it provides a quantitative prediction for star formation given
the basic gas and stellar properties of a galactic disk (see \autoref{fig:obs} for theory/observation comparison).

\paragraph{Effective equation of state.}  From the measured pressure and density averaged over the two-phase gas, we obtain an effective equation of state for star-forming gas as given  in \autoref{eq:EOS_fit}, which has $P \propto \rho^{1.43}$.  The measured relationship encodes the total effective velocity dispersion $\sigma_{\rm eff}^2 = \Ptot/\rho$ for the warm-cold gas.
Based on our fit to the set of TIGRESS simulations presented here, 
this corresponds to 
$\sigma_{\rm eff,2p} = 10 \kms [P/(10^4 k_B \Punit)]^{0.2}$ where $P$ is either the measured 
midplane total pressure $\Ptot$ or the estimated 
gas weight $\PDE$. 
This (or similar) effective equation of state relation can
be combined with \autoref{eq:PDEdef} to obtain $\PDE$ and $\sigma_{\rm eff}$ solely as a function of $\Sgas$ and $\rho_{\rm sd}$.

\subsection{Discussion and prospects}

\paragraph{{Application to galaxy formation modeling}}
As recapitulated above, the individual elements of the PRFM theory are
clearly validated by the numerical results obtained with our suite of
TIGRESS simulations.  We regard this as a success, and on
the basis of this we encourage use of the theory in modeling where the
detailed properties of the ISM and of star formation on $\sim$pc scales 
cannot be directly resolved, but the gaseous and stellar content on
larger ($\sim 10^2 -10^3$pc) scales are known.  Given the gas surface density
$\Sgas$ and the stellar plus dark matter volume density $\rho_{\rm sd}$,
the predicted star formation rate is obtained using \autoref{eq:SFR_pred} for
$\SSFR$
(or \autoref{eq:t_dep} for $t_{\rm dep}=M_{\rm gas}/\dot M_*$)
with \autoref{eq:PDEdef} for $\PDE$ and \autoref{eq:Ups_PDE} for
$\Upsilon_{\rm tot}$.  

Cosmological galaxy
formation simulations and semi-analytic models are an obvious use case for application of the PRFM model and of our numerical calibrations of $\Upstot$ and $\sigma_{\rm eff}$.  In particular,  
while zoom simulations  may resolve the vertical scale 
height and therefore the mean density of the ISM, this is 
generally not true in large-box cosmological simulations.   Nevertheless,
the local gas surface density $\Sgas$ can be computed by integrating through the disk, and this may be combined 
with (resolved) stellar plus dark matter density in the disk to obtain the local 
$\PDE$.  With $\PDE$ in hand, our results provide a prediction
for $\tdep$ that can be used in setting cell-by-cell values of the SFR, given the gas mass.

{Potentially, generalized versions of the effective equation of
  state we derive from our simulations (as reported in
  \autoref{eq:EOS_fit}), could be adopted for star-forming gas in
  cosmological simulations.  The pressure-density relation we report
  here quantitatively lies between the one that has been adopted for
  the Illustris(TNG) simulations
  \citep{Springel_2003,Robertson_2004,Vogelsberger_2013,Pillepich_2018} and the one
  adopted for the EAGLE simulations \citep{Schaye_2008,Schaye_2015}.  Adoption of an equation
  of state derived from TIGRESS would therefore 
  be unlikely to introduce significant differences in overall disk
  stability if it were adopted for cosmological simulations,
  but would have the merit of being calibrated from resolved
  multiphase MHD simulations. To obtain an effective equation of state
  that is suitable for modeling galaxy formation over all of cosmic time,
  however, it
  will be necessary to generalize the results obtained here with lower
  metallicity TIGRESS simulations.}

\paragraph{{Parameter dependence and functional form of $\tdep$ and $\SSFR$
} }
The local star formation rate in a galaxy can be broken down into
available gaseous ``fuel'' and a timescale at which that fuel is converted to
stars; these combine to make $\SSFR=\Sgas/\tdep$ for a disk system. 
We emphasize that the gas and stars are equally
important in setting $\tdep$,
as given in \autoref{eq:t_dep}.  That is,
$\Sgas$ and
$\rho_{\rm sd}=\rho_* + \rho_{\rm dm}$ (dominated by the stars) enter
mathematically and physically 
on an equal footing in $\tdep$ via the vertical dynamical time $\tdyn$.
This can be contrasted
with the commonly adopted assumption that the relevant timescale for
star formation is the free-fall time $t_{\rm ff}$, proportional
to the inverse square root of gas 
density.\footnote{Only in the case where
  the gas dominates the vertical gravity are $t_{\rm ff}$ and $\tdyn$ nearly  the same,
  both
  $\sim 0.5 \sigma_{\rm eff}/(G\Sgas)$.  In  the limit where stars dominate,
  $\tdyn \sim (2 G \rho_*)^{-1/2}$ while
  $t_{\rm ff} \sim 0.5 G^{-3/4}(\sigma_{\rm eff}/\Sgas)^{1/2} \rho_*^{-1/4}$.}
In practice, the gas and stars are usually of comparable
importance in determining the vertical
gravitational field on $\sim$kpc scales within the main ISM layer; even if the global galactic gas fraction is high, the stellar and gas vertical gravity may be comparable. 
Since {the combined gravitational field  from gas, stars, and dark matter}  controls the ISM pressure
(and scale height), 
it determines the necessary star formation and feedback required to
maintain equilibrium.

It is worth noting that 
$\SSFR = \PDE/\Upstot$ does not functionally correspond to
a power-law dependence of
$\SSFR$ on $\Sgas$, or even a product of power laws in $\Sgas$ and $\Sigma_*$,
unless one or the other of the two terms in $\PDE$ (\autoref{eq:PDEdef})
dominates.  Moreover, the thickness of the stellar disk $h_*$
is just as important as $\Sigma_*$, since from \autoref{sec:weight}
it is the midplane
stellar volume density $\rho_*=\Sigma_*/(2h_*)$ rather than $\Sigma_*$
that controls the gas pressure when $h_{\rm gas} < h_*$ (the usual case).  
The effective vertical gas velocity dispersion $\sigma_{\rm eff}$
is similarly important because
it enters in setting the gas disk's half-thickness
$h_{\rm gas}$ (see \autoref{eq:hgas}) and therefore
controls the gas pressure (see \autoref{eq:W_from_h} or \autoref{eq:PDEdef}).
While several different
properties are required to fully describe  the local galactic environment,
the specific parameter combination embodied by the equilibrium
pressure has a special physical significance.  By referring explicitly
to {\text pressure} in the ``PRFM'' moniker, we underscore this point.  

Even though the dependence of $\SSFR$ is 
on the combination of variables in $\PDE$ rather than as a power law $\SSFR \propto \Sgas^{1+p}$, an apparent power law relation can arise observationally for a number of reasons.  For example, if the vertical gravity is dominated 
by the gas and kinetic turbulence dominates the pressure, as may be the case in starburst regions, $\SSFR = \PDE/\Upstot \rightarrow \pi G \Sgas^2/(2 \Upsturb)$ for $\Upsturb$ of a few $100 \kms$ (center panel of \autoref{fig:Ups_vs_PDE}). As previously noted in \citetalias{OS_11} and \citet{Narayanan_2012}, the approximate relation $\SSFR \propto \Sgas^2$ in this regime could appear as a slope between 1 and 2 in $\log \SSFR$  vs. $\log W_{\rm CO}$, since the decrease of $\alpha_{\rm CO}\equiv \Sigma_{\rm mol}/W_{\rm CO}$ in higher-excitation gas \citep[see][and references  therein]{Gong_2020} means that $\Sgas$ increases sublinearly  with $W_{\rm CO}$.  This may explain 
the power law with $p=0.4$ identified 
by \citet{Kennicutt1998}, which adopted constant $\alpha_{\rm CO}$ \citep[see also][]{Kennicutt2021}.  Similarly, a power law dependence 
$\SSFR \propto \Sgas \Sigma_{*}^{0.5}$  
\citep[e.g.][]{Shi2011,Shi2018}
could describe normal galaxies if their
vertical gravity is dominated by 
the stellar component and there is 
limited variation in $h_*$.

\paragraph{{Enhanced observational testing}}
Observational tests to date at $\sim\kpc$-scale resolution, as summarized from previous work in
\autoref{sec:intro} and directly compared with our new simulation results in
\autoref{sec:obs}, show good agreement with the PRFM model.  For the future,
it will be especially valuable to refine the empirical measurements of
parameters that enter in $\PDE$.  As noted above, the midplane density of
the old stellar disk, $\rho_*$, is needed to obtain $\PDE$. This requires
knowledge of both the total surface density of old stars,
$\Sigma_*$, and the effective half-thickness $h_*$ of the stellar layer.
For face-on galaxies, $h_*$ cannot be directly
measured, and a common practice {for normal spiral galaxies} \citep[following][]{vanderKruit_1982}
has been to assume a
constant stellar disk thickness proportional to the radial exponential
scale length $R_s$ of the old stellar disk, with $\rho_*=\Sigma_*/(0.54 R_s)$
\citep[e.g.][]{Leroy2008,OML_10,Herrera-Camus_2017,Sun_2020a,Barrera-Ballesteros_2021}.  However, as noted in \citetalias{OML_10}, the choice {$h_*=0.27 R_s$} may in fact
overestimate $h_*$, leading to an underestimate for $\rho_*$.
Also, stellar disks may flare with radius
\citep{deGrijs_1997,Narayan_Jog2002,Momany_2006,Lopez_2014,Lopez_2020}.  
{An alternative to adopting an empirical relationship between $h_*$ and $R_s$ is to estimate $h_*$ from dynamical considerations.  For example, a very simple estimate adopts $h_*=\sigma_{*,z}^2/(\pi G \Sigma_*)$ for $\sigma_{*,z}$ the effective vertical velocity dispersion of the old stellar disk, based on the \citet{Spitzer_1942} solution for an isothermal plane-parallel system; in this case (and assuming the dark matter contribution is negligible) the second term in \autoref{eq:PDEdef} becomes $\sqrt{\pi} G \Sgas \Sigma_* \sigma_{\rm eff}/\sigma_{*,z}$. However, improved estimates for the vertical stellar distribution can be obtained by including  the influence of gas and dark matter gravity in confining the stellar disk and solving the coupled system of Jeans equation and hydrostatic equation for stars and gas, respectively, and allowing for radial and vertical gradients in velocity dispersions \citep[e.g.][]{Sarkar_Jog2018,Sarkar_Jog2020a,Sarkar_Jog2020b}}.  
A path forward to more accurate values of $h_*$
  would be to seek statistical relationships between observed measures of the stellar disk thickness and
other stellar properties from edge-on disk galaxies, while simultaneously using
synthetic observations of simulated edge-on disks to calibrate the
true $h_*$ in terms of observables (including testing sensitivity to
dust extinction).

Empirical measurements of $\PDE$ require the total gas surface density
$\Sgas$ and the effective vertical velocity dispersion $\sigma_{\rm eff}$.
Especially in the central regions of galaxies, improved calibrations of
$\alpha_{\rm CO}$, e.g. making use of two or more rotational 
lines to allow for varying
excitation \citep{Gong_2020}, will aid in obtaining more accurate $\Sgas$
from CO emission.
The effective vertical velocity dispersion $\sigma_{\rm eff}$ includes magnetic contributions
that are difficult to measure empirically; since these are likely to
scale with the (more easily observable) kinetic turbulence, however,
this can be accounted for at lowest order with a simple multiplicative factor.
It is important to note that
the closest observable proxy of
the kinetic contribution to $\sigma_{\rm eff}$
is a {\it mass-weighted} value.  Thus, it must be derived from observations
of the atomic and molecular gas that together comprise most of the ISM's mass.
While ionized gas velocity dispersions are sometimes
more readily available, especially
for high-redshift galaxies, these sample expanding \ion{H}{2} regions
and diffuse ionized gas; since the motions of ionized gas are not
in general representative of the neutral ISM 
\citep[ionized linewidth are typically larger, e.g.][]{Girard_2021}, linewidths of
H$\alpha$ or other tracers of ionized gas  should not be used as a proxy to estimate 
the kinetic contribution to $\sigma_{\rm eff}$.
Even when CO and \ion{H}{1} lines are spectrally resolved,
it is difficult to correct for spiral-arm streaming
and other in-plane motions that can contaminate measurement of the vertical
velocity dispersion, so face-on systems provide the most reliable targets.

\paragraph{{Further model development}}
The suite of TIGRESS simulations analyzed here represents a significant advance
in resolved modeling of the star-forming, multiphase, magnetized ISM.
Nevertheless, the TIGRESS implementation employed for this suite 
(as described in \citealt{Kim_Ostriker2017}) has
limitations that could affect our results quantitatively, if not qualitatively.
First, a simple fitted cooling function and fixed FUV heating efficiency
are adopted.  Rather than
directly following the FUV photons responsible for photoelectric heating
via radiative transfer, we also adopt a simple analytic attenuation formula.
In addition, we do not follow ionizing radiation (or other
``early feedback'') from young clusters,
which is known to strongly affect the immediate environment of forming stars.  
Extending beyond these simplifying assumptions is an important direction
for future work, {and efforts along these lines have already begun.
Namely, we have extended our TIGRESS framework to include adaptive ray tracing
radiative transfer \citep[as implemented by][]{Kim_JG2017} in  order 
to follow FUV as well as ionizing radiation produced
by the star cluster particles that form, and we have also
implemented more sophisticated heating and cooling functions based on 
nonequilibrium photochemistry
(J.-G. Kim et al 2022, submitted).  Initial simulations with this new
framework produce results that are generally quite
consistent with those reported here, and in particular turning photoionizing
radiation on or off changes $\SSFR$ by $\lesssim 30\%$ (C.-G. Kim et al 2022, in preparation).}
Another potential concern is that with our current implementation
of sink particles -- representing stellar
clusters with a coeval stellar population,
star formation may be somewhat more correlated in space and time than
is realistic, which could quantitatively affect certain results
\citep[e.g. for galactic wind power, as seen in 
  simulations by][]{Smith_2021}.
Testing sensitivity
to this correlation, as well as exploring alternatives to the
current sink particle approach, are also 
important directions for future work.

The TIGRESS simulations analyzed here assume solar metallicity, but
it is of much interest to investigate how higher or lower
metallicity would affect the results.
We have every expectation that the PRFM theory will hold in some generality,
but quantitative calibration of feedback yields at 
low metallicity are needed for realistic application to high-redshift
galaxies.  By extending the range of simulated systems, not
just in metallicity but to environments with much higher and lower gas
surface density, with deeper and shallower stellar potentials,
with global galactic as well as local frameworks (while still resolving
all phases of the ISM and feedback effects), it will be possible to test
the general validity of the PRFM theory.
{In strongly disturbed systems (tidal encounter and merger systems), as
  well as galaxies at early stages before disk rotation dominates other
  kinetic and thermal energy \citep[e.g.][]{Gurvich_2022},
  it remains to be seen whether a modified version of the PRFM theory
  (generalizing to quasi-spherical rather than disk geometry) can be
used to predict the SFR. This could only be expected if feedback is the main
source resupplying losses of the forms of energy that combine to support against gravity.
From dimensional analysis, the expected gas depletion time would be $\tdep \sim 
\Upstot /g$ for $g$ the
relevant gravitational field that is supported by terms with rapid losses ($g_z$ for a disk, or $g_r$ for a quasi-spherical system).} 
Our own analyses of TIGRESS model extensions --
with spiral arm potential perturbations in mid-disk environments
\citep{Kim_WT_etal2020}, and of bar-fed star-forming rings in galactic
center regions \citep{Moon_etal2021,Moon_2022} -- have already corroborated
the PRFM principles. It would be straightforward for other groups
to apply the same kind of analysis to their own simulations with
physics implementations similar to TIGRESS and a fully resolved multiphase ISM,
in order to test these principles further.

\begin{acknowledgements}

This work was supported in part by NASA ATP grant No. NNX17AG26G and by grant No. 510940 from  the Simons Foundation to E.~C. Ostriker. 
We are grateful to Jiayi Sun and Jorge Barrera-Ballesteros for sharing
observational data sets, {and to the referee for helpful comments on the
manuscript.}

\end{acknowledgements}

\bibliographystyle{aasjournal}
\bibliography{ref_revised}

\end{document}